\def\twomat#1#2#3#4{\left(\begin{array}{cc}
 {#1}&{#2}\\ {#3}&{#4}\\
\end{array}
\right)}

\def\eqdef{\mathrel{\mathop{\;=\;}^{\rm def}}}

\let \part\partial
\let \ka\kappa

\def\tr{{\rm Tr}}

\def\o#1#2{{#1\over#2}}

\def\bz{\bar z}

\def\un{{\bf 1}}

\def\pa{\partial}

\def\La{\Lambda}

\def\emp4{e^{-i \o{\pi}{4}}}
\def\cG {{\cal G}}

\def\tr{{\rm Tr}}

\def\pa{\partial}

\def\e2pi{e^{2\pi i}}
\def\em2pi{e^{- 2\pi i}}

\let \part\partial
\let \ka\kappa

\def\um{\o {1}{2}}

\def\o#1#2{{{#1}\over{#2}}}
\def\bz{{\bar z}}

\def\un{{\bf 1}}

\def\cG {{\cal G}}

\def\eq{\,=\,}

\def\um{{\scriptstyle {\o{1}{2}}}}

\def\em{e^{-}}

\def\pa{\partial}

\def\La{{\Lambda}}

\def\Xi{X^{i}}

\def\cAp{\cA_{\ssp}}
\def\cAm{\cA_{\ssm}}
\def\alB{\alpha^{(B)}}
\def\commut#1#2{[#1,#2]}

\def\dim{{\rm dim}}

\def\tr{{\rm Tr}}

\def\cF{{\cal F}}\def\cG{{\cal G}}
\def\cL{{\cal L}}\def\cM{{\cal M}}\def\cN{{\cal N}}
\def\cP{{\cal P}}\def\cQ{{\cal Q}}
\def\cS{{\cal S}}\def\cZ{{\cal Z}}
\def\quat{{\bf H}}
\def\eqdef{\stackrel{\rm def}{=}}
\def\twomat#1#2#3#4{\left(\matrix{#1&#2\cr #3&#4\cr}\right)}

\def\o#1#2{{#1\over#2}}
\def\bz{\bar z}
\def\um{{1 \over 2}}
\def\un{{\bf 1}}
\def\bx{{\bf X}}

\def\r{{\bf R}}
\def\c{{\bf C}}

\def\hika{HyperK\"ahler~}
\def\ka{K\"ahler~}

\newcommand {\cJ}[1]{{\cal J}^{#1}}
\newcommand{\cJt}[3]{{\cal J}^{#1}_{#2#3}}
\newcommand{\epsi}[3]{\varepsilon^{#1#2#3}}

\def\al{\alpha}
\def\alp{\alpha^{'}}
\def\alA{\alpha^{(A)}}
\def\alg{\alpha_g}

\def\f#1#2#3{f^{#1 #2 #3}}
\def\twomat#1#2#3#4{\left(\matrix{#1&#2\cr #3&#4\cr}\right)}

\def\un{{\bf 1}}
\def\tr{{\rm Tr}}
\def\dag#1{{#1}^{\dagger}}
\def\dags#1#2{{#1}^{#2\dagger}}
\def\Zt{{\tilde Z}}

\def\psit{{\tilde\psi}}
\def\ps#1{\psi^{#1}}
\def\pss#1{\psi^{{#1}^*}}
\def\pst#1{{\tilde\psi}^{#1}}
\def\psts#1{{\tilde\psi}^{{#1}^*}}
\def\lap{\lambda^{\scriptscriptstyle +}}
\def\latp{{\tilde\lambda}^{\scriptscriptstyle +}}
\def\lamm{\tilde\lambda^{\scriptscriptstyle -}}
\def\latm{{\tilde\lambda}^{\scriptscriptstyle -}}
\def\mup{\mu^{\scriptscriptstyle +}}
\def\mutp{{\tilde\mu}^{\scriptscriptstyle +}}
\def\mum{\mu^{\scriptscriptstyle -}}
\def\mutm{{\tilde\mu}^{\scriptscriptstyle -}}
\def\latp{\lambda^{\scriptscriptstyle +}}
\def\lap{{\tilde\lambda}^{\scriptscriptstyle +}}
\def\latm{\lambda^{\scriptscriptstyle -}}
\def\lam{{\tilde\lambda}^{\scriptscriptstyle -}}
\def\mutp{\mu^{\scriptscriptstyle +}}
\def\mup{{\tilde\mu}^{\scriptscriptstyle +}}

\def\mum{\mu^{\scriptscriptstyle -}}
\def\mutm{{\tilde\mu}^{\scriptscriptstyle -}}
\def\psu#1{\psi_{\scriptscriptstyle u}^{#1}}
\def\psus#1{\psi_{\scriptscriptstyle u}^{{#1}^*}}
\def\psut#1{{\tilde\psi}_{\scriptscriptstyle u}^{#1}}
\def\psuts#1{{\tilde\psi}_{\scriptscriptstyle u}^{{#1}^*}}
\def\psv#1{\psi_{\scriptscriptstyle v}^{#1}}
\def\psvs#1{\psi_{\scriptscriptstyle v}^{{#1}^*}}
\def\psvt#1{{\tilde\psi}_{\scriptscriptstyle v}^{#1}}
\def\psvts#1{{\tilde\psi}_{\scriptscriptstyle v}^{{#1}^*}}
\def\zep{\zeta^{\scriptscriptstyle +}}
\def\zetp{{\tilde \zeta}^{\scriptscriptstyle +}}
\def\zem{\zeta^{\scriptscriptstyle -}}
\def\zetm{{\tilde \zeta}^{\scriptscriptstyle -}}
\def\cgp{\chi^{\scriptscriptstyle +}}
\def\cgtp{{\tilde \chi}^{\scriptscriptstyle +}}
\def\cgm{\chi^{\scriptscriptstyle -}}
\def\cgtm{{\tilde \chi}^{\scriptscriptstyle -}}
\def\eplus{e^{\scriptscriptstyle +}}
\def\eminus{e^{\scriptscriptstyle -}}
\def\dell{\nabla}
\def\delp{\nabla_{\scriptscriptstyle +}}
\def\delm{\nabla_{\scriptscriptstyle -}}
\def\dep{\partial_{\scriptscriptstyle +}}
\def\dem{\partial_{\scriptscriptstyle -}}
\def\cA{{\cal A}}\def\cD{{\cal D}}
\def\cF{{\cal F}}\def\cL{{\cal L}}\def\cP{{\cal P}}\def\cQ{{\cal Q}}
\def\cW{{\cal W}}
\def\cM{{\cal M}}\def\cN{{\cal N}}\def\cU{{\cal U}}
\def\o#1#2{{#1\over #2}}
\def\eqdef{\stackrel{\rm def}{=}}
\def\twomat#1#2#3#4{\left(\matrix{#1&#2\cr #3&#4\cr}\right)}
\def\coco{{\rm \hskip 2pt c.c.\hskip 2pt}}
\def\ssp{{\scriptscriptstyle +}}
\def\ssm{{\scriptscriptstyle -}}
\def\qu#1#2{q^{#1}_{\hskip 3pt #2}}
\def\etasu#1#2{\eta^{#1{#2}^*}}
\def\etagiu#1#2{\eta_{#1{#2}^*}}
\def\La{\Lambda}\def\Lat{\tilde{\Lambda}}
\def\o#1#2{{#1 \over #2}}
\documentstyle[a4,12pt,titlepage]{article}
\begin{document}
\begin{titlepage}
\begin{flushright}
SISSA/151/93/EP
\end{flushright}
\centerline{\LARGE N=4 VERSUS N=2 PHASES}
\centerline{\LARGE HYPERK\"AHLER QUOTIENTS}
\centerline{{\LARGE and THE 2D TOPOLOGICAL TWIST}
\footnote{Research supported in part by MURST and by the EEC
under the SCIENCE project contract named "Gauge Theories,
Applied Supersymmetry and Quantum Gravity"}}
\vspace{1truecm}
\centerline{\Large Marco Bill\'o and  Pietro Fr\`e}
\centerline{International School for Advanced Studies (SISSA/ISAS)}
\centerline{Via Beirut 2, 34014 Trieste, Italy}
\centerline{and INFN, Sezione di Trieste.  }
\vspace{0.3truecm}
\centerline{\large {\bf Abstract}}
\vspace{0.3truecm}

{\small We consider the rheonomic construction of N=2 and N=4 supersymmetric
gauge theories in two-dimensions,
coupled to matter multiplets. In full analogy with the N=2 case studied by
Witten, we show that
also in the N=4 case one can introduce Fayet-Iliopoulos terms for each of the
abelian factors of the
gauge group. The three-parameters of the N=4 Fayet-Iliopoulos term have the
meaning of momentum-map
levels in a HyperK\"ahler quotient construction just as the single parameter
of the N=2 Fayet-Iliopoulos term has the meaning of momentum map level
in a \ka quotient construction. Differently from the N=2 case, however, the N=4
has a single
phase corresponding to an effective $\sigma$-model. The Landau-Ginzburg phase
possible in the
N=2 case seems to be deleted in those N=2 theories that have an enhanced N=4
supersymmetry.
The main application of our N=4 model is to an effective Lagrangian
construction of a
$\sigma$-model on ALE-manifolds or other gravitational instantons.
\par
 We discuss in detail the
topological twists of these theories (A and B models) emphasizing the role of
R-symmetries and
clarifying some subtleties, not yet discussed in the literature, related with
the redefinition
of the ghost number and the identification
of the topological systems  after twisting. In the A-twist, we show that one
obtains
a topological matter system (of the topological $\sigma$-model type)
coupled to a topological gauge
theory.
In the B-twist, instead, we show that the theory describes
a topological matter system (of the topological Landau-Ginzburg
type) coupled to an ordinary (non-topological) gauge-theory:
in addition one has a massive
topological vector, which decouples from the other fields. Applying our results
to the case of
ALE-manifolds we indicate how one can use the topologically twisted theories to
study
the K\"ahler class and complex structure deformations of these gravitational
instantons.
\par
Our results are also preparatory
for a study of matter coupled topological 2D-gravity as the twist of matter
coupled N=2, D=2
supergravity.}

\end{titlepage}

\section{Introduction}
\label{intro}
In this paper, we make a detailed comparison of two-dimensional N=2 and N=4
supersymmetric
gauge theories coupled to matter multiplets, discussing their phase-structure
and their relation
both with the geometrical constructions known as \ka or \hika quotients and
with two-dimensional
topological field-theories \cite{topolfield}. Our interest in the subject was
motivated by a recent
paper
by Witten \cite{Wittenphases}, who analyzed the N=2 case and showed how, by
means of such a gauge
theory, one can interpolate between an N=2 Landau-Ginzburg model
\cite{topolLGliterature} and
an N=2 $\sigma$-model on
a compact Calabi-Yau manifold. This
interpolation has important implications for the understanding of the
corresponding topological
versions of the two N=2 theories and for the structure of the N=2
superconformal model that
emerges at their critical point.
\par
We wanted to investigate the same problem in the N=4 case, having in particular
in mind the N=4
$\sigma$-model on those \hika manifolds that can be obtained as \hika quotients
of flat-manifolds.
This situation was naturally suggested by our previous investigation
\cite{newalenostro}
of the N=4 superconformal
field theories associated with four-dimensional gravitational instantons, in
particular the ALE
manifolds \cite{gravinstanton}. Specifically a natural question that arose in
connection with these
spaces was the following: {\it what is the Landau-Ginzburg phase for a
$\sigma$-model on such
non-compact \hika manifolds that can be described as affine algebraic varieties
in ${\bf C}^N$,
rather than projective algebraic varieties in ${\bf CP}^N$ ?}. The answer that
we found is
that there is no Landau-Ginzburg phase: indeed the N=4 theories are special
instances of
N=2 theories with such a superpotential that it admits only one type of
extremum: the $\sigma$-model
phase. The would-be Landau-Ginzburg phase disappears.
\par
To obtain these results  and make our detailed comparison between the general
N=2 case and that
which actually corresponds to an enhanced N=4 supersymmetry, we considered the
formulation of
the N=2 and N=4 theories in the set-up of the rheonomy framework
\cite{CastellaniDauriaFre}.
This laborious technical work, presented in the central sections of our paper,
had an
additional motivation, besides that of providing a unified framework for the
N=4 and N=2 cases.
This is our intention to study the coupling of matter systems to N=2
2D-supergravity and
use this as a starting point for an approach to topological matter-coupled 2D
gravity based on
a systematic use of the topological twist, in complete analogy with the results
previously obtained
in the D=4 case \cite{FreAnselmi}.
Indeed, once a rigid supersymmetric theory is recast into the rheonomy
framework, its coupling
to supergravity is already almost achieved, since,
by construction, the rheonomic action already
contains all the couplings of the matter fields to the bare vielbein and
gravitino fields.
Just the possible couplings to the gravitino-curl and to the bosonic curvature
are missing
and they can be easily retrieved in a second step.
\par
In the present paper, we shall present the rheonomic curvature parametrizations
and the rheonomic
action for both the N=2 and the N=4  gauge theories coupled to chiral
Wess-Zumino multiplets or hypermultiplets (in the N=4 case). In addition to our
study of the
phase structure, we shall present a careful analysis of the
R-symmetries and a critical discussion of the formal structure for the
A and B topological twists \cite{Wittenmirror}.
This discussion clarifies, along the same lines of thought followed by us in
the D=4 case
\cite{FreAnselmi} some points that were, in our opinion, not completly
clear in the existing literature. It also provides the basis for the study of
the topological
twists in the gravitational case which is postponed to a future publication.
\par
To get into the heart of our topic we begin with a review of the \hika quotient
construction.
\vskip 0.2cm
\underline{\sl DEFINITION of \hika manifolds}
\vskip 0.2cm
{\sl On a \hika manifold $\cM$, which is necessarily $4n$-dimensional, there
exist three
covariantly
constant complex structures $\cJ i: T\cM\rightarrow T\cM$, $i=1,2,3$,
Bthe metric is hermitean with respect to all of them and they satisfy the
quaternionic algebra: $\cJ i \cJ j = - \delta^{ij} + \epsi ijk \cJ k$.}
\par
In a vierbein basis $\{V^A\}$, hermiticity of the metric is equivalent
to the statement that  the matrices $\cJt iAB$ are
antisymmetric. By covariant constancy, the three \hika
two-forms $\Omega^i=\cJt iAB V^A\wedge V^B$ are closed: $d\Omega^i=0$.
In the four-dimensional case, because of the quaternionic algebra
constraint, the $\cJt iAB$
can  be either selfdual or
antiselfdual; if we take them to be antiselfdual: $\cJt iAB =-{1\over 2}
\epsilon_{ABCD} \cJt iCD$, then the integrability condition for the
covariant constancy of $\cJ i$ forces the curvature two-form $R^{AB}$ to
be selfdual: thus, in the four-dimensional case, \hika manifolds are particular
instances
of gravitational instantons.
\par
A \hika manifold is a \ka manifold with respect to each of
its complex structures.
\par
Consider a compact Lie group $G$ acting on a \hika manifold $\cS$ of
real dimension $4n$ by means of Killing vector fields $\bx$ that are
holomorphic
with respect to the three complex structures of $\cS$; then these vector
fields preserve also the \hika forms:
\begin{equation}
\left.\begin{array}{l}
{\cal L}_{\scriptscriptstyle\bx}g = 0 \leftrightarrow
\nabla_{(\mu}X_{\nu)}=0 \\
{\cal L}_{\scriptscriptstyle\bx}\cJ i = 0 \,\, , \,i =1,2,3\\
\end{array}\right\} \,\,\Rightarrow\,\, 0={\cal L}_{\scriptscriptstyle\bx}
\Omega^i = i_{\scriptscriptstyle\bx} d\Omega^i+d(i_{\scriptscriptstyle\bx}
\Omega^i) = d(i_{\scriptscriptstyle\bx}\Omega^i)\, .
\label{holkillingvectors}
\end{equation}

Here ${\cal L}_{\scriptscriptstyle\bx}$ and
$i_{\scriptscriptstyle\bx}$ denote respectively the Lie derivative along
the vector field $\bx$ and the contraction (of forms) with it.

If $\cS$ is simply connected, $d(i_{\bx}\Omega^i)=0$ implies the existence
of three functions ${\cal D}_i^{\bx}$ such that $d{\cal D}_i^{\bx}=
i_{\scriptscriptstyle\bx}\Omega^i$. The functions ${\cal D}_i^{\bx}$ are
defined up to a constant,
which can be arranged so to make them equivariant: $\bx {\cal D}_i^{\bf Y} =
{\cal D}_i^{[\bx,{\bf Y}]}$.

The $\{{\cal D}_i^{\bx}\}$ constitute then a {\it momentum
map}. This can be regarded as a map ${\cal D}: \cS \rightarrow \r^3\otimes
\cG^*$, where $\cG^*$ denotes the dual of the Lie algebra $\cG$ of the
group $G$. Indeed let $x\in \cG$ be the Lie algebra element corresponding to
the Killing
vector $\bx$; then, for a given $m\in \cS$, ${\cal D}_i(m) :x\longmapsto
{\cal D}_i^{\bx}(m)\in\c$ is a linear functional on $\cG$. In practice,
expanding $\bx =X_a {\bf k}^a$ in a basis of Killing vectors ${\bf k}^{a}$
such that $[{\bf k}^a,{\bf k}^b]=\f abc {\bf k}^c$, where
$\f abc $ are the structure constants of $\cG$,
we also have  ${\cal D}_i^{\bx}=X_a \, {\cal D}_i^a$, $i=1,2,3$;
the $\{{\cal D}_i^a\}$ are the components of the momentum map.

The {\it \hika quotient} \cite{hklr} is a procedure that provides a way to
construct from
$\cS$ a
lower-dimensional \hika manifold $\cM$, as follows.
Let ${\cZ}^*\subset \cG^*$ be
the dual of the centre of $\cG$. For each $\zeta\in \r^3\otimes \cZ^*$ the
level set of the momentum map
\begin{equation}
\cN \equiv\bigcap_{i} {\cal D}_i^{-1}(\zeta^i) \subset \cS ,
\label{levelsetdef}
\end{equation}
which has dimension $\,\dim\,\cN=\dim\,\cS -3\,\,\dim\,G$,
is invariant under the action of $G$, due to the equivariance of
${\cal D}$. It is thus possible to take the quotient
\[\cM =\cN/G.\]
$\cM$ is a smooth manifold of dimension ${\rm dim}\cM={\rm dim}\cS
-4\, {\rm dim}G$ as long as the action of $G$ on $\cN$ has no fixed points.
The three two-forms $\omega^i$ on $\cM$, defined  via
the restriction to $\cN\subset\cS$ of the $\Omega^i$ and the quotient
projection from $\cN$ to $\cM$, are closed and satisfy the quaternionic algebra
thus
providing $\cM$ with a \hika structure.
\par
For future use, it is important to note that, once $\cJ 3$ is chosen as
the preferred complex structure, the momentum maps ${\cal D}_{\pm}={\cal
D}_1\pm i{\cal D}_2$
are holomorphic (resp. antiholomorphic) functions.
\par
The standard use of the \hika quotient is that of obtaining non trivial
\hika manifolds starting from a
flat $4n$ real-dimensional manifold $\r^{4n}$ acted on by a suitable
group G generating triholomorphic isometries \cite{hklr,lr}.
For instance this is the way it was utilized by Kronheimer \cite{kronheimer} in
its exhaustive
construction of all self-dual asymptotically locally Euclidean four-spaces (ALE
manifolds).
We reviewed this construction in the already quoted
discussion of the N=4 conformal
field theories describing string propagation on gravitational instantons
\cite{newalenostro}.
Indeed the manifold $\r^{4n}$ can be given a quaternionic structure, and the
corresponding quaternionic notation is sometimes convenient. For $n=1$
one has the flat quaternionic space \label{quater}
$\quat\stackrel{\rm def}{=}\left(\r^4,\left\{J^i\right\}\right)$ . We represent
its elements
\[ q\in\quat=x+i y+j z+k t=x^0+x^i J^i,\hskip 15pt x,y,z,t\in\r\]
realizing the quaternionic structures $J^i$ by means of Pauli matrices:
 $J^i=i\left(\sigma^i\right)^T$. Thus
\begin{equation}
q=\twomat{u}{iv^*}{iv}{u^*} \hskip 1cm\longrightarrow\hskip 1cm
\dag q=\twomat{u^*}{-iv^*}{-iv}{u}
\label{singlequaternion}
\end{equation}
where $u=x^0+ix^3$ and $v=x^1+ix^2$.
The euclidean metric on $\r^4$ is retrieved as $d\dag q\otimes dq=ds^2
\un$. The \hika forms are grouped into a quaternionic two-form
\begin{equation}
\Theta=d\dag q\wedge dq\,\,\stackrel{def}{=}\,\,\Omega^i J^i=
\twomat{i\Omega^3}{i\Omega^+}{i\Omega^-}{-i\Omega^3}\,\, .
\label{thetaform}
\end{equation}
For generic $n$, we have the space $\quat^n$, of elements
\begin{equation}
q=\twomat{u^A}{iv^{A^*}}{i v^A}{u^{A^*}} \hskip 1cm
\longrightarrow\hskip 1cm \dag q=\twomat{u^{A^*}}{-iv^{A^*}}
{-iv^A}{u^A}\hskip 1cm
\begin{array}{l}u^A,v^A\in\c^n\\A=1,\ldots n\end{array}
\label{multiquaternion1}
\end{equation}
Thus $d\dag q\otimes dq=ds^2 \un$ gives $ds^2=d
u^{A^*}\otimes du^A+dv^{A^*}\otimes dv^A$ and the \hika forms are grouped
into the obvious generalization of the
quaternionic two-form in eq.(\ref{thetaform}):
$\Theta=\sum_{A=1}^n d\dags qA \wedge dq^A=\Omega^i J^i$, leading to
$\Omega^3=2i\partial\bar\partial K$
where the \ka potential $K$ is $K=\um\left(u^{A^*} u^A + v^{A^*}
v^A\right)$,
and to $\Omega^+=2 i du^A\wedge dv^A$,
$\Omega^-=\left(\Omega^+\right)^*$.
\par
Let $\left(T_a\right)^A_B$ be the antihermitean generators of a compact
Lie group G in its  $n\times n$ representation. A triholomorphic action
of $G$ on $\quat^n$ is realized by the Killing vectors of components
\begin{equation}
X_a=\left(\hat T_a\right)^A_B q^B {\partial\over\partial q^A}+\dags qB
\left(\hat T_a\right)^B_A {\partial\over\partial \dags qA}\hskip 1cm ;
\hskip 1cm \left(\hat T_a\right)^A_B=\twomat{\left(T_a\right)^A_B}{{\bf
0}}{{\bf 0}}{\left(T^*_a\right)^A_B}\,\, .
\label{triholoaction}
\end{equation}
Indeed one has $\cL_{\scriptscriptstyle \bx}\Theta=0$.
The corresponding components of the momentum map are:
\begin{equation}
{\cal D}^a=\dags qA\twomat{\left(T_a\right)^A_B}{{\bf
0}}{{\bf 0}}{\left(T^*_a\right)^A_B}q^B +\twomat{c}{\bar b}{b}{-i c}
\label{momentumcomponents}
\end{equation}
where $c\in\r,\,b\in\c$ are constants.
\vskip 0.2cm\noindent
\par
As we have already anticipated
the geometrical \hika quotient construction is intimately related with N=2
supersymmetry in
four-dimensions or with N=4 supersymmetry in two-dimensions. The relation
occurs through the
auxiliary-field structure of the N=2 vector multiplet in D=4 or of the N=4
vector multiplet in
D=2. In both cases, in addition to the physical fields, the vector multiplet
contains a triplet
of auxuliary scalars, specifically a real scalar $\cP={\cP}^*$ and a complex
scalar
$\cQ \ne {\cQ}^*$.  When the vector multiplet is utilized to gauge an isometry
of an N=2
$\sigma$-model in D=4 or of an N=4 $\sigma$-model in D=2, the auxiliary fields
$\left \{ \cP , \cQ
\right \}$ are identified with the momentum-map functions $\left \{ {\cal D}^3
\left ( m \right )
\, , \, {\cal D}^\pm \left ( m \right ) \, \right \}$ of the $\sigma$-model
target-space ${\cal S}$.
Indeed, in both cases, $\cM_4 \, \longrightarrow \, {\cal S}$ or
$\cM_2 \, \longrightarrow \, {\cal S}$, the condition for the $\sigma$-model to
possess either
N=2 or N=4 supersymmetry is that the target space ${\cal S}$ be endowed with a
\hika structure.
\par
In view of this fundamental property, the \hika quotient offers a natural way
to construct an
N=2, D=4 or N=4, D=2 $\sigma$-model on a non-trivial manifold $\cM$ starting
from a free
$\sigma$-model on a trivial flat-manifold ${\cal S}={\bf H}^n$.
It suffices to gauge appropriate triholomorphic
isometries by means of non-propagating gauge multiplets.
Omitting the kinetic term of these gauge multiplets and performing the gaussian
integration of
the corresponding fields one realizes the \hika quotient in a Lagrangian way.
In the four-dimensional case, this fact was
fully exploited,  long time ago by Hitchin, K\"arlhede, Lindstrom and
Rocek \cite{hklr}, was further discussed by Galicki \cite{Galicki} and was
applied,
in the context of string theory by Ferrara, Girardello, Kounnas and
Porrati \cite{FerPorKounGirar}. Actually the \hika quotient is a generalization
of a similar
\ka quotient procedure, where the momentum map ${\cal D}: \cS \rightarrow
\r\otimes
\cG^*$ consists  just of one hamiltonian function, rather than three. The \ka
quotient is
related with either N=1,D=4 or N=2,D=2 supersymmetry, the reason being that, in
these cases
the vector multiplet contains just one real auxiliary field  $\cP$.
\par
Recently, Witten has reconsidered the \ka quotient construction of an N=2
two-dimensional
$\sigma$-model in
\cite{Wittenphases}. His point of view was that of regarding the \ka quotient
as
an effective low-energy phenomenon rather than as a mere trick to implement the
geometrical quotient
construction in a Lagrangian field-theory language. In other words he included
the kinetic
terms of the vector multiplet and also a Fayet-Iliopoulos term for each of the
abelian factors in
the gauge group; then he considered the whole system as a {\it bona fide}
gauge-theory
spontaneously broken via an ordinary Higgs mechanism by the extrema of the
scalar potential.
Integrating on the massive modes, that include the gauge vectors, the effective
Lagrangian
of the massless modes turns out to be that of an N=2 $\sigma$-model on a
K\"ahler target manifold
that is obtained as a hypersurface in a \ka quotient. This, however, happens
only in one phase, namely
in a certain range of the parameters contained by the superpotential. When the
parameters are
in another range we fall in a Landau-Ginzburg phase, namely the low-energy
effective theory
of the massless modes is a Landau-Ginzburg model with superpotential equal to
the polynomial
constraint $\cW(X)$ that defines the target-manifold as a hypersurface in the
$\sigma$-model phase.
\par
Following the same line of thought, after reconstructing Witten's theory in a
rheonomic framework,
we construct the N=4 analogue of this model. We introduce N=4 gauge multiplets
and the N=4
analogues of the Wess-Zumino multiplets, namely the quaternionic
hypermultiplets. We show that
an N=4 counterpart of the Fayet-Iliopoulos term does indeed exist and involves
three real
parameters. After coupling to the hypermultiplets these parameters play the
role of triholomorphic
momentum-map levels, in the same way as, in Witten's case, the single parameter
introduced by
the N=2 Fayet Iliopoulos term plays the role of momentum-map level for the
holomorphic isometry.
 What is different
in the N=4 case is the absence of auxiliary fields for the hypermultiplets.
This implies that
besides the interaction introduced by the gauge coupling, no other arbitrary
quaternionic
superpotential can be introduced. This is the essential reason why, at the end
of the day
we do not find any analogue of the Landau-Ginzburg phase. It must be noted,
however, that
when we apply our construction to the ALE manifolds, the Fayet-Iliopoulos
parameters have
a deep geometrical meaning: they are the the moduli of the self-dual metric.
\par
In the last part of the paper, after an analysis of the R-symmetries, that in
the N=4 case
from $U(1)_L \otimes U(1)_R$ are promoted to $U(2)_L \otimes U(2)_R$,  we
discuss the A and B
topological twists, clarifying, as we have already
anticipated, some delicate formal aspects of the procedure.
In particular we discuss the subtleties related with the redefinition of the
ghost number, which has to be performed simultaneously with the redefinition of
the spin.

\section{The N=2 and N=4 rheonomic set up for globally supersymmetric
field theories}
\label{n2n4}
The rheonomy approach to the construction of both locally and globally
supersymmetric
field-theories is almost fifteen years old and it has been extensively applied
to all
supergravity models in all space-time dimensions. A complete exposition of the
method
is contained in the book \cite{CastellaniDauriaFre}: we refer to it for the
basic
concepts and we just begin with the specific definitions and conventions needed
in our
case.
\par
The starting point for the whole construction is the definition of the
curvatures of the
(2,2) and (4,4) extended 2D-superspace. We denote by $e^{\pm}$ the two
components of the
world-sheet zweibein (in the flat case $e^+ \,= \, dz \, + \, \theta-terms$,
$e^- \, = \, d{\bz}\, + \, \theta-terms$), by $\omega$
the world-sheet spin-connection 1-form (in the flat-case we can choose $\omega
\, =\, 0$)
and by $\zeta^{\pm} \, , \, {\tilde \zeta}^{\pm}$ the four fermionic one-forms
gauging
the (2,2) supersymmetries, namely the 4 components of the 2 gravitinos. In the
flat case
we  have $\zeta^{\pm} \, =\, d\theta^{\pm}$, ${\tilde \zeta}^{\pm} \, = \,
d{\tilde \theta}^{\pm}$).
In the (4,4) case, in addition to $\zeta^{\pm} \, , \, {\tilde \zeta}^{\pm}$,
we have four
other fermionic 1-forms $\chi^{\pm} \, , \, {\tilde \chi}^{\pm}$, that complete
the eight components
of the four gravitinos. Furthermore, in the N=2 case there is a bosonic 1-form
$A^{\bullet}$
gauging the $U(1)$ central charge, while in the N=4 case, in addition to
$A^{\bullet}$, we have
two others bosonic 1-forms $A^{\pm}$ gauging the other two central charges.
 \par
In terms of these 1-forms the superspace curvatures are:
\begin{eqnarray}
de^+ + \omega\wedge e^+ -\o{\rm i}{2} \zeta^{+} \wedge \zeta^{-}& = & T^+
\nonumber\\
 de^- - \omega\wedge e^- -\o{\rm i}{2} \tilde\zeta^{+} \wedge \tilde\zeta^{-}&
= &T^-\nonumber\\
d\zeta^+ +\o{1}{2} \omega \wedge \zeta^+ & = & \rho^+ \nonumber\\
d \tilde\zeta^+ -\o{1}{2} \omega \wedge \tilde\zeta^+ & = &{\tilde
\rho}^+\nonumber\\
d \zeta^- +\o{1}{2} \omega \wedge \zeta^- & = & \rho^-\nonumber\\
d \tilde\zeta^- -\o{1}{2} \omega \wedge \tilde\zeta^- & = & {\tilde
\rho}^-\nonumber\\
d\omega & = &R \nonumber\\
dA^{\bullet} - \zeta^- \wedge \tilde \zeta^+ +\tilde \zeta^+ \wedge \zeta^-& =
&F^{\bullet}
\label{ntwo1}
\end{eqnarray}
Flat superspace is described by the equations
\begin{equation}
T^{\pm} \, = \, \rho^{\pm} \, = \, {\tilde \rho}^{\pm} \, = \, R \, = \,
F^{\bullet} \, = \, 0
\label{flatn2superspace}
\end{equation}
In the background of these flat superspace 1-forms we are supposed to solve the
Bianchi identities
for the matter fields, spanning the various matter multiplets and to construct
the associated
rheonomic actions. In this way we determine the SUSY rules and the world-sheet
supersymmetric
actions for all the theories under consideration. If we remove eq.s
(\ref{flatn2superspace})
and we introduce a rheonomic parametrization for the curvatures (\ref{ntwo1})
then we are
dealing with N=2 2D-supergravity and the solution of Bianchi identities in this
curved
background constitutes the coupling of matter to supergravity. This programme
is left
for a future publication: in this paper we concentrate on the flat case.

For convenience we also recall the rule for complex conjugation. Let $\psi_1 ,
\psi_2$ be
two forms of degree $p_1 , p_2$ and statistics $F_{1} , F_{2}$ ( $F=0,1$ for
bosons or fermions)
so that $\psi_1 \psi_2 \, = \, (-1)^{p_1p_2+F_1F_2}\, \psi_2  \psi_1$, then we
have:
\begin{equation}
\left (\,  \psi_1 \, \psi_2 \, \right )^* \, = \, \left ( -1 \right )^{F_1F_2}
\, \psi_1 ^{*} \, \psi_2^* = (-1)^{p_1 p_2}\psi^*_2 \,\psi^*_1
\label{ntwo2}
\end{equation}
Thus, for example, for the gravitinos we have:
\begin{equation}
\left (\, \zep \,\wedge \zem  \,\right )^* \, = -\,\left (\, \zep \,\right )^*
\wedge
\,\left (\, \zem \, \right )^* \, = \, - \, \zem \wedge \zep\, = \, - \, \zep
\wedge \zem
\label{ntwo3}
\end{equation}

We proceed next to write the curvatures of the N=4 extended two-dimensional
superspace, namely:
\begin{eqnarray}
de^+ + \omega\wedge e^+ -\o{\rm i}{2} \zeta^{+} \wedge \zeta^{-}
-\o{\rm i}{2} \chi^{+} \wedge \chi^{-}& = &T^{+}\nonumber\\
 de^- - \omega\wedge e^- -\o{\rm i}{2} \tilde\zeta^{+} \wedge \tilde\zeta^{-}
-\o{\rm i}{2} \tilde\chi^{+} \wedge \tilde\chi^{-}& = &T^{-}\nonumber\\
d\chi^+ +\o{1}{2} \omega \wedge \chi^+ & = &\tau^{+}\nonumber\\
d \tilde\chi^+ -\o{1}{2} \omega \wedge \tilde\chi^+ & = & {\tilde
\tau}^{+}\nonumber\\
d \chi^- +\o{1}{2} \omega \wedge \chi^- & = &\tau^{-}\nonumber\\
d \tilde\chi^- -\o{1}{2} \omega \wedge \tilde\chi^- & = &{\tilde
\tau}^{-}\nonumber\\
d\omega & = &R\nonumber\\
d\zeta^+ +\o{1}{2} \omega \wedge \zeta^+ & = & \rho^+ \nonumber\\
d \tilde\zeta^+ -\o{1}{2} \omega \wedge \tilde\zeta^+ & = &{\tilde
\rho}^+\nonumber\\
d \zeta^- +\o{1}{2} \omega \wedge \zeta^- & = & \rho^-\nonumber\\
d \tilde\zeta^- -\o{1}{2} \omega \wedge \tilde\zeta^- & = & {\tilde
\rho}^-\nonumber\\
dA^{\bullet} - \zeta^- \wedge \tilde \zeta^+ +\tilde \zeta^+ \wedge \zeta^- +
\chi^-\wedge \tilde \chi^+ -\tilde \chi^+ \wedge \chi^-& = &
F^{\bullet}\nonumber\\
dA^+ - \chi^- \wedge \tilde \zeta^+ +\tilde \chi^+ \wedge \zeta^-& = &
F^{+}\nonumber\\
dA^- - \zeta^- \wedge \tilde \chi^+ +\tilde \zeta^+ \wedge \chi^-&=&F^{-}
\label{nfour1}
\end{eqnarray}
Also in this case flat superspace is described by
\begin{equation}
T^{\pm} \, = \, \rho^{\pm} \, = \, {\tilde \rho}^{\pm} \, = \, R \, = \,
F^{\bullet} \, = \,
F^{\pm} \, = \, 0
\label{flatn4superspace}
\end{equation}
For both the N=2 and the N=4 case, the determination of the  globally
supersymmetric
field theories is done by solving the Bianchi identities of the matter fields
in the background of the flat superspace 1-forms and then by constructing the
associated
rheonomic actions. In this way, for each matter multiplet we can determine the
SUSY rules and
the world-sheet supersymmetric actions. The convention for complex conjugation
is the same
in the N=4 and  in the N=2 case.

\section{ The N=2  abelian gauge multiplet}
\label{n2abel}
In this section we discuss the rheonomic construction of an N=2 abelian gauge
theory
in two-dimensions. This study will provide a basis for our subsequent coupling
of
the N=2 gauge multiplet to  an N=2 Landau-Ginzburg system invariant under the
action
of one or several $U(1)$ gauge-groups or even of some non
abelian gauge group $G$.
\par
In the  N=2 case a vector multiplet is composed of a gauge boson ${\cal A}$,
namely a world-sheet 1-form,
two spin 1/2 gauginos, whose four components we denote by
$\lambda^+$,$\lambda^-$,
$\tilde\lambda^+$,$\tilde\lambda^-$, a complex physical scalar $M \ne M^*$ and
a real
auxiliary scalar ${\cal P}^*={\cal P}$. Each of these fields is in the adjoint
representation
of the gauge group $G$ and carries an index of that representation that we have
not written.

In the abelian case, defining the field strength
\begin{equation}
F \, = \, d{\cal A}
\label{ntwo4}
\end{equation}
the rheonomic parametrizations that solve the Bianchi identities:
\begin{equation}
dF\, = \, d^2\lamm
\, = \, d^2 \lap \, =\, d^2 \latp \, = \, d^2 \latm \, = \,d^2  M \,
 = \, d^2 {\cal P} \, =\, 0
\label{ntwo5}
\end{equation}
 are given by
\begin{eqnarray}
F & = & \cF\,\eplus\eminus -\o i2\bigl(\lap\zem +\lamm\zep\bigr)\,\eminus
+\o i2\bigl(\latp\zetm +\latm\zetp \bigr)\,\eplus
+ M\,\zem\zetp - M^*\,\zep\zetm\nonumber  \\
dM & = & \dep M\,\eplus +\dem M\,\eminus -\o 14 \bigr(\latm\zep
-\lap\zetm\bigl) \nonumber\\
d\lap & = & \dep\lap\,\eplus +\dem\lap\,\eminus +\bigl(\o {\cF}2 +
i\cP\bigr)\,\zep -2i\,\dem M\,\zetp\nonumber \\
d\lamm & = & \dep \lamm \, \eplus +\dem\lamm\,\eminus + \bigl (\o{\cF}2 -
i\cP\bigr)\,\zem
+ 2i \dem M^{*} \zetm\nonumber \\
d\latp & = & \dep\latp\,\eplus +\dem\latp\,\eminus +\bigl(\o {\cF}2 -
i\cP\bigr)\,\zetp -2i\,\dep M^*\,\zep \nonumber\\
d\latm & = & \dep\latm\,\eplus +\dem\latm\,\eminus +\bigl(\o {\cF}2 +
i\cP\bigr)\,\zetm +2i\,\dep M\,\zem\nonumber\\
d\cP & = & \dep\cP\,\eplus +\dem\cP\,\eminus -\o 14 \bigl( \dep\lap\zem
-\dep\lamm\zep -\dem\latp\zetm +\dem\latm\zetp\bigr)
\label{ntwo6}
\end{eqnarray}
Given these parametrizations, we next write the rheonomic action whose
variation yields the
above parametrizations as field equations in superspace, together with the
world-sheet equations
of motion.

\begin{eqnarray}
\cL^{(rheon)}_{gauge}&=&\cF\biggl [F +\o i2\left (\lap\zem +\lamm\zep\right )\,
\eminus -\o i2\left(\latp\zetp +\latm\zetp \right)\,\eplus \nonumber\\
&&\mbox{}- M\,\zem\zetp - M^*\,\zep\zetm\biggr ] -\o 12 \cF^2\,\eplus\eminus
\nonumber\\
&&\mbox{}-\o 12 \bigl(\lap\,d\lamm +\lamm\, d\lap\bigr)\,\eminus +\o
i2\bigl(\latp\,d\latm +\latm\,
d\latp\bigr)\,\eplus  \nonumber\\
&&\mbox{}- 4 \biggl[dM^* -\o 14\bigl(\latp\zem - \lamm\zetp\bigr)\biggr]\bigl(
\cM_{\ssp}\eplus -\cM_{\ssm}\eminus\bigr)\nonumber\\
&&\mbox{} -4 \biggl[dM +\o 14\bigl(\latm\zep
- \lap\zetm\bigr)\biggr]\bigl(\cM^*_{\ssp}\eplus -\cM^*_{\ssm}\eminus\bigr) \\
&&\mbox{}- 4\bigl(\cM^*_{\ssp}\cM_{\ssm}
+\cM^*_{\ssm}\cM_{\ssp}\bigr)\,\eplus\eminus -dM\bigl(\lamm\zetp
+\latp\zem\bigr) + dM^*\bigl(\lap\zetm +\latm\zep\bigr)\nonumber\\
&&\mbox{}-\o 14 \bigl(\lap\latp\,\zem\zetm +\lamm\latm\,\zep\zetp\bigr)+
 2\cP^2\,\eplus\eminus + 4i\,\o{\partial\cU}{\partial
M}\,\biggl(\o F2 + i \cP\,\eplus\eminus\biggr)\nonumber\\
&&\mbox{}- 4i\,\o{\partial\cU^*}{\partial
M^*}\,\biggl(\o F2 - i \cP\,\eplus\eminus\biggr)
-i\,\biggl(\o{\partial^2\cU}{\partial M^2}\,\lap\latm +
\o{\partial^2\cU^*}{{\partial M^*}^2}\,\lamm\latp
\biggr)\,\eplus\eminus \nonumber\\
&&\mbox{} +\biggl(\o{\partial\cU}{\partial M}
+\o{\partial\cU^*}{\partial M^*}\biggr)\biggl[\bigl(\lap\zem
-\lamm\zep\bigr)\,\eminus +\bigl(\latp\zetm
-\latm\zetp\bigr)\,\eplus\biggr] \nonumber\\
&&\mbox{}+2i\,\biggl[2\cU-M\,\biggl(\o{\partial\cU}{\partial M}
-\o{\partial\cU^*}{\partial M^*}\biggr)\biggr]\,\zem\zetp +
2i\,\biggl[2\cU^*-M^*\,\biggl(\o{\partial\cU}{\partial M}
-\o{\partial\cU^*}{\partial M^*}\biggr)\biggr]\,\zep\zetm\nonumber\\
\label{ntwo7}
\end{eqnarray}
The symbol ${\cal U}$ denotes a holomorphic function ${\cal U}\left (M \right)$
of the physical scalar $M$ that is named the superpotential. It induces a self
interaction
of the scalar $M$ field and an interaction of this field with the gauge-vector.
The existence of an arbitrariness
in the choice of  the vector multiplet dynamics is a consequence of the
existence of the auxiliary field ${\cal P}$ in the solution of the Bianchi
identities
(\ref{ntwo5}) and hence in the determination of the SUSY rules for this type of
N=2
multiplet. In the superspace formalism the inclusion in the action of the terms
containing the
superpotential is effected by means of the use of the so called twisted chiral
superfields.
In the rheonomic framework there is no need of these distinctions: we just have
an
interaction codified by an arbitary holomorphic superpotential.

Note that in eq.s (\ref{ntwo6}) and (\ref{ntwo7}) we have suppressed the wedge
product symbols
for differential forms. This convention will be often adopted  also in the
sequel to avoid
clumsiness. From the rheonomic action (\ref{ntwo6}) we easily obtain the
world-sheet action
of the N=2 globally supersymmetric abelian vector multiplet, by deleting all
the terms containing
the gravitino 1-forms, replacing the first order fields ${\cal F}, {\cal
M}_{\pm}$ with
their values following from their own field equations, namely ${\cal
F}=\o{1}{2}
 \left (\partial_+ {\cal A}_{-} \, - \, \partial_{-}{\cal A}_{+} \right )$,
${\cal M}_{\pm}
= \partial_{\pm}M$, and by replacing $e^{+} \wedge e^{-} $ with $d^2 z$ that is
factored out.
In this way we get:
\begin{eqnarray}
{\cL^{(ws)}_{gauge}} & = & \o 12 \cF^2 -i\,\bigl(\lap\dep\lamm +\latp\dem\latm
\bigr) -4\bigl(\dep M^*\dem M +\dem M^*\dep M\bigr) + 2\cP^2 \nonumber\\
&&\mbox{}+4i\,\o{\partial \cU}{\partial M}\biggl(\o{\cF}2 +i\cP\biggr)
-4i\,\o{\partial \cU^*}{\partial M^*}\biggl(\o{\cF}2 -i\cP\biggr)-
i \biggl(\o{\partial^2\cU}{\partial M^2}\,\lap\latm +
\o{\partial^2\cU^*}{{\partial M^2}^*}\,\lamm\latp\biggr) \nonumber\\
\label{ntwo8}
\end{eqnarray}
In the particular case of a linear superpotential
\begin{equation}
\cU=\o t4\,M ~,~ t\in
{\bf C}
\label{ntwo8bis}
\end{equation}
 setting
\begin{equation}
t=r - i \theta/2\pi~,~r\in {\bf R}~, ~\theta \in [0,2\pi]
\label{ntwo8tris}
\end{equation}
the above expression reduces to
\begin{eqnarray}
\cL_{ws} & = &\o 12 \cF^2 -i\,\bigl(\lap\dep\lamm +\latp\dem\latm
\bigr) -4\bigl(\dep M^*\dem M +\dem M^*\dep M)\nonumber\\
&&\mbox{}+ 2\cP^2 -2r\cP +\o{\theta}{2\pi}\cF
\label{ntwo9}
\end{eqnarray}
The meaning of the parameters $r$ and $\theta$ introduced in the above
lagrangian is clear.
Indeed $r$, giving a vacuum expectation value ${\cal P}=\o r2$ to the auxiliary
field ${\cal P}$
induces a spontaneous breaking of supersymmetry and shows that the choice
${\cal U} =-\o r4 M$
corresponds to the insertion  of a Fayet-Iliopoulos term into the action. On
the other hand
the parameter $\theta$ is clearly a theta-angle multiplying the first
Chern class $\o{1}{2\pi} {\cal F}$ of the gauge connection.

\section{ $N=2$  Landau Ginzburg models with an abelian gauge symmetry}
\label{Landauginzburg}
 As stated above, our interest in the N=2 vector multiplet was instrumental to
the study of
an N=2 Landau-Ginzburg system possesing in addition to its own self interaction
a minimal
coupling to a gauge theory. This is the system studied by Witten in
\cite{Wittenphases}, using
superspace techniques, rather than the rheonomy framework.
 By definition a Landau Ginzburg system is a collection of
N=2 chiral multiplets self-interacting via an analytic superpotential $W(X)$.
Each chiral multiplet is composed of a complex scalar field $\left ( X^i \right
)^{*} =
X^{i^*} $ ( $i=1,....,n$), two spin 1/2 fermions, whose four components we
denote by $\psi^{i} ,
 \pst i$
and $\ps {i^*} = \left ( \ps i\right )^* , \pst {i^*} = \left ( \pst i\right
)^*$,
together with a complex auxiliary field ${\cal H}^{i}$ which is identified with
the derivative
of the holomorphic superpotential ${\bar W}\left ( X \right )$, namely
 ${\cal H}^{i}=\etasu ij \partial_{j^*}W^*$, $\etasu ij$ being the flat
K\"ahlerian metric
on the complex manifold ${\bf C}^n$ of which the complex scalar fields $X^{i}$
are interpreted
as the coordinates. Using this system of fields, we could construct a rheonomic
solution of
the superspace Bianchi identities, a rheonomic action and a world-sheet action
invariant under
the supersymmetry transformations induced by the rheonomic parametrizations. In
this action
the kinetic terms are the canonical ones of a free field theory and the only
interaction is
that induced by the superpotential. Rather than doing this we prefer to study
the same system
in presence of a minimal coupling to the gauge system studied in the previous
section. In
practice this amounts to solve the Bianchi identities for the gauge covariant
derivatives
rather than for the ordinary derivatives, using as a background the rheonomic
parametrizations
of the gauge mulitiplet determined above. At the end of the construction, by
setting the
gauge coupling constant to zero, we can also recover the formulation of the
ordinary
Landau-Ginzburg theory, later referred to as the {\it rigid Landau-Ginzburg
theory}.
\par
Indeed
the coupling of the chiral multiplets to the gauge multiplet is defined through
the
covariant derivative
\begin{equation}
\dell X^i \,\eqdef \, dX^i+i \cA\qu ij X^j
\label{ntwo9bis}
\end{equation}
where the hermitean matrix $\qu ij $ is the generator of the $U(1)$ action on
the chiral matter.
As a consequence, the Bianchi identities are of the form $\dell^2 X^i=i\,F\qu
ij
X^j$.

 Let $W(X^i)$ be the holomorphic the superpotential: then the rheonomic
solution of the
Bianchi identities is given by the following parametrizations:
\begin{eqnarray}
\dell X^i & = & \delp X^i\,\eplus +\delm X^i\,\eminus +\ps i\zem + \pst
i\zetm\nonumber \\
\dell X^{i^*} & = & \delp X^{i^*}\,\eplus +\delm X^{i^*}\,\eminus -\ps {i^*}
\zep - \pst {i^*}\zetp\nonumber \\
\dell\ps i & = & \delp\ps i\,\eplus +\delm\ps i\,\eminus -\o i2\delp
X^i\,\zep + \etasu ij \partial_{j^*}W^*\,\zetm + i\, M\qu ij
X^j\,\zetp\nonumber\\
\dell\ps {i^*} & = & \delp\ps {i^*}\,\eplus +\delm\ps {i^*}\,\eminus +\o
i2\delm
X^{i^*}\,\zem + \etasu ji \partial_{j}W\,\zetp -
 i\, M^* \qu ji X^{j^*}\,\zetm\nonumber\\
\dell \pst i & = & \delp\pst i\,\eplus +\delm\pst i\,\eminus -\o i2\delm
X^i\,\zetp - \etasu ij \partial_{j^*}W^*\,\zem - i\, M^*\qu ij
X^j\,\zep\nonumber\\
\dell \pst {i^*} & = & \delp\pst {i^*}\,\eplus +\delm\pst {i^*}\,\eminus +\o
i2\delp
X^{i^*}\,\zetm - \etasu ji \partial_{j}W\,\zep +i\, M\qu ji X^{j^*}\,\zem
\label{ntwo10}
\end{eqnarray}
{}From the consistency of the above parametrizations with the
Bianchi identities one also gets the following fermionic  world-sheet equations
of motion:
\begin{eqnarray}
\o i2 \delm \ps i -\etasu ij \partial_{l^*}\partial_{j^*}W^*\,
\psts l + \o i4 \lap\qu ij X^j +i\, M \qu ij \pst j & = & 0\nonumber\\
\o i2 \delp \pst i +\etasu ij \partial_{l^*}\partial_{j^*}W^*\,
\pss l - \o i4 \latp\qu ij X^j -i\, M^* \qu ij \ps j & =& 0
\label{ntwo11}
\end{eqnarray}
and their complex conjugates for the other two fermions. Applying to eq.s
(\ref{ntwo11})
a supersymmetry transformation,
as it is determined by the parametrizations (\ref{ntwo10}),  we obtain the
bosonic
field equation:
\begin{eqnarray}
&&\o 18\bigl(\delp\delm X^i+\delm\delp X^i\bigr) -\etasu
ik\partial_{k^*}\partial_{j^*}\partial_{l^*}W^*\,\pss j\psts l + \etasu
ik \partial_{k^*}\partial_{j^*}W^*\,\etasu lj \partial_l W \nonumber \\
&&\mbox{}- \o i4 \lamm\qu ij\ps j + \o i4 \latm\qu ij\pst j + M^*M
(q^2)^i_j X^j - \o 14 \cP \qu ij X^j = 0
\label{ntwo12}
\end{eqnarray}
Equipped with this information, we can easily derive the rheonomic action from
which
the parametrizations (\ref{ntwo10}) and the field equations
(\ref{ntwo11}),(\ref{ntwo12})
follow as variational equations: it is the following one:

\begin{eqnarray}
\lefteqn{ \cL^{(rheon)}_{chiral}\, =\, \etagiu ij\biggl(\dell X^i -\ps i\zem
-\pst i
\zetm\biggr)
\,\biggl(\Pi^{j^*}_{\ssp}\eplus-\Pi^{j^*}_{\ssm}\eminus\biggr)}\nonumber\\
&&\mbox{}+
\etagiu ij\bigl(\dell X^{j^*} -\pss j\zep -\psts j
\zetp\bigr)\,
 \bigl(\Pi^i_{\ssp}\,\eplus-\Pi^i_{\ssm}\,\eminus\biggr)\nonumber\\
&&\mbox{}+\etagiu ij\bigl(\Pi^i_{\ssp}\Pi^{j^*}_{\ssm}
+\Pi^i_{\ssm}\Pi^{j^*}_{\ssp}\bigr)\,\eplus\eminus -4i\,\etagiu ij\bigl(\ps
i\dell \pss j\,\eplus - \pst i\dell \psts j\,\eminus\bigr) \nonumber\\
&&\mbox{}+4i\bigl(\ps k\partial_k W\,\zetp\eplus -\coco\bigr)
-4i\bigl(\pst k\partial_k W\,\zep\eminus -\coco\bigr)\nonumber\\
&&\mbox{} +\etagiu ij\bigl(\ps i\pss j\,\zep\zem -
\pst i\psts j\,\zetp\zetm +\coco\bigr)\nonumber\\
&&\mbox{}+ 8\biggl(\bigl(\partial_i\partial_j W\ps i\pst j +\coco\bigr)
+ \etasu ij \partial_i W\partial_{j^*}W^*\biggr)\,\eplus\eminus \nonumber\\
&&\mbox{}- \etagiu ij \bigl(\dell X^i\pss j\,\zep -\dell X^i\psts
j\,\zetp +\coco\bigr) -\bigl(4 M\pss j\etagiu ij\qu ik X^k\,\zetp\eplus
+\coco\bigr) \nonumber\\
&&\mbox{}-\bigl(4 M^* \psts j\etagiu ij\qu ik X^k\,\zep\eminus
+\coco\bigr)
 -\biggl(8i\bigl(M^*\psts j\etagiu ij\qu ik\ps k -\coco\bigr) \nonumber \\
&&\mbox{}+ 2i \bigl(\lap\pss j\etagiu ij\qu ik X^k -\coco\bigr)
+2i \bigl(\latp\psts j\etagiu ij\qu ik X^k -\coco\bigr) \nonumber\\
&&\mbox{}-2\cP \etagiu ij\,X^{j^*}\qu ik X^k +8\, M^* M\,\etagiu ij
X^{j^*}(q^2)^i_k X^k\biggr)\,\eplus\eminus
\label{ntwo13}
\end{eqnarray}
The world-sheet lagrangian for this system is now easily obtained  through the
same steps
applied in the previous case. To write it, we introduce the following
semplifications in
our notation: a) we use
a diagonal form for the flat ${\bf C}^n$ metric $\etagiu ij X^i X^{j^*} \equiv
X^i X^{i^*}$,
b) we diagonalise the $U(1)$ generator, by setting $\qu
ij \equiv q^i \delta^i_j$ ($q^i$ being the charge of the field $X^i$). Then
we have:
\begin{eqnarray}
\lefteqn{\cL^{(ws)}_{chiral}\, =\, -\bigl(\delp X^{i^*}\delm X^i + \delm
X^{i^*}\delp X^i
\bigr) + 4i\bigr(\ps i\delm\pss i +\pst i\delp\psts i\bigr) }\nonumber \\
&&\mbox{}+ 8\biggl(
\bigl(\ps i\pst j\partial_i\partial_j W +\coco\bigr) +
\partial_i W\partial_{i^*}W^*\biggr) +2i\,\sum_i q^i\bigl(
\ps i\lamm X^{i^*} -\pst i\latm X^{i^*} -\coco\bigr)\nonumber\\
&&\mbox{}+8i\,\biggl(M^*\sum_i q^i\ps i\psts i -\coco\biggr) +8
M^*M\,\sum_i (q^i)^2\,X^{i^*} X^i -2\cP\sum_i q^i X^{i^*}X^i
\label{ntwo14}
\end{eqnarray}

\section{ Structure of the scalar potential in the N=2
  Landau-Ginzburg model with an abelian gauge symmetry }
\label{sec5}
We consider next the coupled system, whose lagrangian, with our conventions, is
the difference of the
two lagrangians we have just described:
\begin{equation}
\cL\,=\,\cL_{gauge}-\cL_{chiral}
\label{ntwo15}
\end{equation}
the relative sign being fixed by the requirement of positivity of the energy.
The world-sheet form of the action (\ref{ntwo15}) is the same, modulo trivial
notation differences
as the action (2.19)+(2.23)+(2.27) in Witten's paper \cite{Wittenphases}.
We focus our attention
on the  potential energy of the bosonic fields: it is  given by the following
expression
\begin{eqnarray}
-U& = & 2\cP^2-4\cP\biggl(\o{\partial\cU}{\partial M} +\o{\partial\cU^*}
{\partial M^*}\biggr)+2\cP \sum_i q^i |X^i|^2\nonumber\\
&&\mbox{}-8\partial_i W\partial_{i^*}W^* -8 |M|^2\sum_i (q^i)^2|X^i|^2
\label{ntwo16}
\end{eqnarray}
The variation in the auxiliary field $\cP$ yields the expression of $\cP$
itself in terms
of the physical scalars:
\begin{equation}
\cP\,=\,\o{\partial\cU}{\partial M} +\o{\partial\cU^*}{\partial M^*} - \o
12 \sum_i q^i |X^i|^2
\label{ntwo17}
\end{equation}
In the above equation the expression ${\cal D}^{\bf X}\left ( X,X^*\right )
\,= \,\sum_i q^i |X^i|^2$ is the momentum map function for the holomorphic
action of the
gauge group on the matter multiplets. Indeed if we denote by
${\bf X}= i \sum_i q^{i} \left ( X^{i}\partial_i - X^{i^*}\partial_{i^*} \right
)$
the killing vector and by $\Omega\, = \,\sum_i \, dX^{i}\, \wedge \, dX^{i^*}$,
then
we have $i d{\cal D}^{\bx}=
i_{\scriptscriptstyle\bx}\Omega$. As anticipated the auxiliary field $\cP$ is
identified
with the momentum-map function,
plus  the term $\,\o{\partial\cU}{\partial M} +\o{\partial\cU^*}{\partial M^*}$
due to the
self interaction of the vector-multiplet.
In the case of the linear superpotential of eq.s (\ref{ntwo8bis}) and
(\ref{ntwo8tris}),
the auxiliary field is identified with:
\begin{equation}
\cP\, = \, -\o 12 (D^{\bf X}(X,X^*) -r)
\label{inpiu}
\end{equation}
Eliminating ${\cal P}$ through eq. (\ref{ntwo17}) , we obtain the final form
for the scalar
field potential in this kind of models, namely:
\begin{equation}
U\,=\,2\biggl[\biggl(\o{\partial\cU}{\partial M} +\o{\partial\cU^*}
{\partial M^*}\biggr)-\o 12 \sum_i q^i |X^i|^2\biggr]^2 +|\partial_i
W|^2 +8 |M|^2\sum_i (q^i)^2|X^i|^2
\label{ntwo18}
\end{equation}
In the case of the linear superpotential
 this reduces to
\begin{equation}
U\,=\,\o 12 \biggl[r - \sum_i q^i |X^i|^2\biggr]^2 +8|\partial_i
W|^2 +8 |M|^2\sum_i (q^i)^2|X^i|^2
\label{ntwo19}
\end{equation}
The theory characterized by the above scalar potential exhibits a two phase
structure
as the parameter $r$ varies on the right line.
This is the essential point in Witten's paper that allows an interpolation
between an
N=2 $\sigma$-model on a Calabi-Yau manifold and a rigid Landau-Ginzburg theory.
 The review of these two regimes
is postponed to later sections. Here we note that the above results can be
generalized
to the case of a non abelian vector-multiplet or to the case of several
abelian gauge multiplets.

\section{ Extension to the case where the gauge symmetry of the N=2
Landau-Ginzburg model
is non abelian}
\label{n2nonab}
We fix our notations and conventions.\par
Consider a Lie algebra $\cG$ with structure constants $\f abc$:
\begin{equation}
\commut{t^a}{t^b}=i\f abc \,t^c
\label{ntwo20}
\end{equation}
in every representation the hermitean generators $t^a=\dag {(t^a)}$ are
normalized in such a way  that $\tr\,(t^a t^b) =\delta^{ab}$. Let us name $T^a$
the
generators of the adjoint representation, defined by $\f abc=i(T^a)^{bc}$.\par
Let us introduce the gauge vector field as a $\cG$-valued one-form:
\begin{equation}
\cA=\cA^{a}_{\mu} T^{a} dx^{\mu}
\label{ntwo21}
\end{equation}
In the case we are interested, the index $\mu$  takes two values and
we can write $\cA=\cA^a_{\ssp}\eplus +\cA^a_{\ssm}\eminus$.
Note that $\dag \cA=\cA$.
The {field strength} is defined as the two-form
\begin{equation}
F=d\cA+i A\wedge \cA
\label{ntwo21bis}
\end{equation}
The {Bianchi Identities} read
\begin{equation}
\nabla F \eqdef dF + i(\cA\wedge F -F\wedge \cA) = 0
\label{ntwo22}
\end{equation}
The component expression of the field strength and of its associated Bianchi
identity is:
\begin{eqnarray}
&&F^a_{\mu\nu}\, =\,\partial_{[\mu} \cA^a_{\nu]} -\o 12 \f abc \cA^b_{\mu}
\cA^c_{\nu}\nonumber\\
&&\partial_{[\rho}F^a_{\mu\nu]} -\f abc \cA^b_{[\mu} F^c_{\rho\nu]}\,=\, 0
\label{ntwo23}
\end{eqnarray}
Note that the Bianchi identity for a field $ M=M^a T^a$ transforming in
the adjoint representation is:
\begin{equation}
\nabla^2 M =i\commut FM
\label{ntwo24}
\end{equation}
The non-abelian analogue of the rheonomic parametrizations (\ref{ntwo6}) is
obtained in the following
way: first we write the $\cG$-valued parametrization of $F$:
\begin{equation}
F = \cF\,\eplus\eminus -\o i2\bigl(\lap\zem +\lamm\zep\bigr)\,\eminus
+\o i2\bigl(\latp\zetp +\latm\zetp \bigr)\,\eplus + M\,\zem\zetp
- M^*\,\zep\zetm
\label{ntwo25}
\end{equation}
In this way we have introduced the gauge scalars $M=M^a T^a$ and the
gauginos
$\lambda^{\scriptscriptstyle\pm}=\lambda^{\scriptscriptstyle\pm}_a T^a$,
$\tilde{\lambda}^{\scriptscriptstyle\pm} =\tilde{\lambda}^{\scriptscriptstyle
\pm}_a T^a$; their parametrizations are obtained by implementing the
Bianchis for $F$, $\nabla F=0$.  One must also take into account the Bianchi
identies for
these fields: $\nabla^2 M= i\commut FM$ and $\nabla^2
\lambda^{\scriptscriptstyle
\pm} = i\commut F{\lambda^{\scriptscriptstyle\pm}}$ ( analogously for the
tilded gauginos).
The rheonomic parametrizations fulfilling all these constraints turn out to be
the
following ones:
\begin{eqnarray}
\label {nonabpar}
F& = &\cF\,\eplus\eminus -\o i2\bigl(\lap\zem +\lamm\zep\bigr)\,\eminus
+\o i2\bigl(\latp\zetp +\latm\zetp \bigr)\,\eplus + M\,\zem\zetp
- M^*\,\zep\zetm\nonumber\\
\dell M & = & \delp M\,\eplus +\delm M\,\eminus -\o 14 \bigr(\latm\zep
-\lap\zetm\bigl)\nonumber \\
\dell \lap & = & \delp\lap\,\eplus +\delm\lap\,\eminus +\biggl(\o {\cF}2
-2i\commut {\dag M}M+i\cP\biggr)\,\zep -2i\,\delm M\,\zetp\nonumber \\
\dell\latp & = & \delp\latp\,\eplus +\delm\latp\,\eminus +\biggl(\o {\cF}2
-2i\commut {\dag M}M-i\cP\biggr)\,\zetp -2i\,\delp \dag M\,\zep\nonumber \\
\dell\cP & = & \delp\cP\,\eplus +\delm\cP\,\eminus -\o 14 \biggl[
\biggl(\delp\lap
-2\commut {\latp}M\biggr)\zem -\biggl(\delp\lamm +2\commut {\latm}{\dag M}
\biggr)
\zep \nonumber\\
&&\mbox{}-\biggl(\delm\latp +2\commut {\lamm}{\dag M}\biggr)\zetm +\biggl(
\delm\latm -2\commut {\lap}M \biggr)\zetp\biggr]
\label{ntwo26}
\end{eqnarray}
We obtain the rheonomic action for the N=2 non-abelian gauge multiplet
in two steps, setting:
\begin{equation}
{\cal L}^{(rheon)}_{non-abelian} \, = \, {\cal L}_0 \, + \, \Delta {\cal
L}_{int}
\label{ntwo27}
\end{equation}
where ${\cal L}_0 \,$ is the free part of the Lagrangian whose associated
equations
of motion would set the auxiliary fields to zero: $\cP=\cP_a T^a = 0$
The insertion of the interaction
term $\Delta {\cal L}_{int}$ corrects the equation of motion of the
auxiliary fields, depending on a holomorphic function $\cU (M)$
of the physical gauge scalars $M^b$, just as in the abelian case.
The form of ${\cal L}_0$ is given below, where the trace is performed over the
indices
of the adjoint representation:
 \begin{eqnarray}
\lefteqn{\cL=\tr\,\biggl\{\cF\biggl[F +\o i2\bigl(\lap\zem +\lamm\zep\bigr)\,
\eminus -\o i2\bigl(\latp\zetp +\latm\zetp \bigr)\,\eplus - M\,\zem\zetp
+ \dag M\,\zep\zetm\biggr] } \nonumber\\
&&\mbox{} -\o 12 \cF^2\,\eplus\eminus -\o i2 \bigl(
\lap\,\dell\lamm +\lamm\, \dell\lap\bigr)\,\eminus +\o
i2\bigl(\latp\,\dell\latm +
\latm\, \dell\latp\bigr)\,\eplus\nonumber\\
&&\mbox{} - 4 \biggl[\dell\dag M -
\o 14\bigl(\latp\zem - \lamm\zetp\bigr)\biggr]
\bigl(\cM_{\ssp}\eplus -\cM_{\ssm}\eminus\bigr)\nonumber\\
&&\mbox{} -4 \biggl[\dell M +\o 14\bigl(\latm
\zep - \lap\zetm\bigr)\biggr]\bigl(\dag{\cM}_{\ssp}\eplus -\dag{\cM}_{\ssm}
\eminus\bigr)\nonumber\\
&&\mbox{}- 4\bigl(\dag{\cM}_{\ssp}\cM_{\ssm}
+\dag{\cM}_{\ssm}\cM_{\ssp}\bigr)\,\eplus\eminus -\dell  M\bigl(\lamm\zetp
+\latp\zem\bigr) + \dell\dag M\bigl(\lap\zetm +\latm\zep\bigr)  \nonumber\\
&&+2\commut{\dag M}{M}\left((\latp\,\zetm -\latm\zetp)\eplus
-(\lap\zem-\lamm\zep)\eminus\right) \nonumber\\
&&\mbox{}-\o 14 \bigl(\lap\latp\,\zem\zetm +\lamm\latm\,\zep\zetp\bigr)+
 2\cP^2\,\eplus\eminus \biggr\}
\label{ntwo28}
\end{eqnarray}
As stated above, the variational equations associated with this action yield
the rheonomic parametrizations (\ref{ntwo26}) for the particular value
$\cP^{a}=0$ of the
auxiliary field. Furthermore they also imply $\cP^{a}=0$ as a field equation.

To determine
the form of $\Delta{\cal L}_{int}$ we suppose that in presence of this
interaction the
new field equation of $\cP^{a}$ yields
\begin{equation}
\cP^a=\o{\partial \cU(M)}{\partial M^a}+\left(\o{\partial\cU(M)}
{\partial M^a}\right)^* =\o{\partial \cU(M)}{\partial M^a}
+\o{\partial\cU^*(M^*)}{\partial {\dag M}^a}.
\end{equation}
$\cU$ is a holomorphic function of the scalars $M^a$ that characterizes
their self-interaction.
Then we can express $\dell \cP^a$ through
the chain rule: $\dell\cP^a =\o {\partial^2\cU}{\partial M^a\partial M^b}\,
\dell M^b +\o {\partial^2 \cU^*}{\partial{\dag M}^a\partial {\dag M}^b }
\,\dell {\dag M}^b $. Using the rheonomic parametrizations (\ref {ntwo26})
for $\dell M^b$ and comparing with the parametrization of $\dell\cP^a$ in
the same eq.(\ref{nonabpar}) we get the fermionic equations of motion
that the complete interacting lagrangian
should imply as variational equations:
\begin{eqnarray}
\label {fermeq}
\delp\lap_a -2i\f abc\,\latp_b\,M_c &=& -\o{\partial^2\cU}
{\partial {\dag M}^a\partial {\dag M}^b}\,\latp_b\nonumber\\
\delm\latp_a -2i\f abc\,\lap_b\,M^*_c &=& \o{\partial^2\cU}
{\partial M^a\partial M^b }\,\lap_b
\label{ntwo30}
\end{eqnarray}
plus, of course, the complex conjugate equations.
Furthermore also the parametrization of $\dell\cF$ is affected by having
$\cP^a$ a non-zero function of $M$. This can be seen from the parametrizations
(\ref{nonabpar}). Taking the covariant derivative
of  $\dell \lap_a$ and focusing on the $\zep\zep$ sector, one can extract
$\dell_{\bullet\ssp }\cF^a$, the component of $\dell\cF^a$ along $\zep$:

\begin{equation}
\label {ntwo32}
\dell_{\bullet\ssp }\cF^a=\f abc \dag M_b\dell_{\bullet\ssp } M_c
+\o i2 \o{\partial^2\cU}{\partial M^a\partial M^b}\,\latm_b
\end{equation}
Analogously one gets the other fermionic components of $\dell\cF^a$.\par
Summarizing, in order to obtain  $\cP^a=\o{\partial\cU(M)}{\partial M^a
\phantom{(M)}} +\o{\partial\cU*(\dag M)}{\partial {\dag M}^a}$,
to reproduce the fermionic field equations  (\ref{ntwo30}) and the last terms
 in the fermionic components of the
parametrization (\ref{ntwo32}) of $\dell\cF^a$, we have to set:
\begin{eqnarray}
\lefteqn {\Delta\cL_0\,=\, 4i\o{\partial\cU}{\partial M^a}
(\o{F^a}2 +i\cP^a\,\eplus\eminus) -4i\o{\partial\cU^*}{\partial {\dag M}^a}
(\o{F^a }2 -i\cP^a\,\eplus\eminus)}\nonumber\\
&&\mbox{} + i\biggl(
\o {\partial^2\cU} {\partial M^a\partial M^b } \latm_a\lap_b +
\o {\partial^2\cU^*} {\partial {\dag M}^a\partial {\dag M}^b} \latp_a\lamm_b
\biggr)\eplus\eminus\nonumber\\
&&\mbox{}+\biggl(\o{\partial\cU}{\partial M^a} =\o{\partial\cU^*}{\partial
{\dag M}^a} \biggr)\biggl[(\latp_a\zetm -\latm_a\zetp)\eplus +
(\lap_a\zem -\lamm_a\zep)\eminus\biggr]\nonumber\\
&&\mbox{}2i\biggl[2\cU -M^a\biggl(\o{\partial\cU}{\partial M^a} -
\o{\partial\cU^*}{\partial {\dag M}^a}\biggr)\biggr]\zem\zetp +
2i\biggl[2\cU^* -{\dag M}^a\biggl(\o{\partial\cU^*}{\partial {\dag M}^a} -
\o{\partial\cU}{\partial M^a}\biggr)\biggr]\zep\zetm\nonumber\\
\label{ntwo34}
\end{eqnarray}
Note that $\cU$ must be a gauge singlet. A linear potential of the
type $\cU =\sum_a c^a M^a$ with $c^a ={\rm const}$ does not satisfy this
requirement. Hence the "linear potential'' of the abelian case,
corresponding to the insertion of a  Fayet-Iliopoulos term has no non-abelian
counterpart.
Similarly a $\theta$-term is also ruled out in the non-abelian case.
Indeed a term like $\o {\theta^a} {2\pi} F^a$
would not be gauge-invariant, with a constant $\theta^a$. Also in this case, a
term of this
type would be implied by a linear superpotential ${\cal U}$, which is therefore
excluded.
The problem is that no linear function of the gauge scalars $M^{a}$ can be
gauge-invariant.
\par
In conclusion, if the Lie
algebra $\cG$ is not semisimple, then for each of its $U(1)$ factors we can
introduce
a  Fayet-Iliopoulos and a $\theta$-term. As we are going to see, the same
property will
occur in the N=4 case. Fayet-Iliopoulos terms are associated only with abelian
factors of the
gauge-group, namely with the center ${\cal Z} \subset {\cal G}$ of the gauge
Lie-algebra.
This yield of supersymmetry perfectly matches with the properties of the
K\"ahler or
HyperK\"ahler quotients. Indeed we recall from section I that the level set of
the momentum
map (see eq.(\ref{levelsetdef}) is well-defined only for $\zeta \, \in \, R^3
\otimes {\cal Z}^*$
in the HyperK\"ahler case and for $\zeta \, \in \, R \otimes {\cal Z}$ in the
K\"ahler
case, ${\cal Z}^*$ being the center of the dual Lie-algebra ${\cal G}^*$. Now
the
level parameters $\zeta$ are precisely identified with the parameters
introduced into the
Lagrangian by the Fayet-Iliopoulos terms.

\section{R-symmetries of the N=2 Landau-Ginzburg model with and without
local gauge symmetries}
\label{n2rsym}
As we stated in the introduction, a crucial role in the topological twist of
the N=2 and
N=4 theories is played by the so called R-symmetries. These are global
symmetries of the
rheonomic parametrizations (namely automorphisms of the supersymmetry algebra)
and of the
action (both the rheonomic one and that concentrated on the bosonic
world-sheet) that have
a non trivial action also on the gravitino one-forms (in the global theories
this means on the
supersymmetry parameters, but when extending  the analysis to the locally
supersymmetric case
this means also on the world-sheet gravitinos). In the N=2 theories the
R-symmetry group
is $U(1)_L \otimes U(1)_R$, the first $U(1)_L$ acting as a phase rotation
$\zeta^{\pm}
\, \longrightarrow \, \zeta^{\pm} \, e^{\pm i \alpha_L}$ on the left-moving
gravitinos,
and leaving the right-moving gravitinos invariant, the second $U(1)_R$ factor
rotating
in the same way the right-moving gravitinos ${\tilde \zeta}^{\pm}
\, \longrightarrow \,{\tilde \zeta}^{\pm} \, e^{\pm i \alpha_R}$ and leaving
the left-moving
ones invariant. In the N=4 case, as we are going to see the R-symmetry extends
to an
$U(2)_L \otimes U(2)_R$ group each $U(2)$-factor acting on a doublet of complex
gravitinos
 $(\zeta^{\pm} ,\chi^{\pm})$ with or without the tildas.
\par
 We begin by considering the R-symmetries of the N=2 Landau-Ginzburg model with
abelian
gauge symmetries discussed in the previous sections.
\par
Let us assume that the superpotential $W(X)$ of the gauge invariant
Landau-Ginzburg model
is quasi-homogeneous of degree $d \in {\bf R}$ with scaling weights
$\omega_i\in {\bf R}$
for the chiral scalar fields $X^{i}$. This means that if we rescale each
$X^{i}$ according
to the rule:
\begin{equation}
X^{i} \, \longrightarrow \,  \exp \bigl [ \omega_i \lambda \bigr ] \,  X^{i}
\label{scaleweights}
\end{equation}
where $\lambda \in {\bf C}$ is some constant complex parameter, then the
superpotential rescales
as follows:
\begin{equation}
W\,\left ( \, e^{\omega_i \lambda } \,X^{i}\, \right ) ~=~\exp \bigl [ d
\lambda \bigr ]
\, W\left ( X^{i}\right )
\label{rescalesuperpotential}
\end{equation}
Under these assumption, we can easily verify that
the rheonomic parametrizations, the rheonomic and world-sheet action of the N=2
locally gauge
invariant Landau-Ginzburg model are also invariant under the following global
$U(1)_L \otimes U(1)_R$ transformations:

\begin{center}
\begin{tabular}
{c c c c c c c}

$\zeta^{\pm}$ & $\longrightarrow$ & $\exp[ \pm i\alpha_L] \, \zeta^{\pm}$ &
$~~~~~~$ &
${\tilde\zeta}^{\pm}$ & $\longrightarrow$ & $\exp[\pm i\alpha_R]
{\tilde\zeta}^{\pm}$\\
$\lambda^{\pm}$ & $\longrightarrow$ & $\exp[\pm i\alpha_R] \, \lambda^{\pm}$ &
$~~~~~~$ &
${\tilde\lambda}^{\pm}$ & $\longrightarrow$ & $\exp[\pm i\alpha_L]
{\tilde\lambda}^{\pm}$\\
$M$ & $\longrightarrow$ & $\exp[i(\alpha_L -\alpha_R)] \, M$ & $~~~~~~$ &
$M^{*}$ & $\longrightarrow$ & $\exp[-i(\alpha_L - \alpha_R)] M^{*}$\\
$~$&$~$&$\cP$&$\longrightarrow$&$\cP$&$~$&$~$\\
$~$&$~$&${\cal A}$&$\longrightarrow$&${\cal A}$&$~$&$~$\\
$X^{i}$ & $\longrightarrow$ & $\exp[-i\o{\omega_i \,(\alpha_L +
\alpha_R)}{d}]\, X^{i}$ &
 $~~~~~~$
& $X^{i*}$ & $\longrightarrow$ & $\exp[i\o{\omega_i \,(\alpha_L +
\alpha_R)}{d}]\,X^{i^*}$\\
$\ps i$ & $\longrightarrow$ & $\exp[i\o{(d-\omega_i)\alpha_L
-\omega_i\alpha_R}{d}] \, \ps i$ &
$~~~~~~$ &
$\pst i$ & $\longrightarrow$ & $\exp[i\o{(d-\omega_i)\alpha_R
-\omega_i\alpha_L}{d}]  \, \pst i$ \\
$\pss i$ & $\longrightarrow$ & $\exp[-i\o{(d-\omega_i)\alpha_L -
\omega_i\alpha_R}{d}] \, \pss i$
& $~~~~~~$ &
$\psts i$ & $\longrightarrow$ & $\exp[-i\o{(d-\omega_i)\alpha_R -
\omega_i\alpha_L}{d}] \, \psts i$\\
\end{tabular}
\end{center}
\begin{equation}
{}~~~~\label{n2gaugeRsym}
\end{equation}

If we define the R-symmetry charges of a field $\varphi$ by means of the
formula
\begin{equation}
\varphi \, \longrightarrow \,
\exp \bigl [ \, i \, \left ( q_L \, \alpha_L \, +\,q_R \, \alpha_R \right ) \,
\bigr ]\,Ê\varphi
\label{rcharges}
\end{equation}
then the charge assigments of the locally gauge invariant N=2 Landau-Ginzburg
model are
displayed in table I.
\par
We can also consider a {\sl rigid N=2 Landau-Ginzburg} model. By this we mean a
Landau-Ginzburg
theory of the type described in the previous sections, where the coupling to
the gauge fields
has been suppressed. The structure of such a theory is easily retrieved from
our general
formulae (\ref{ntwo10}) , (\ref{ntwo13}), (\ref{ntwo14}) by setting the
gauge-coupling constant
to zero: redefine
$q^{i}_{j} \, \longrightarrow \, g \, {\bar q}^{i}_{j}$ and then let $g\,
\longrightarrow \, 0$.
In this limit the matter fields decouple from the gauge fields and we obtain
the following world-sheet lagrangian:
\begin{eqnarray}
\lefteqn{\cL^{(ws)}_{chiral}\, =\, -\bigl(\partial_{+} X^{i^*}\partial_{-} X^i
+
\partial_{-} X^{i^*}\partial_{+} X^i
\bigr)}\nonumber \\
&&\mbox{} + 2i\bigr(\ps i\partial_{-}\pss i +\pst i\partial_{+}\psts i\bigr)
+2i\bigr(\pss i\partial_{-}\ps i +\psts i\partial_{+}\pst i\bigr) \nonumber \\
&&\mbox{}+
8\Biggl\{ \bigl ( \ps i\pst j\partial_i\partial_j {\cal W} +\coco\bigr )
+\partial_i {\cal W}\partial_{i^*}{\cal W}^* \Biggr\}
\label{rigidLGaction}
\end{eqnarray}
where to emphasize that we are discussing a different theory we have used a
curly letter
${\cal W}(X)$ to denote the superpotential. The action (\ref{rigidLGaction})
defines a
model extensively studied in the literature both for its own sake
\cite{LGliterature}
and in its topological version \cite{topolLGliterature}.
 This action  is invariant
against the supersymmetry transformations that we derive from the rheonomic
parametrizations
(\ref{ntwo10}) upon suppression of the gauge coupling ($g\, \longrightarrow \,
0$), namely from:
\begin{eqnarray}
\dell X^i & = & \partial_{+} X^i\,\eplus +\partial_{-} X^i\,\eminus +\ps i\zem
+ \pst
i\zetm\nonumber \\
\dell X^{i^*} & = & \partial_{+} X^{i^*}\,\eplus +\partial_{-} X^{i^*}\,
\eminus -\ps {i^*}
\zep - \pst {i^*}\zetp\nonumber \\
\dell\ps i & = & \partial_{+}\ps i\,\eplus +\partial_{-}\ps i\,\eminus -\o
i2\partial_{+}
X^i\,\zep + \etasu ij \partial_{j^*}{\cal W}^*\,\zetm \nonumber\\
\dell\ps {i^*} & = & \partial_{+}\ps {i^*}\,\eplus +\partial_{-}\ps {i^*}\,
\eminus +\o i2\partial_{-}
X^{i^*}\,\zem + \etasu ji \partial_{j}{\cal W}\,\zetp
 \nonumber\\
\dell \pst i & = & \partial_{+}\pst i\,\eplus +\partial_{-}\pst i\,\eminus -\o
i2\partial_{-}
X^i\,\zetp - \etasu ij \partial_{j^*}{\cal W}^*\,\zem \nonumber\\
\dell \pst {i^*} & = & \partial_{+}\pst {i^*}\,\eplus +\partial_{-}\pst {i^*}\,
\eminus +\o i2\partial_{+}
X^{i^*}\,\zetm - \etasu ji \partial_{j}{\cal W}\,\zep
\label{rigidLGparam}
\end{eqnarray}
Assuming that  under the rescalings (\ref{scaleweights}) the superpotential
${\cal W}(X)$ has
the scaling property (\ref{rescalesuperpotential}) with an appropriate
$d=d_{{\cal W}}$ then
the rigid Landau-Ginzburg model admits a $U(1)_L \otimes U(1)_R$ group of
R-symmetries whose
action on the fields is {\sl formally} the restriction, to the matter fields of
the R-symmetries
(\ref{n2gaugeRsym}), namely:
\begin{center}
\begin{tabular}
{c c c c c c c}

$X^{i}$ & $\longrightarrow$ & $\exp[-i\o{\omega_i \,
(\alpha_L + \alpha_R)}{d_{\cal W}}]\, X^{i}$ &
 $~~~~~~$
& $X^{i*}$ & $\longrightarrow$ & $\exp[i\o{\omega_i \,
(\alpha_L + \alpha_R)}{d_{\cal W}}]\,X^{i^*}$\\
$\ps i$ & $\longrightarrow$ & $\exp[i\o{(d_{\cal W}-\omega_i)\alpha_L
-\omega_i\alpha_R}
{d_{\cal W}}] \, \ps i$ &
$~~~~~~$ &
$\pst i$ & $\longrightarrow$ & $\exp[i\o{(d_{\cal W}-\omega_i)\alpha_R
-\omega_i\alpha_L}
{d_{\cal W}}]  \, \pst i$ \\
$\pss i$ & $\longrightarrow$ & $\exp[-i\o{(d_{\cal W}-\omega_i)\alpha_L -
\omega_i\alpha_R}
{d_{\cal W}}] \, \pss i$
& $~~~~~~$ &
$\psts i$ & $\longrightarrow$ & $\exp[-i\o{(d_{\cal W}-\omega_i)\alpha_R -
\omega_i\alpha_L}
{d_{\cal W}}] \, \psts i$\\
\end{tabular}
\end{center}
\begin{equation}
{}~~~~\label{rigidLGRsym}
\end{equation}
One, however, has to be careful that the parameter $d_{\cal W}$ in eq.s
(\ref{rigidLGRsym})
is the scale dimension of the superpotential ${\cal W}(X^{i})$ and not $d$, the
scale dimension
of the original $W(X)$ of the gauge coupled model. This discussion is relevant
in view of
the N=2 gauge model considered by Witten \cite{Wittenphases}
as an interpolation between a rigid N=2 Landau-Ginzburg theory and an N=2
$\sigma$-model, that
appear as the low energy effective actions in two different phases of the same
gauge theory.
In Witten's case the superpotential of the locally gauge invariant
Landau-Ginzburg theory
is chosen as follows:
\begin{equation}
W\left ( X^{I} \right) ~=~X^{0} \, {\cal W} \left ( X^{i} \right )
\label{wittensuperpotential}
\end{equation}
where the index $i$ runs on $n$ values $i=1,...,n$, the index $I$ runs on $n+1$
values
$I=0,1,.....,n$ and ${\cal W}(X^{i})$ is a quasi-homogeneous holomorphic
function of degree
$d_{\cal W}$ under the rescalings (\ref{scaleweights}) with appropriate choices
of the
$\omega_i$. Then choosing arbitrarily a scale weight $\omega_0$ for the field
$X^{0}$, the
complete superpotential $W(X^{I})$ becomes a quasi-homogeneous function of
degree
$d=d_{\cal W} \,+\,\omega_0$. Now in Witten's model, as we are going to see
later in our
discussion of the N=2 phases, there is a phase where the gauge multiplet
becomes massive,
together with the multiplet of $X^{0}$, while all the $X^{i}$-multiplets are
massless and have
 vanishing vacuum expectation values. In this phase the low energy effective
action is
a rigid Landau-Ginzburg model with superpotential ${\cal W}(X^{i})$. In this
case, if we want
to identify the R-symmetries of the effective action with those of the original
theory, something
which is important in the discussion of the topological twists, we have to be
careful to choose
$\omega_0 = 0$. Only in this case $d=d_{\cal W}$  and eq.s (\ref{rigidLGRsym})
are truely
the restriction of eq.s (\ref{n2gaugeRsym}).
\par
An extremely opposite case  occurs in the N=2 reinterpretations of the N=4
models.
As we are going to see, also there the superpotential of the gauge model has
the structure
(\ref{wittensuperpotential}) but, in this case, the holomorphic function is not
quasi-homogeneous,
a fact that can be retold by saying that $d_{\cal W}=0$ with $\omega_i = 0$. In
this case the
R-symmetries  of the rigid Landau-Ginzburg model (\ref{rigidLGRsym}) are
undefined and loose meaning.
Hovever, from the N=4 structure of the model we deduce the existence of an
R-symmetry where
the fields $X^{i}$ have $q_L =q_R=0$ , their fermionic partners $\ps i$ and
$\pst i$ have
$(q_L = 1,q_R=0)$ and $(q_L=0,q_R=1)$ respectively,  while $X^{0}$ has charges
$(q_L=-1,q_R=-1)$,
 its partners $\ps 0 , \pst 0$ being assigned the charges $(q_L=-1,q_R=0)$ and
$(q_L=0,q_R=-1)$,
respectively. This result is reconciled with general N=2 formulae if we declare
that $\omega_0 =1$
which implies $d=1$. With this choice the above charge assignements, are the
same as those
following from formulae (\ref{n2gaugeRsym}). The reason why in this case the
formulae of
the rigid Landau-Ginzburg model (\ref{rigidLGRsym}) become meaningless is
simple: in this
case differently from Witten's case there is no rigid Landau-Ginzburg phase.
For all value
of the parameters we end up in a $\sigma$-model phase. Indeed the above
assignments of the
R-charges is just the one typical of the $\sigma$-model. This will become clear
after we
have discussed the N=2 $\sigma$-model and its global symmetries.

\section{N=2 sigma-models}
\label{sigmod}
As a necessary term of comparison for our subsequent discussion of the
effective low energy
lagrangians of the N=2 matter coupled gauge models and of their topological
twists, in the present
section we consider the rheonomic construction of  the N=2 $\sigma$-model. By
definition,
this is a theory of maps:
 \begin{equation}
X ~~~~: ~~\Sigma ~~~\longrightarrow~~~{\cal  M}
\label{sigmod1}
\end{equation}
from a two-dimensional world sheet $\Sigma$ that, after Wick rotation, can be
identified with
 a Riemann surface, to a K\"ahler manifold ${\cal M}$, whose first
Chern number  $c_1({\cal M})$ is not necessarily vanishing. In the specific
case when
 ${\cal M}$ is a Calabi-Yau n-fold ($c_1 \, = \, 0$) the $\sigma$-model leads
to an
 N=2 superconformal field theory with central charge $c\, = \, 3n$ but, as far
as ordinary N=2 supersymmetry is concerned, the Calabi-Yau condition is not
required,
 the only restriction on the target manifold  being that it is
K\"ahlerian.
\par
Our notation is as follows.  The holomorphic coordinates of the K\"ahlerian
target
manifold ${\cal M}$ are denoted by $X^{i}$ ($i=1,....,n)$, their complex
conjugates by
$X^{i^*}$. The field content of the N=2 $\sigma$-model is identical with
 that of the $N=2$ Landau-Ginzburg theory:
in addition to the $X$-fields, that  transform
 as world-sheet scalars, the spectrum contains four sets of
of spin $1/2$ fermions, $\psi^{i} \, , \, {\tilde \psi}^{i} \, , \,
\psi^{i^{*}} \, ,
\, {\tilde \psi}^{i^{*}}$, that appear in the N=2 rheonomic parametrizations of
$dX^{i}$
 and $dX^{i^{*}}$ :
\begin{eqnarray}
dX^{i} &=& {\Pi}^{i}_{+} \, e^{+} \, + {\Pi}^{i}_{-} \, e^{-} \, + \, \psi^{i}
\, \zeta^{-}
\, + \, {\tilde \psi}^{i} \, {\tilde \zeta}^{-}\nonumber\\
dX^{i^{*}} &=& {\Pi}^{i^{*}}_{+} \, e^{+} \, + {\Pi}^{i^{*}}_{-} \, e^{-} \,
 -\, \psi^{i^{*}} \, \zeta^{-}
\, -\, {\tilde \psi}^{i^{*}} \, {\tilde \zeta}^{-}
\label{twosigmod1}
\end{eqnarray}
The equations above are identical with the homologous rheonomic
parametrizations of the
Landau-Ginzburg theory (the first two of eq.s \ref{rigidLGparam}).  The
difference with the
 Landau-Ginzburg case appears at the level of the rheonomic parametrizations of
the fermion
 differentials. Rather than the last four of eq.s (\ref{rigidLGparam} we write:
\begin{eqnarray}
\nabla \, \psi^{i} &=& \nabla_{+} \psi^{i} \, e^{+} \, + \, \nabla_{-} \psi^{i}
\,e^{-} - \,
\o{i}{2} \, \Pi^{i}_{+} \, \zeta^{+} \nonumber\\
\nabla \, {\tilde \psi}^{i} &=& \nabla_{+} {\tilde \psi}^{i} \, e^{+} \, + \,
\nabla_{-}
 {\tilde \psi}^{i} \,e^{-} - \,
\o{i}{2} \, \Pi^{i}_{-} \, {\tilde \zeta}^{+} \nonumber\\
\nabla \, \psi^{i^{*}} &=& \nabla_{+} \psi^{i^{*}} \, e^{+} \, + \, \nabla_{-}
 \psi^{i^{*}} \,e^{-}+\,
\o{i}{2} \, \Pi^{i^{*}}_{+} \, \zeta^{-} \nonumber\\
\nabla \, {\tilde \psi}^{i^{*}} &=& \nabla_{+} {\tilde \psi}^{i^{*}} \, e^{+}
\,
+ \, \nabla_{-} {\tilde \psi}^{i^{*}} \,e^{-} + \,
\o{i}{2} \, \Pi^{i^{*}}_{-} \, {\tilde \zeta}^{-}
\label{twosigmod2}
\end{eqnarray}
where the symbol $\nabla$ denotes the covariant derivative with respect to the
target space
 Levi-Civita connection:
\begin{eqnarray}
\nabla \, \psi^{i} &=& d \psi^{i} \, - \, \Gamma^{i}_{jk} \, dX^{j} \, \psi^{k}
\nonumber\\
\nabla \, {\tilde \psi}^{i} &=& d {\tilde \psi}^{i}  \, - \, \Gamma^{i}_{jk} \,
dX^{j} \,
{\tilde \psi}^{k}\nonumber\\
\nabla \, \psi^{i^{*}} &=& d \psi^{i^{*}} \, - \,
\Gamma^{i^{*}}_{j^{*}k^{*}} \, dX^{j^{*}} \, \psi^{k^{*}}\nonumber\\
\nabla \, {\tilde \psi}^{i^{*}} &=& d {\tilde \psi}^{i^{*}} \, - \,
\Gamma^{i^{*}}_{j^{*}k^{*}} \, dX^{j^{*}} \, {\tilde \psi}^{k^{*}}
\label{twosigmod3}
\end{eqnarray}
In agreement with standard conventions the metric, connection and curvature of
the
 K\"ahlerian target manifold are given by:
\begin{eqnarray}
g_{ij^{*}}&=& \o{\partial}{\partial X^{i}} \, \o {\partial}{\partial X^{j^{*}}}
 {\cal K}\nonumber\\
\Gamma^{i}_{jk} &=& - g^{il^{*}} \, \partial_{j} g_{kl^{*}}\nonumber\\
\Gamma^{i^{*}}_{j^{*}k^{*}} &=& - g^{i^{*}l} \, \partial_{j^{*}}
g_{k^{*}l}\nonumber\\
\Gamma^{i}_{~j}&=&\Gamma^{i}_{jk} \, dX^{k}\nonumber\\
R_{i^{*}jk^{*}l}&=&g_{ip^{*}} \, R^{p}_{~jk^{*}l}\nonumber\\
R^{p}_{~jk^{*}l}&=& \partial_{k^{*}} \, \Gamma^{p}_{jl}\nonumber\\
R^{i}_{j}&=& R^{i}_{~jk^{*}l} \, dX^{k^{*}} \, \wedge \, dX^{l}
\label{twosigmod4}
\end{eqnarray}
where ${\cal K}(X^*,X)$ denotes the K\"ahler potential.
The parametrizations (\ref{twosigmod1}) and (\ref{twosigmod2}) are the unique
solution to the
 Bianchi identities:
\begin{eqnarray}
d^2 \, X^{i} &=& d^2 \, X^{i^{*}} \, = \, 0 \nonumber\\
\nabla^2 \, \psi^{i} &=& -R^{i}_{~j} \, \psi^{j}\nonumber\\
\nabla^2 \, {\tilde \psi}^{i} &=& -R^{i}_{~j} \, {\tilde \psi}^{j}\nonumber\\
\nabla^2 \, \psi^{i^{*}} &=& -R^{i^{*}}_{~j^{*}} \, \psi^{j^{*}}\nonumber\\
\nabla^2 \, {\tilde \psi}^{i^{*}} &=& -R^{i^{*}}_{~j^{*}} \, {\tilde
\psi}^{j^{*}}
\label{twosigmod5}
\end{eqnarray}
The complete rheonomic action that yields these parametrizations as outer field
equations
is given by the following expression:
\begin{eqnarray}
S_{{\rm rheonomic}}~& =&~ \int \,\Big [ g_{ij^{*}} \, \left ( \, d X^{i} \, -
\psi^{i}
 \zeta^{-} \,
 - \, {\tilde \psi}^{i} {\tilde \zeta}^{-} \, \right ) \wedge \, \left (
\Pi^{j^{*}}_{+} \,
 e^{+} \, - \, \Pi^{j^{*}}_{-} e^{-} \, \right ) \nonumber\\
&+& \, g_{ij^{*}} \, \left ( \, d X^{j^{*}} \,+ \psi^{j^{*}} \zeta^{+} \, + \,
{\tilde \psi}^{j^{*}} {\tilde \zeta}^{+} \, \right ) \wedge \, \left (
\Pi^{i}_{+} \,
 e^{+} \, - \, \Pi^{i}_{-} e^{-} \, \right )\nonumber\\
&+& \, g_{ij^{*}} \, \left ( \Pi^{i}_{+} \, \Pi^{j^{*}}_{-} \, + \,
\Pi^{i}_{-} \, \Pi^{j^{*}}_{+} \right ) \, e^{+} \, \wedge \, e^{-}\nonumber\\
&-&\, 2 i\, g_{ij^{*}} \, \left ( \, \ps i \, \nabla \, \psi^{j^{*}} \,
 \wedge \, e^{+} \, - \, {\tilde \psi}^{i} \, \nabla \, {\tilde \psi}^{j^{*}}
\,
 \wedge \, e^{-} \right )\nonumber\\
&-&\, 2 i\, g_{ij^{*}} \, \left ( \, \psi^{j^{*}} \, \nabla \, \psi^{i} \,
 \wedge \, e^{+} \, - \, {\tilde \psi}^{j^{*}} \, \nabla \, {\tilde \psi}^{i}
\,
\wedge \, e^{-} \right )\nonumber\\
&-& \, g_{ij^{*}} \, \left ( \, dX^{i} \, \psi^{j^{*}} \, \wedge \, \zeta^{+}
\, - \,
dX^{i} \, {\tilde \psi}^{j^{*}} \, \wedge \, {\tilde \zeta}^{+}
\right )\nonumber\\
&+& \, g_{ij^{*}} \, \left ( \, dX^{j^{*}} \, \psi^{i} \, \wedge \, \zeta^{-}
\, - \,
dX^{j^{*}} \, {\tilde \psi}^{i} \, \wedge \, {\tilde \zeta}^{-}
\right )\nonumber\\
&+& \, g_{ij^{*}} \, \left ( \, \ps i \, \psts j \, \zem\,
\wedge \, \zetp \, -\, \pss j \, \pst i \,
 \zep \, \wedge \, \zetm \, \right )\nonumber\\
&+& \, 8 \, R_{ij^{*}kl^{*}} \, \psi^{i} \, \psi^{j^{*}} \, {\tilde \psi}^{k}
\, {\tilde \psi}^{l^*} \, e^{+} \, \wedge \, e^{-} \Big ]
\label{twosigmod6}
\end{eqnarray}
{}From eq. (\ref {twosigmod6}) we immediately obtain the world-sheet action in
second order formalism,
 by deleting the terms containing the fermionic vielbein
$\zeta$.s and by substituting back the value of the auxiliary fields $\Pi$.s
determined by their
 own field equations. The result is:
\begin{eqnarray}
S_{\rm {world-sheet}} &=& \int \, \Big [ \, - \, g_{ij^{*}} \, \left (
\partial_{+} X^{i} \, \partial_{-} X^{j^{*}} \, + \,
\partial_{-} X^{i} \, \partial_{+} X^{j^{*}} \, \right )\nonumber\\
&+& 2 i \, g_{ij^{*}}  \, \left ( \, \psi^{i} \, \nabla_{-} \psi^{j^{*}} \, +
\,
\psi^{j^{*}} \, \nabla_{-} \, \psi^{i} \right )\nonumber\\
&+& 2 i \, g_{ij^{*}}  \, \left ( \, {\tilde \psi}^{i} \, \nabla_{+} {\tilde
\psi}^{j^{*}}
 \, + \,
{\tilde \psi}^{j^{*}} \, \nabla_{+} \, {\tilde \psi}^{i} \right )\nonumber\\
&+& 8 \, R_{ij^{*}kl^{*}} \psi^{i} \, \psi^{j^{*}} \, {\tilde \psi}^{k}
\, {\tilde \psi}^{l^{/star}}  \, \Big ] \, d^2 z
\label{twosigmod7}
\end{eqnarray}
where we have denoted by
\begin{eqnarray}
\nabla_{\pm} \, \psi^{i} &=& \partial_{\pm} \, \psi^{i} \, - \, \Gamma^{i}_{jk}
\, \partial_{\pm} X^{j} \, \psi^{k}\nonumber\\
\nabla_{\pm} \, \psi^{i^{*}} &=& \partial_{\pm} \, \psi^{i^{*}} \, - \,
\Gamma^{i^{*}}_{j^{*}k^{*}} \, \partial_{\pm} X^{j^{*}} \, \psi^{k^{*}}
\label{twosigmod8}
\end{eqnarray}
the world-sheet components of the target-space covariant derivatives: identical
equations hold for the tilded fermions.
The world-sheet action (\ref{twosigmod7}) is invariant against the
supersymmetry transformation rules
 descending from the rheonomic parametrizations (\ref{twosigmod1}) and
(\ref{twosigmod2}), namely:
\begin{eqnarray}
\delta \, \psi^{i} &=& - \, \o{i}{2} \, \partial_{+} X^{i} \, \varepsilon^{+}
\,
- \, {\tilde \varepsilon}^{-} \, \Gamma^{i}_{jk} \, {\tilde \psi}^{j} \,
\psi^{k} \nonumber\\
\delta \,{\tilde \psi}^{i} &=& - \, \o{i}{2} \, \partial_{-} X^{i}
\,{\tilde \varepsilon}^{+} \, - \, \varepsilon^{-} \, \Gamma^{i}_{jk} \,
 \psi^{j} \, {\tilde \psi}^{k} \nonumber\\
\delta \, \psi^{i^{*}} &=& +\, \o{i}{2} \, \partial_{+} X^{i^{*}} \,
 \varepsilon^{-} \, + \, {\tilde \varepsilon}^{+} \,
\Gamma^{i^{*}}_{j^{*}k^{*}}
 \, {\tilde \psi}^{j^{*}} \, \psi^{k^{*}} \nonumber\\
\delta \,{\tilde \psi}^{i^{*}} &=&+ \, \o{i}{2} \, \partial_{-} X^{i^{*}}
\,{\tilde \varepsilon}^{-} \, + \, \varepsilon^{+} \,
\Gamma^{i^{*}}_{j^{*}k^{*}}
 \,  \psi^{j^{*}} \, {\tilde \psi}^{k^{*}}
\label{twosigmod9}
\end{eqnarray}
Comparing with the transformation rules defined by eq.s (\ref{rigidLGparam})
we see that in the variation of
 the fermionic fields, the term proportional to the derivative of the
superpotential has been
 replaced with a fermion bilinear containing the Levi-Civita
connection of the target manifold.  Indeed one set of rules can be obtained
from the other
by means of the replacement:
\begin{eqnarray}
\eta^{ij^{*}} \, \partial_{j^{*}} \, W^* &\longrightarrow & - \Gamma^{i}_{jk}
\, {\tilde \psi}^{j} \, \psi^{k}\nonumber\\
\eta^{i^{*}j} \, \partial_{j} \, {W}&\longrightarrow &
\Gamma^{i^{*}}_{j^{*}k^{*}} \, {\tilde \psi}^{j^{*}} \, \psi^{k^{*}}
\label{twosigmod10}
\end{eqnarray}
This fact emphasizes that in the
$\sigma$-model the form  of the interaction and hence all the quantum
properties of the theory
 are dictated by the K\"ahler structure, namely by the real, non holomorphic
K\"ahler potential
 ${\cal K}(X,X^*)$, while in the Landau-Ginzburg case the structure of the
interaction and
 the resulting quantum
properties are governed by the holomorphic superpotential ${\cal W}(X)$.
In spite of these differences, both type of models can yield
 at the infrared critical point an N=2 superconformal theory and can be related
to the same
 Calabi-Yau manifold.  In the case of the $\sigma$-model, the relation is most
direct: it suffices to take, as target manifold ${\cal M}$, the very
Calabi-Yau $n$-fold
one is interested in and to choose for the K\"ahler metric $g_{ij^{*}}$ one
representative
 in one of the available K\"ahler classes:
\begin{equation}
K \, = \, i \, g_{ij^{*}} \, dX^{i} \, \wedge \, dX^{j^{*}} \, \in \,
 \Big [ \, K \, \Big ] \, \in \, H^{(1,1)} \, \left ( \, {\cal M } \, \right )
\label{twosigmod11}
\end{equation}
If $c_1 ({\cal M})=0$, within each K\"ahler class we can readjust the choice of
the representative
 metric $g_{ij^{*}}$ , so that at each perturbative order the beta-function is
made
 equal to zero. In this way we  obtain conformal invariance and we associate an
N=2 superconformal
 theory with any N=2 $\sigma$-model on a Calabi-Yau n-fold ${\cal M}$. The N=2
gauge model
discussed in the previous sections interpolates between the $\sigma$-model and
the Landau-Ginzburg
theory with, as superpotential, the very function ${\cal W}(X)$ whose vanishing
defines ${\cal M}$ as
a hypersurface in a (weighted) projective space.
\par
As a matter of comparison a very important issue are the left-moving and
right-moving R-symmetries
of the $\sigma$-model. Indeed, also in this case,
the rheonomic parametrizations, the rheonomic and world-sheet actions
are invariant under a global $U(1)_L \otimes U(1)_R$ group. The action of this
group
on the $\sigma$-model fields, however, is different from that on the
Landau-Ginzburg fields, namely
we have:
\begin{center}
\begin{tabular}
{c c c c c c c}
$\zeta^{\pm}$ & $\longrightarrow$ & $\exp[\pm i\alpha_L] \, \zeta^{\pm}$ &
$~~~~~~$ &
${\tilde\zeta}^{\pm}$ & $\longrightarrow$ & $\exp[\pm i\alpha_R]
{\tilde\zeta}^{\pm}$\\
$X^{i}$ & $\longrightarrow$ & $X^{i}$ & $~~~~~~$
 & $X^{i*}$ & $\longrightarrow$ & $X^{i^*}$\\
$\ps i$ & $\longrightarrow$ & $\exp[i\alpha_L] \, \ps i$ & $~~~~~~$ &
$\pst i$ & $\longrightarrow$ & $\exp[i\alpha_R] \, \pst i$ \\
$\pss i$ & $\longrightarrow$ & $\exp[-i\alpha_L] \, \pss i$ & $~~~~~~$ &
$\psts i$ & $\longrightarrow$ & $\exp[-i\alpha_R] \, \psts i$ \\
\end{tabular}
\end{center}
\begin{equation}
{}~~~~\label{sigmodRsym}
\end{equation}
where $\alpha_L$ and $\alpha_R$ are the two constant phase parameters.
The crucial difference of eq.s (\ref{sigmodRsym}) with respect to eq.s
(\ref{rigidLGRsym})
resides in the R-invariance of the scalar fields $X^{i}$ that applies to the
$\sigma$-model
case, but not the Landau-Ginzburg case. As a consequence, in the $\sigma$-model
case
the fermions have fixed integer R-symmetry charges, while in the
Landau-Ginzburg case
they acquire fractional R-charges depending on the homogeneity degree of the
corresponding
scalar field and of the superpotential.
\par

\section{Extrema of the N=2 scalar potential, phases of the theory and
reconstruction of the
effective  N=2 $\sigma$-model }
\label{cpn}
Now we focus on the effective low-energy theory emerging
from the $N=2$ gauge plus matter systems  described in the above sections.
Our considerations remain
at a classical level. We are mostly interested in the case where the
effective theory is an $N=2$ $\sigma$-model.
We show how the $N=2$ $\sigma$-model Lagrangian  is
technically retrieved, in a
manner that is intimately related with the momentum map construction.
Indeed this
latter is just the geometrical counterpart of the physical concept of
low-energy effective Lagrangian. To be simple we perform our computations
in the case where the
target space of the low-energy $\sigma$-model is
the manifold ${\bf C}{\bf P}^N$.

First of all we need to recall  the structure of
the classical vacua for a system decribed by the Lagrangian (\ref{ntwo15}),
referring to the linear superpotential case:
$\cU =(\o r4-i\o{\theta}{8\pi}) M$; this structure was studied in Witten's
paper \cite{Wittenphases}. We set the fermions to zero and we have to
extremize the scalar potential (\ref{ntwo19}). Since $U$ is given by a sum of
moduli squared, this amounts to equate each term in
(\ref{ntwo19}) separately to zero. A particularly interesting situation arises
when the
Landau-Ginzburg potential has the form
\begin{equation}
\label{9.1}
W = X^0 \, {\cal W}(X^i)
\end{equation}
Here ${\cal W}(X^i)$ is a quasihomogeneous function of degree $d$ of the fields
$X^i$
that are assigned the weigths $q^i$, i.e. their charges with respect to
the abelian gauge group.
In the case all the charges $q^i$ are equal (say all equal to $1$, for
simplicity) $\cW(X^i)$ is homogeneous.
$X^0$ is a scalar field of charge $-d$. ${\cal W}(X^i)$
must moreover be {\it transverse}: $\partial_i{\cal W} =0 \hskip 3pt\forall i$
iff $X^i =0\hskip 3pt\forall i$.

In this case we have:
\begin{eqnarray}
\label{9.2}
U & = & \o 12 \biggr( r + d |X^0|^2 -\sum_i q^i |X^i|^2\biggl)^2 +8|{\cal
W}(X^i)|^2
+8|X^0|^2 |\partial_i {\cal W}|^2 \nonumber\\
&& \mbox{} +8|M|^2 \biggl(d^2|X^0|^2 + \sum_i (q^i)^2|X^i|^2\biggl),
\end{eqnarray}
and two possibilities emerge.
\begin{itemize}
\item $r>0$.\hskip 12pt In this case some of the $X^i$ must be different from
zero. Due to the transversality of ${\cal W}$ it follows that $X^0=0$. The
space
of classical vacua is characterized not only by having $X^0=0$ and
$M=0$, but also by the condition $\sum_i q^i|X^i|^2 = r$.
When $q^i =1\hskip 6pt \forall i$ this condition,
together with the $U(1)$ gauge invariance, is equivalent to the statement
that the $X^i$ represent coordinates on ${\bf C}{\bf P}^N$.
In general, the $X^i$'s are coordinates on the weighted projective space
${\bf WCP}^N_{q^1\ldots q^N}$ The last
requirement, ${\cal W}(X^i)=0$, defines the space of classical
vacua as a transverse hypersurface embedded in ${\bf C}{\bf P}^N$ or,
in general, in ${\bf WCP}^N_{q^1\ldots q^N}$.
The low energy theory around these vacua is expected to correspond
to the $N=2$ $\sigma$-model on such a hypersurface. Indeed, studying the
quadratic fluctuations one sees that the gauge field
${\cal A}$ acquires a mass due to a Higgs phenomenon; the gauge scalar $M$
becomes
massive together with those modes of the matter fields that are not
tangent to the hypersurface. The only massless degrees of freedom, i.e.
those described by the low energy theory, are the excitations tangent to
the hypersurface. The fermionic partners behave consistently. We are in
the {\it ``$\sigma$-model phase''}.
\item $r<0$. \hskip 12pt In this case $X^0$ must be different from zero.
Then it is necessary that $\partial_i {\cal W} =0\hskip 3pt\forall i$; this
implies by transversality that all the $X^i$ vanish. The space of
classical vacua is just a point.Indeed utilizing the gauge invariance we can
reduce $X^0$ to be real, so that it is fixed to have the constant value
$X^0=\sqrt{\o{-r}d}$. $M$ vanishes together with the $X^i$. The low energy
theory can now be recognized to be a theory of massles fields, the $X^i$'s,
governed by a Landau Ginzburg potential which is just ${\cal W}(x^i)$. We are
in the {\it ``Landau-Ginzburg phase''}.
\end{itemize}
Now we turn our attention to the ${\bf C}{\bf P}^N$-model, which corresponds to
the particular case in which all the charges are equal to $1$ and $W=0$.
As it is easy to see from the above
discussion, in this case the only possible vacuum phase is the $\sigma$-model
phase, i.e one must have $r>0$. We start by writing the complete
rheonomic lagrangian of the system consisting of $N+1$ chiral
multiplets with no selfinteraction $(X^A,\ps A,\pst A), \hskip 3pt
A=1,\ldots N$, coupled to an abelian gauge multiplet, each with
charge one. Differently to what we did in the previous sections,
in this section we make the dependence on the gauge coupling
constant $g$ explicit. To reinstall $g$ appropriately, after reinserting it
into the
covariant derivatives, $\nabla X^A=dx^A+ig\cA X^A$, we redefine the fields of
the gauge multiplet as follows:
\begin{equation}
\label{9.3}
\cA\hskip 3pt\longrightarrow \o 1g \cA\hskip 12pt ;\hskip 12pt
M\hskip 3pt\longrightarrow \o 1g M\hskip 12pt ;\hskip 12pt
\lambda\hskip 3pt\longrightarrow \o 1g \lambda\hskip 12pt
\end{equation}
so that at the end no modification occurs in the matter lagrangian, while
the gauge kinetic lagrangian is multiplied by $\o 1{g^2}$.  Altogether we have:
\begin{eqnarray}
\label{9.4}
\lefteqn{\cL=\o {\cF}{g^2}\biggl[F +\o i2\bigl(\lap\zem
+\lam\zep\bigr)\,\eminus
-\o i2\bigl(\latp\zetp +\latm\zetp \bigr)\,\eplus - M\,\zem\zetp
- M^*\,\zep\zetm\biggr] }\nonumber \\
&&\mbox{} -\o 1{2g^2} \cF^2\,\eplus\eminus -\o i{2g^2} \bigl(
\lap\,d\lam +\lam\, d\lap\bigr)\,\eminus +\o i{2g^2}\bigl(\latp\,d\latm
+\latm\,
d\latp\bigr)\,\eplus \nonumber \\
&&\mbox{} - \o 4{g^2} \biggl[dM^*- \o 14\bigl(\latp\zem -
\lam\zetp\bigr)\biggr]\bigl(
\cM_{\ssp}\eplus -\cM_{\ssm}\eminus\bigr)\nonumber\\
&&\mbox{} - \o 4{g^2} \biggl[dM +\o 14\bigl(\latm\zep
- \lap\zetm\bigr)\biggr]\bigl(\cM^*_{\ssp}\eplus -\cM^*_{\ssm}\eminus\bigr)
\nonumber \\
&&\mbox{}- \o 4{g^2}\bigl(\cM^*_{\ssp}\cM_{\ssm}
+\cM^*_{\ssm}\cM_{\ssp}\bigr)\,\eplus\eminus - \o 1{g^2}dM\bigl(\lam\zetp
+\latp\zem\bigr) + \o 1{g^2}dM^*\bigl(\lap\zetm +\latm\zep\bigr)\nonumber \\
&&\mbox{}-\o 1{4g^2} \bigl(\lap\latp\,\zem\zetm +\lam\latm\,\zep\zetp\bigr)+
\o 2{g^2}\cP^2\,\eplus\eminus - 2 r \cP \eplus\eminus +\o{\theta}{2\pi}
F \nonumber\\
&&\mbox{} +\o r{2g^2} \biggl[\bigl(\lap\zem
-\lam\zep\bigr)\,\eminus +\bigl(\latp\zetm
-\latm\zetp\bigr)\,\eplus\biggr] + i\o r{g^2} \biggl( M\zem\zetp
+M^*\zep\zetm\biggr)\nonumber\\
&&\mbox{} -(\dell X^A -\ps A\zem -\pst A\zetm)(\Pi^{A^*}_{\ssp}\eplus
-\Pi^{A^*}_{\ssm}\eminus)\nonumber\\
&&\mbox{} - (\dell X^{A^*} +\pss A\zep +\psts A\zetp)
(\Pi^A_{\ssp}\eplus -\Pi^A_{\ssm}\eminus)\nonumber\\
&&\mbox{}+(\Pi^{A^*}_{\ssp}\Pi^A_{\ssm} - \Pi^{A^*}_{\ssm}\Pi^A_{\ssp})
\eplus\eminus + 2i(\ps A\dell\pss A +\pss a\dell\ps A)\eplus\nonumber\\
&&\mbox{} -2i(\pst A\dell \psts A +\psts A\dell\pst A)\eminus -\ps A\pss
A\zem\zep +\pst A\psts A\zetm\zetp \nonumber\\
&&\mbox{}-\ps A\psts A \zem\zetp -\pss A\pst A\zep\zetm +\dell X^A(\pss
A\zep -\psts A\zetp) \nonumber\\
&&\mbox{} -\dell X^{A^*}(\ps A\zem -\pst A\zetm) +4MX^A\pss A\zetp\eplus
-4M^*X^{A^*}\ps A\zetm\eplus \nonumber\\
&&\mbox{} +4M^*X^A\psts A\zep\eminus -4MX^{A^*}\pst A\zem\eminus
\nonumber\\
&&\mbox{} +\biggl\{8iM^*\psts A\ps A+8iM\pst A\pss A +2i\latp\pss A X^A
+2i \latm \ps A X^{A^*}\nonumber\\
&&\mbox{} -2i\lap\psts A X^A -2i\lam\pst A X^{A^*} +2\cP X^{A^*}X^A
-8M^*MX^{A^*}X^A\biggr\}\eplus\eminus
\end{eqnarray}
The procedure that we follow  to extract the effective lagrangian is
the following. We let   the gauge coupling constant go
to infinity and we are left with a gauge invariant lagrangian describing
matter coupled to gauge fields that  have no kinetic terms. Varying the
action in these fields, the resulting equations of motion  express
the gauge fields in terms of the matter fields. Substituting back
their expressions into the lagrangian we end up with a $\sigma$-model
having as target manifold the quotient of the  manifold spanned
by the matter fields with respect to the action of the gauge group \cite{hklr}.
This procedure is nothing else, from the functional integral viewpoint,  but
the
gaussian integration over the gauge multiplet in the limit $g \,
\longrightarrow \,
\infty$.
As already pointed out in the  introduction,  to consider a gauge coupled
lagrangian without gauge kinetic terms is not a mere trick to implement
the quotient procedure in a Lagrangian formalism . It  rather amounts to
deriving the low-energy effective action around the classical vacua of the
complete,
gauge plus matter  system. Indeed we have seen that around these vacua the
oscillations of the gauge fields are massive, and thus decouple from the
low-energy point of view. So we integrate over them: furthermore all masses
are proportional to $\o {1}{g}$ and the integration makes sense for
energy-scales $E \, <<  \, \o {1}{g}$, namely in the limit $g \,
\longrightarrow \,
\infty$.
\par
Here we show in detail how the above-sketched procedure works  at the level of
the rheonomic
approach. In this way we retrieve the  rheonomic lagrangian and
the rheonomic parametrizations  of the $N=2$ $\sigma$-model, as described
in section (\ref{sigmod}), the target space being
${\bf C}{\bf P}^N$, equipped with the standard Fubini-Study metric.  The whole
procedure amounts geometrically to realize ${\bf CP}^N$ as a K\"ahler
quotient \cite{hklr}.

Let us consider the lagrangian (\ref{9.4}), {\it in the limit
$g\longrightarrow\infty$} and let us perform the variations in the
gauge fields.
\vskip 0.2cm
The variations in $\lam ,\lap ,\latm ,\latp$ give the following {\it
``fermionic constraints''}:
\begin{equation}
\label{9.5}
X^A\pss A =X^{A^*}\ps A = X^A\psts A =X^{A^*}\psts A =0
\end{equation}
Here the summation on  the capital index $A$ is understood. In the following we
use simplified notations, such as $X\psi^*$ for
$X^A\pss A$, and the like, everywhere  it  is possible without generating
confusion.

The fermionic constraints (\ref{9.5}) are explained by the bosonic
 constraint $X^*X = r$, for which the auxiliary field $\cP$,  in the limit
 $g \, \longrightarrow \,
\infty$ becomes
a Lagrange multiplier. Indeed taking the exterior derivative of this bosonic
constraint we obtain
$0=d(X^*X)=X^*dX+XdX^*$ and substituting the rheonomic parametrizations
(\ref{ntwo10}) in the gravitino sectors this implies
\begin{equation}
\label{9.6}
X^*(\psi\zem+\psit {\tilde \zeta}^{-})-X(\psi^*\zep+\psit^*\zetp)=0
\end{equation}
from which (\ref{9.5}) follows.
\vskip 0.2cm
The variation of the action with respect to  $M^*$ in the gravitino sectors
implies again the
fermionic constraints (\ref{9.5}). In the $\eplus\eminus$ sector we get
the following equation of motion:
\begin {equation}
\label{9.7}
M=\o{i\psit^*\psi}{X^*X}
\end{equation}
\vskip 0.2cm
The terms in the lagrangian (\ref{9.4}) containing the connection ${\cal A}$
are hidden in the covariant derivatives. Explicitely they are:
\begin{eqnarray}
\label{9.8}
\lefteqn{-i \cA X^A(\Pi^{A^*}_{\ssp}\eplus -\Pi^{A^*}_{\ssm}\eminus)
+i\cA
X^{A^*}(\Pi^A_{\ssp}\eplus -\Pi^A_{\ssm}\eminus) +2i \ps A (-i)\cA\pss
A\eplus }\nonumber\\
&&\mbox{}+2i\pss A i \cA\ps A\eplus -2i\pst A(-i) \cA\psts A\eminus
-2i\psts A i A\pst A\eminus\nonumber\\
&&\mbox{}+i\cA X^A (\pss A\zep -\psts A\zetp) +i\cA X^{A^*}(\ps A\zem -\pst
A\zetm) +\o{\theta}{2\pi}d\cA
\end{eqnarray}
In the gravitino sector we again retrieve the constraints (\ref{9.5}).
In the $\eplus$ and $\eminus$ sector we respectively obtain:
\begin{eqnarray}
\label{9.9}
iX^A\Pi^{A^*}_{\ssp} -iX^{A^*}\Pi^A_{\ssp} -4\ps A\pss A & =0& \nonumber\\
-iX^A\Pi^{A^*}_{\ssm} +iX^{A^*}\Pi^A_{\ssm} +4\pst A\psts A & =0&
\end{eqnarray}
At this point we take into account the variations with respect to the first
order
fields $\Pi$, that give $\Pi^A_{\ssp}=\nabla_{+} X^A=\nabla_{+}  X^A
+i\cA_{\ssp} X^A$,
and so on.  Substituting into eq.s (\ref{9.9}) and solving for
$\cA_{\ssp}, \cA_{\ssm}$ we get:
\begin{eqnarray}
\label{9.10}
\cA_{\ssp} &=& \o{-i (X\dep X^* -X^*\dep X) +
4\psi\psi^*}{2X^*X}\nonumber\\
\cA_{\ssm} &=& \o{-i (X\dem X^* -X^*\dem X) +
4\psit\psit^*}{2X^*X}
\end{eqnarray}
\vskip 0.2cm
Substituting back the expression (\ref{9.7}) for $M$ into the  lagrangian
(\ref{9.4}) in the $g\longrightarrow\infty$ limit we have
\begin{eqnarray}
\label{9.11}
\lefteqn{\cL = -\biggl[dX^A +iX^A (\cA_{\ssp}\eplus +\cA_{\ssm}\eminus)-\ps
A\zem
-\pst a\zetm\biggr] (\Pi^{A^*}_{\ssp}\eplus -\Pi^{A^*}_{\ssm}\eminus)}
\nonumber\\
&&\mbox{} -\biggl[dX^{A^*} -iX^{A^*} (\cA_{\ssp}\eplus +\cA_{\ssm}\eminus)+
\pss A\zep +\psts a\zetp\biggr] (\Pi^A_{\ssp}\eplus
-\Pi^A_{\ssm}\eminus)\nonumber\\
&&\mbox{}
-(\Pi^{A^*}_{\ssp}\Pi^A_{\ssm} +\Pi^{A^*}_{\ssm}\Pi^A_{\ssp})\eplus\eminus
+2i(\ps A d\pss A +\pss A d\ps A -2i\cA_{\ssm}\eminus\ps A\pss A)\eplus
\nonumber\\
&&\mbox{}-2i(\pst A d\psts A +\psts A d\pst A -2i \cA_{\ssp}\eplus \psit A
\psts A)\eminus - \ps A\pss A\zem\zep\nonumber\\
&&\mbox{} - \pst A\psts A\zetm\zetp -\ps A\psts A\zem\zetp -\pss A\pst A
\zep\zetm\nonumber\\
&&\mbox{}+dX^A(\pss A\zep - \psts A\zetp) -dX^{A^*}(\ps A\zem -\pst A\zetp)
\nonumber\\
&&\mbox{}-8\o {\ps A\pss B\pst B\psts A}{X^*X} \eplus\eminus +2\cP (r -
X^*X)\eplus\eminus
\end{eqnarray}
where  $\cA_{\ssp}$ and $\cA_{\ssm}$ are to be identified with their
expressions (\ref{9.10}). To obtain this expression we have also used the
``fermionic constraints'' (\ref{9.5}).
The $U(1)$ gauge invariance of the above lagrangian can be extended to
a ${\bf C}^*$-invariance, where ${\bf C}^*\equiv {\bf C} - \{0\}$ is the
complexification of the $U(1)$ gauge group, by introducing an extra
scalar field $v$ transforming appropriately. Consider the ${\bf C}^*$ gauge
transformation given by
\begin{equation}
\label{9.12}
   \begin{array}{ccc}
   X^A \hskip 3pt &\longrightarrow &\hskip 3pt e^{i\Phi} X^A\\
   X^{A^*} \hskip 3pt &\longrightarrow& \hskip 3pt e^{-i\Phi^*} X^{A^*}
   \end{array}
\hskip 12pt ; \hskip 12 pt
   \begin{array}{c}
   \ps A \hskip 3pt \longrightarrow\hskip 3pt e^{i\Phi} \ps A \\
   \ldots
   \end{array}
\hskip 12pt ;\hskip 12pt
   \ldots
\hskip 12 pt ;\hskip 12pt
   (\Phi\in {\bf C})
\end{equation}
which is just the complexification of the $U(1)$ transformation, the latter
corresponding  to the case $\Phi\in {\bf R}$, supplemented with
\begin{equation}
\label{9.13}
v\hskip 3pt\longrightarrow\hskip 3pt v+\o i2 (\Phi -\Phi^*)
\end{equation}
One realizes that under the transformations (\ref{9.12},
\ref{9.13}) the combinations $e^{-v} X^A$ (and similar ones) undergo just a
$U(1)$ transformation:
\begin{eqnarray}
\label{9.14}
e^{-v} X^A \hskip 3pt & \longrightarrow &\hskip 3pt e^{i {\rm Re}\Phi} e^{-v}
X^A \nonumber\\
e^{-v} X^{A^*} \hskip 3pt & \longrightarrow &\hskip 3pt e^{-i {\rm Re}\Phi}
e^{-v} X^{A^*}
\end{eqnarray}
By substituting
\begin{equation}
\label{9.10bis}
X^A, \ps A,\pss A,\pst A, \psts A, \Pi^A_{\ssp},\ldots
\hskip 3pt\longrightarrow\hskip 3pt e^{-v}X^A, e^{-v}\ps A,
e^{-v}\pss A,\ldots
\end{equation}
into the lagrangian (\ref{9.11}) we obtain an expression which is invariant
with respect to the ${\bf C}^*$-transformations (\ref{9.12},\ref{9.13}).\par
In particular the last term of (\ref{9.15}) becomes
\begin{equation}
-2\cP (r-e^{-2v}X^*X)
\label{9.15}
\end{equation}
If  at this point we perform  the so far delayed  variation with respect  to
the auxiliary field $\cP$,  the resulting equation of motion
identifies the extra scalar field $v$ in terms of the matter fields.
Introducing $\rho^2 \equiv r$ the result is that
\begin{equation}
\label{9.16}
e^{-v}=\o {\rho}{\sqrt{X^*X}}
\end{equation}

What is the geometrical meaning of the above ``tricks'' (introduction
of the extra field $v$, consideration of the complexified gauge group)?
The answer relies on the properties of the K\"ahler quotient construction;
extensively  discussed in \cite{hklr}, \cite{kronheimer}. Let us recall
few concepts, keeping always in touch with the example we are dealing with.
We use the notions and notations introduced in section I
\par
Let ${\bf Y}(s) =Y^a{\bf k}_a (s)$ be a Killing vector on $\cS$ (in our case
${\bf C}^{N+1}$), belonging to $\cG$ (in our case ${\bf R}$), the algebra of
the gauge group. In our case ${\bf Y}$ has a single component: ${\bf Y}=i\Phi
(X^A \o{\partial \phantom{X^A}}{\partial X^A} -X^{A^*}\o{\partial
\phantom{X^{A^*}}}{\partial X^{A^*}})$ ($\Phi\in{\bf R}$). The $X^A$'s are the
coordinates on $\cS$.
Consider the vector field $I{\bf Y}\in\cG^c$ (the complexified algebra), $I$
being the complex structure acting on $T\cS$. In our case $I{\bf Y}=
\Phi(X^A \o{\partial\phantom{X^A}}{\partial X^A} + X^{A^*} \o{\partial
\phantom{X^{A^*}}}{\partial X^{A^*}} )$.  This vector field is orthogonal to
the hypersurface
$\cD^{-1}(\zeta)$, for any level $\zeta$; that is, it generates transformations
that change the level of the surface. In our case the surface $\cD^{-1}
(\rho^2)\in {\bf C}^{N+1}$ is defined by the equation $X^{A^*}X^A =\rho^2$.
The infinitesimal transormation generated by $I{\bf Y}$ is $X^A\rightarrow
(1+\Phi)X^A$, $X^{A^*}\rightarrow (1+\Phi)X^{A^*}$ so that the transormed
$X^A$'s satisfy $X^{A^*}X^A = (1+2\Phi)\rho^2$.
As recalled in section I, the K\"ahler quotient consists in starting
from $\cS$, restricting to $\cN =\cD^{-1}(\zeta)$ and taking the quotient
$\cM =\cN /G$. The above remarks about the action of the complexified gauge
group suggest that this is equivalent (at least if we skip the problems
due to the non-compactness of $G^c$) to simply taking the quotient $\cS /
G^c$, the so-called ``algebro-geometric'' quotient \cite{hklr}, \cite{Galicki}.
\par
The K\"ahler quotient allows,  in principle to determine the expression of the
K\"ahler form on $\cM$ in terms of the original one on $\cS$. Schematically,
let $j$ be the inclusion map of $\cN$ into $\cS$, $p$ the projection from
$\cN$ to the quotient $\cM=\cN /G$, $\Omega$ the K\"ahler form on $\cS$ and
$\omega$ the K\"ahler form on $\cM$. It can be shown \cite{hklr} that
\begin{eqnarray}
\label{kq1}
\cS \hskip 3pt \stackrel{j}{\longleftarrow}\hskip 3pt & \cN =\cD^{-1}(\zeta) &
\hskip 3pt\stackrel{p}{\longrightarrow}\hskip 3pt \cM =\cN /G\nonumber\\
\Omega \hskip 3pt \longrightarrow \hskip 3pt & j^*\Omega =p^*\omega &
\hskip 3pt\longleftarrow\hskip 3pt \omega
\end{eqnarray}
In the algebro-geometric setting, the holomorphic map that associates to a
point
$s\in\cS$ (for us, $\{X^A\}\in{\bf C}^{N+1}$) its image $m\in\cM$ is obtained
as follows:\par\noindent
{\it i)} Bringing $s$ to $\cN$ by means of the finite action infinitesimally
generated by a vector field of the form ${\bf V} =I{\bf Y} = V^a{\bf k}_a$
\begin{equation}
\label{kq2}
\pi:\hskip 10pt s\in\cS\hskip 3pt\longrightarrow e^{-V} s\in\cD^{-1}(\zeta)
\end{equation}
{\it ii)} Projecting $e^{-V}$ to its image in the quotient $\cM =\cN /G$.\par
Thus we can consider  the pullback of the K\"ahler form $\omega$ through
the map $p\cdot\pi$:
\begin{eqnarray}
\label{kq3}
\cS\hskip 3pt\stackrel{\pi}{\longrightarrow}\hskip 3pt & \cN =\cD^{-1}(\zeta)
&\hskip 3pt\stackrel{p}{\longrightarrow} \hskip 3pt\cN /G\nonumber\\
\pi^* p^*\omega \hskip 3pt \longleftarrow \hskip 3pt & p^*\omega &
\hskip 3pt \longleftarrow\hskip 3pt \omega
\end{eqnarray}
Looking at (\ref{kq1}) we see that $\pi^* p^*\omega =\pi^* j^*\Omega$ so that
at the end of the day, in order to recover the pullback of $\omega$ to $\cS$
it is sufficient:
\par\noindent
{\it \phantom{i}i)} to restrict $\Omega$ to $\cN$\par\noindent
{\it ii)} to pull back this restriction to $\cM$ with respect to the map $\pi
=e^{-V}$.\par
We see from (\ref{kq2}) that the components of the vector field ${\bf V}$
must be determined by requiring
\begin{equation}
\label{kq2bis}
\cD(e^{-V}s)=\zeta
\end{equation}
But this is precisely effected in the lagrangian context by the term
having as Lagrange multiplier the auxiliary field $\cP$, see eq. (\ref{9.16}),
through the equation of motion of $\cP$, once we have introduced the extra
field
$v$ (which is now interpreted as the unique component of the vector field
${\bf V}$) to make the lagrangian invariant under the complexified gauge
group ${\bf C}^*$. The lagrangian formalism of
$N=2$ supersymmetry perfectely matches the key points of the momentum map
construction. This allows us to determine the form of the map $\pi$ : it
corresponds to the transformations (\ref{9.10bis}). The steps that we are
going to discuss in treating the lagrangian just consist in implementing
the K\"ahler quotient as in (\ref{kq3}). Thus it is  clear why at the end
we obtain the $\sigma$-model on the target space $\cM$ (in our case
${\bf CP}^N$) endowed with the K\"ahler metric corresponding to the
K\"ahler form $\omega$. In our example such metric is the Fubini-Study
metric. Indeed one can show in full generality  \cite{hklr}
that the  K\"ahler potential $\hat K$ for the manifold $\cM$, such that
$\omega= 2i\partial\bar\partial\hat K$ is given by
\begin{equation}
\label{kq4}
\hat K = K|_{\cN} + V^a \zeta_a
\end{equation}
Here $K$ is the K\"ahler potential on $\cS$; $K|_{\cN}$ is
the restriction of $K$ to $\cN$, that is, it is computed after acting on the
point $s\in\cS$ with the transformation $e^{-V}$ determined by eq.
(\ref {kq2bis}); $V^a$ are the components of the vector field ${\bf V}$
along the $a^{\rm th}$ generator of the gauge group, and $\zeta_a$ those
of the level $\zeta$ of the momentum map. In our case we have the single
component $v$ given by eq. (\ref{9.16}), and we named $\rho^2$ the
single component of the level. The original K\"ahler potential on $\cS =
{\bf C}^{N+1}$ is $K = \o 12 X^{A^*}X^A$ so that when restricted to $\cD^{-1}
(\rho^2)$ it takes an irrelevant constant value $\o{\rho^2}2$. Thus
we deduce from (\ref{kq4}) that the K\"ahler potential for $\cM =
{\bf CP}^N$ that we obtain is $\hat K =\o 12\rho^2 \log (X^*X)$.
Fixing a particular gauge to perform the quotient with respect to
${\bf C}^*$ (see later),  this  potential can be rewritten as $\hat K =\o
12\rho^2\log (1+
x^*x)$, namely the Fubini-Study potential.
\vskip 0.2cm
Let us now procede with  our manipulations  of the lagrangian.
It is a trivial algebraic matter to rewrite the lagrangian (\ref{9.11}) after
the substitutions (\ref{9.10bis}) with $e^{-v}$ given by eq. (\ref{9.16}). For
convenience we divide the resulting expressions into three parts to be
separately handled.
\vskip 0.2cm
First we have what we can call the ``bosonic kinetic terms'':
\begin{eqnarray}
\label{9.17}
\cL_1 &=& -\o {\rho^2}{X^*X} \sum_A\biggl\{\sum_B\biggl[ \biggr( \delta_{AB}
-\o{X^A X^{B^*}}{2X^*X}\biggr)dX^B -\o{X^A X^B}{2X^*X} dX^{B^*}\biggr]
\nonumber\\
&&\mbox{}+i X^A \o{-i(X\dep X^* -X^*\dep X)+4\psi\psi^*}{2X^*X}\eplus
+i X^A \o{-i(X\dem X^* -X^*\dem X)+4\psit\psit^*}{2X^*X}\eminus
\nonumber\\
&&\mbox{}-\ps A\zem -\pst A\zetm \biggr\}(\Pi^{A^*}_{\ssp}\eplus -
\Pi^{A^*}_{\ssm}\eminus) +\coco -\o{\rho^2}{X^*X}\sum_A
(\Pi^{A^*}_{\ssp}\Pi^A_{\ssm} + \Pi^{A^*}_{\ssm}\Pi^A_{\ssp})\eplus
\eminus\nonumber\\
\end{eqnarray}
We would like  to recognize in the above expressions the
bosonic kinetic terms of an $N=2$ $\sigma$-model. By looking at the
$\sigma$-model rheonomic lagrangian (\ref{twosigmod6})
we are inspired to perform a series of
manipulations. \par
Collecting  some suitable terms we  can rewrite
\begin{eqnarray}
\label{9.18}
X^A (X\dep X^*\eplus +X\dem X^*\eminus) \hskip 3pt &\longrightarrow &
\hskip 3pt X^A X dX^* \nonumber\\
\mbox{} X^A (X^*\dep X\eplus +X^*\dem X\eminus)\hskip 3pt &\longrightarrow &
\hskip 3pt X^A X^* dX
\end{eqnarray}
due to the fact that the further terms in the rheonomic parametrizations
of $dX, dX^*$, proportional to the gravitinos, give here a vanishing
contribution in force of  the constraints (\ref{9.5}).\par
We introduce the following provisional notation:
\def\Gi#1#2{G_{#1 #2^*}}
\begin{equation}
\label{9.19}
\Gi AB = \o{\rho^2}{X^*X}\biggl(\delta_{AB} -\o{X^{A^*}X^B}{X^*X}\biggr).
\end{equation}
Noting that, because of the constraints (\ref{9.5}),
\begin{equation}
\label{9.20}
\Gi AB \ps A= \o{\rho^2}{X^*X}\ps A
\end{equation}
we can write
\begin{eqnarray}
\label{9.21}
\cL_1 &=&-\biggl[\Gi AB (dX^A-\ps A\zem-\pst A\zetm) +2i\o{\rho^2}{X^*X}
X^B (\psi\psi^*\eplus\nonumber\\
&&\mbox{}+\psit\psit^*\eminus)\biggr](\Pi^{A^*}_{\ssp}\eplus
\Pi^{A^*}_{\ssm}\eminus) - \coco - \o{\rho^2}{X^*X}(\Pi^{A^*}_{\ssp}
\Pi^A_{\ssm} + \Pi^{^*}_{\ssm} \Pi^A_{\ssp})\eplus\eminus\nonumber\\
\end{eqnarray}
In order to eliminate the terms containing the first order fields $\Pi$'s
multiplied by fermionic expressions we redefine the $\Pi$'s:
\begin{equation}
\label{9.22}
   \begin{array}{ccc}
   \Pi^A_{\ssm}\hskip 2pt &\rightarrow &\hskip 2pt \Pi^A_{\ssm} +2i X^A
   \o{\psit\psit^*}{X^*X}\\
   \Pi^A_{\ssp}\hskip 2pt &\rightarrow &\hskip 2pt \Pi^A_{\ssp} +2i X^A
   \o{\psi\psi^*}{X^*X}
   \end{array}
\hskip 1cm
   \begin{array}{ccc}
   \Pi^{A^*}_{\ssm}\hskip 2pt &\rightarrow &\hskip 2pt\Pi^{A^*}_{\ssm}
   -2i X^{A^*}\o{\psit\psit^*}{X^*X}\\
   \Pi^{A^*}_{\ssp}\hskip 2pt &\rightarrow &\hskip 2pt\Pi^{A^*}_{\ssp}
   -2i X^{A^*}\o{\psi\psi^*}{X^*X}
   \end{array}
\end{equation}
Then we perform a second redefinition of the $\Pi$'s:
\def\sspm{{\scriptscriptstyle \pm}}
\begin{eqnarray}
\label{9.23}
\Pi^A_{\sspm}\hskip 2pt &\rightarrow &\hskip 2pt \biggl(\delta_{AB} \pm
\o{X^A X^{B^*}}{X^*X}\biggr) \Pi^B_{\sspm}\nonumber\\
\Pi^{A^*}_{\sspm}\hskip 2pt &\rightarrow &\hskip 2pt \biggl(\delta_{AB}
\pm \o{X^{A^*}X^B}{X^*X}\biggr) \Pi^{B^*}_{\sspm}
\end{eqnarray}
in such a way that the quadratic term in the first order fields takes
the form
\begin{equation}
-\Gi AB (\Pi^A_{\ssp} \Pi^{B^*}_{\ssm} +\Pi^A_{\ssm} \Pi^{B^*}_{\ssp})
\eplus\eminus
\end{equation}
After the redefinitions (\ref{9.22}) and (\ref{9.23}) we can rewrite
the part $\cL_1$ of the Lagrangian in the following way;  we take into account,
besides the constraints (\ref{9.5}), the fact that
\begin{equation}
\label{9.24}
\Gi AB a^A X^{B^*} \propto \biggl(\delta_{AB} -\o{X^{A^*}X^B}{X^*X}
\biggl) a^A X^{B^*} = 0
\end{equation}
and we obtain:
\begin{eqnarray}
\label{9.25}
\cL_1 &=& -\Gi AB (dX^A - \ps A\zem -\pst A\zetm)(\Pi^{B^*}_{\ssp}\eplus
-\Pi^{B^*}_{\ssm}\eminus)\nonumber\\
&&\mbox{}-\Gi AB (dX^{B^*}+\pss B\zep +\psts B\zetp)(\Pi^A_{\ssp}\eplus
-\Pi^A_{\ssm}\eminus)\nonumber\\
&&\mbox{} -\Gi AB (\Pi^A_{\ssp}\Pi^{B^*}_{\ssm} +\Pi^A_{\ssm}
\Pi^{B^*}_{\ssp})\eplus\eminus +\o{8\rho^2}{(X^*X)^2}\psi\psi^*\psit
\psit^*\eplus\eminus
\end{eqnarray}
\vskip 0.2cm
Next we consider the fermionic kinetic terms in eq. (\ref{9.11}).
Performing the substitutions (\ref{9.10bis}) with $v$ given by eq.
(\ref{9.16}) and using the fact that, for instance,
\begin{equation}
\o{\rho^2}{X^*X}\ps A d\pss A = \Gi AB \ps A d\pss B
\end{equation}
these terms are
\begin{eqnarray}
\label{9.26}
\cL_2 &=& 2i\biggl\{ \Gi AB (\ps A d\pss B +\pss B d\ps A) -
\o{\rho^2}{(X^*X)^2}\ps A\pss A (X\dem X^* -X^*\dem X)\eminus\biggr\}\eplus
\nonumber\\
&&\mbox{} -2i\biggl\{\Gi AB(\pst A d\psts B + \psts B d \pst A) +
\o{\rho^2}{(X^*X)^2}\pst A\psts A (X\dep X^* -X^*\dep X)\eplus\biggr\}
\eminus\nonumber\\
&&\mbox{}-16 \o{\rho^2}{(X^*X)^2}\psi\psi^*\psit\psit^*\eplus\eminus
\end{eqnarray}
Let us  introduce another provisional notation:
\def\gatre#1#2#3{\gamma^{#1}_{#2 #3}}
\def\gatres#1#2#3{\gamma^{#1^*}_{#1^* #2^*}}
\begin{equation}
\label{9.27}
\gatre ABC = \o 1{X^*X}(\delta^A_B X^{C^*} +\delta^A_C X^{B^*})
\end{equation}
It is not difficult to check that the expression (\ref{9.26}) can be
rewritten as follows:
\begin{eqnarray}
\label{9.28}
\cL_2 &=& 2i\biggl\{\Gi AB\ps A (d\pss B -\gatres BCD\pss C d
X^{D^*}) +\Gi AB\pss B (d \ps A -\gatre ACD\ps C d X^D)\biggr\}\eplus
\nonumber\\
&&\mbox{}-2i\biggl\{\Gi AB\pst A( d\psts B -\gatres BCD\psts C d X^{D^*})
+\Gi AB\psts B(d\pst A -\gatre ACD\pst C d X^D)\biggr\}\eminus
\nonumber\\
&&\mbox{}-16\o{\rho^2}{(X^*X)^2}(\psi\psi^*)(\psit\psit^*)
\end{eqnarray}
\vskip 0.2cm
The remaining terms in the lagrangian (\ref{9.11}) become, after the
substitutions (\ref{9.10bis})
\begin{eqnarray}
\label{9.29}
\lefteqn{\cL_3 = -\o{8\rho^2}{(X^*X)^2}\ps A\pss B\pst B\psts A -\Gi AB
(\ps A\pss A\zem\zep -\pst A\psts B\zetm\zetp +\ps A\psts B\zem\zetp}
\nonumber\\
&&\mbox{}+\pss B\pst A\zep\zetm) -\Gi AB dX^A(\pss B\zep -\psts B\zetp) +
\Gi AB d X^{B^*}(\ps A\zem -\pst A\zetm)\nonumber\\
\end{eqnarray}
\vskip 0.2cm
We have succeded so far in  making  the lagrangian (\ref{9.11}) invariant
under the ${\bf C}^*$-transformations (\ref{9.12}) , and to write it in a
nicer form consisting of the sum of the three parts $\cL_1, \cL_2,
\cL_3$ as given in eqs. (\ref{9.25},\ref{9.28},\ref{9.29}), respectively:
\begin{eqnarray}
\label{9.29bis}
\cL &=& -\Gi AB (dX^A - \ps A\zem -\pst A\zetm)(\Pi^{B^*}_{\ssp}\eplus
-\Pi^{B^*}_{\ssm}\eminus)\nonumber\\
&&\mbox{}-\Gi AB (dX^{B^*}+\pss B\zep +\psts B\zetp)(\Pi^A_{\ssp}\eplus
-\Pi^A_{\ssm}\eminus)\nonumber\\
&&\mbox{} -\Gi AB (\Pi^A_{\ssp}\Pi^{B^*}_{\ssm} +\Pi^A_{\ssm}
\Pi^{B^*}_{\ssp})\eplus\eminus \nonumber\\
&&\mbox{}+2i\biggl\{\Gi AB\ps A (d\pss B -\gatres BCD\pss C d
X^{D^*}) +\Gi AB\pss B (d \ps A -\gatre ACD\ps C d X^D)\biggr\}\eplus
\nonumber\\
&&\mbox{}-2i\biggl\{\Gi AB\pst A( d\psts B -\gatres BCD\psts C d X^{D^*})
+\Gi AB\psts B(d\pst A -\gatre ACD\pst C d X^D)\biggr\}\eminus
\nonumber\\
&&\mbox{}-\Gi AB(\ps A\pss A\zem\zep -\pst A\psts B\zetm\zetp +\ps A\psts
B\zem\zetp +
\pss B\pst A\zep\zetm\nonumber\\
&&\mbox{}-\Gi AB dX^A(\pss B\zep -\psts B\zetp) +
\Gi AB d X^{B^*}(\ps A\zem -\pst A\zetm)\nonumber\\
&&\mbox{}-8 \o{\rho^2}{(X^*X)^2}(\ps A\pss A\pst B\psts B +\ps A\pss B\pst
B\psts A)
\eplus\eminus
\end{eqnarray}
We can now utilize the gauge invariance to fix for instance
(in the coordinate patch where $X^0 \neq 0$) $X^0 = 1$, fixing completely
the gauge. In practice we perform the transformation
\begin{equation}
\label{9.30}
X^A\hskip 2pt\rightarrow \hskip 2pt e^{-\Phi} X^A = \o 1{X^0} X^A
\end{equation}
that is we go from the homogeneous coordinates $(X^0,X^i)$ to the
inhomogeneous coordinates $(1, x^i=X^i/X^0)$ on ${\bf CP}^N$.\par
Having chosen our gauge, we rewrite the lagrangian (\ref{9.29bis})
in terms of the fields $x^i$ (and of their fermionic partners $\ps i,
\pst i)$. Note that now $dx^0 =0$ implies (because of the rheonomic
parametrizations) $\ps 0 =0$ and $\pst 0 =0$. The expression $X^*X \equiv
X^{A^*}X^A$ becomes $1+x^{i^*}x^i\equiv 1+x^*x$. Of the expressions $\Gi AB$
and $\gatre ABC$ only the components not involvig the index zero survive. We
introduce the following notations:
\def\gi#1#2{g_{#1 #2^*}}
\def\Gatre#1#2#3{\Gamma^{#1}_{#2 #3}}
\def\Gatres#1#2#3{\Gamma^{#1^*}_{#2^* #3^*}}
\begin{equation}
\label{9.31}
   \begin{array}{ccc}
   \Gi AB &\equiv &\o{\rho^2}{X^*X}\biggl(\delta_{AB}
-\o{X^{A^*}X^B}{X^*X}\biggr)\\
   \gatre ABC &\equiv &\o 1{X^*X}(\delta^A_B X^{C^*}+\delta^A_C X^{B^*})
   \end{array}
\hskip 8pt \longrightarrow \hskip 8pt
   \begin{array}{ccc}
   \gi ij &\equiv &\o{\rho^2}{1+x^*x}\biggl(\delta_{ij} -\o{x^{i^*}x^j}{1+
   x^*x}\biggr)\\
   \Gatre ijk &\equiv &\o 1{1+x^*x}(\delta ^i_j x^{k^*} +\delta ^i_k x^{j^*})
   \end{array}
\end{equation}
We see that $\gi ij$ is just the standard Fubini-Study metric on
${\bf CP}^N$, which is a K\"ahler metric of K\"ahler potential $K = \rho^2
\log (1+x^*x)$; $\Gatre ijk$ is just the purely holomorphic part of its
associated Levi-Civita connection. Moreover the Riemann tensor for the
Fubini-Study metric is given by:
\begin{eqnarray}
\label{9.32}
R_{ij^*kl^*} &=& \o{\rho^2}{(1+x^*x)}\biggl\{\delta^i_j \delta^k_l +
\delta^i_l \delta^k_j -\o 1{1+x^*x}\biggl[(\delta^i_j x^l +\delta^i_l x^j)
x^{k^*}\nonumber\\
&&\mbox{}+\delta^k_j x^l +\delta^k_l x^j)x^{i^*})\biggr] +2\o{x^{i^*}x^j
x^{k^*}x^l}{(1+x^*x)^2}\biggr\}
\end{eqnarray}
and we see that using once more the fermionic constraints (\ref{9.5}) the
four-fermion terms in (\ref{9.29bis}) can be rewritten as follows:
\begin{equation}
\label{9.33}
\o{\rho^2}{(1+x^*x)^2}(\ps i\pss i\pst j\psts j +\ps i\pss j\pst j\psts
i)
= R_{ij^*kl^*} \ps i\pss j\pst k\psts l
\end{equation}
\vskip 0.2cm
Thus at the end of the above manipulations, corresponding to the procedure
of obtaining ${\bf CP}^N$ as the K\"ahler quotient of ${\bf C}^{N+1}$, we
have reduced our initial rheonomic lagrangian (\ref{9.4}),  in the
limit $g\rightarrow\infty$, to a form which is that of the $N=2$
$\sigma$-model as given in eq. (\ref{twosigmod6}). The target space is ${\bf
CP}^N$
equipped with the K\"ahlerian Fubini-Study metric.

\section{The $N=4$ Abelian gauge multiplet,}
\label{gaugen4}
\noindent
Having exhausted our rheonomic reconstruction of the N=2 models we now turn our
attention
to the N=4 case. We start with the gauge multiplet.
The N=4 vector multiplet, in addition to the gauge boson, namely the 1-form
${\cal A}$,
contains four spin 1/2 gauginos whose eight components are denoted by
$\lambda^+$,$\lambda^-$,
$\tilde\lambda^+$,$\tilde\lambda^-$,$\mu^+$,$\mu^-$,
$\tilde\mu^+$,$\tilde\mu^-$, two complex physical scalars $M\ne M^*$, $N\ne
N^*$, and
three  auxiliary fields arranged into a real scalar $\cP = {\cP}^*$ and a
complex
scalar $\cQ \ne{\cQ}^*$. The rheonomic parametrization of the abelian
field-strength
$F=d{\cal A}$
and of the exterior derivatives of the scalars, gauginos and auxiliary fields
is given
below. It is uniquely determined from the Bianchi identities:

\begin{eqnarray}
F & = &\cF\eplus\eminus -\o i2\bigl(\lap\zem +\lamm\zep +\mup\cgm +\mutm\cgp
\bigr)\eminus +\o i2\bigl(\latp\zetp +\latm\zetp \nonumber\\
&&\mbox{}+\mutp\cgtm +\mum\cgtp \bigr)\,\eplus +
M\bigl(\zem\zetp +\cgp\cgtm \bigr)-M^*\bigl(\zep\zetm +\cgm\cgtm \bigr)
\nonumber\\
&&\mbox{}+ N\bigl(\zep\cgtm -\cgm\zetp\bigr) -N^*\bigl(\zem\cgtp -
\cgp\zetm\bigr)\nonumber\\
dM & = &\dep M\,\eplus+\dem M\,\eminus -\o 14\bigl(\latm\zep -\lap\zetm
+\mutp\cgm -\mutm\cgtp\bigr)\nonumber\\
dN & = &\dep N\,\eplus+\dem N\,\eminus -\o 14\bigl(\mutp\zem +\mup\zetm
-\latm\cgp -\lamm\cgtp\bigr)\nonumber\\
d\lap & = &\dep\lap\,\eplus +\dem\lap\,\eminus +\bigl(\o{\cF}2
+i\cP\bigr)\,\zep -2i\,\dem M\,\zetp +\cQ\,\cgm +2i\,\dem N^*\,\cgtp\nonumber\\
d\latp & = &\dep\latp\,\eplus +\dem\latp\,\eminus +\bigl(\o{\cF}2
-i\cP\bigr)\,\zetp -2i\,\dep M^*\,\zep -\cQ\,\cgtm +2i\,\dep
N^*\,\cgp\nonumber\\
d\mup & = &\dep\mup\,\eplus +\dem\mup\,\eminus +\bigl(\o{\cF}2
+i\cP\bigr)\,\cgp +2i\,\dem M^*\,\cgtp -\cQ\,\zem +2i\,\dem N\,\zetp\nonumber\\
d\mutp & = &\dep\mutp\,\eplus +\dem\mutp\,\eminus +\bigl(\o{\cF}2
-i\cP\bigr)\,\cgtp +2i\,\dep M\,\cgp +\cQ\,\zetm +2i\,\dep N\,\zep\nonumber\\
d\cP & = &\dep\cP\,\eplus +\dem\cP\,\eminus -\o 14\bigl(\dep\lap\,\zem
-\dep\lamm\,\zep -\dem\latp\,\zetm +\dem\latm\,\zetp \nonumber\\
&&\mbox{}+\dep\mup\,\cgm -\dep\mutm\,\cgp -\dem\mutp\,\cgtm
+\dem\mum\,\cgtp\bigr)\nonumber\\
d\cQ & = &\dep\cQ\,\eplus +\dem\cQ\,\eminus +\o i2\bigl(\dep\mutp\,\zep
-\dem\mup\zetp -\dep\lap\,\cgm +\dem\latp\,\cgtm\bigr)
\label{nfour2}
\end{eqnarray}
The rheonomic parametrizations of the complex
conjugate fields $d\lamm$, $d\latm$, $d\mum$, $d\mutm$ ,$dM^*$, $d\cQ^*$
are immediately obtained by applying the rules of complex conjugation.

Using these results, by means of lengthy but straightforward algebra we can
derive the
rheonomic action of the N=4 abelian gauge multiplet. The result is given below
\begin{eqnarray}
\lefteqn{\cL^{(rheon)}_{gauge}\left ( N=4 \right)
=\cF\biggl[F +\o i2\bigl(\lap\zem +\lamm\zep +\mup\cgm
+\mutm\cgp\bigr)\,\eminus
-\o i2\bigl(\latp\zetp +\latm\zetp }\nonumber\\
&&\mbox{}+\mutp\cgtm +\mum\cgtp\bigr)\,\eplus -
M\bigl(\zem\zetp +\cgp\cgtm \bigr)+M^*\bigl(\zep\zetm +\cgm\cgtm \bigr)
- N\bigl(\zep\cgtm -\cgm\zetp\bigr) \nonumber\\
&&\mbox{}+N^*\bigl(\zem\cgtp -\cgp\zetm\bigr)\biggr]
-\o 12 \cF^2\,\eplus\eminus -\o 12 \bigl(
\lap\,d\lamm +\lamm\, d\lap + +\mup\, d\mutm +\mutm\, d\mup
\bigr)\,\eminus \nonumber\\
&&\mbox{}+\o i2\bigl(\latp\,d\latm +\latm\,
d\latp +\mutp\, d\mum +\mum\, d\mutp\bigr)\,\eplus -
4 \biggl[dM^*- \o 14\bigl(\latp\zem - \lamm\zetp \nonumber\\
&&\mbox{}+\mum\cgp -\mup\cgtm\bigr)\biggr]\bigl(
\cM_{\ssp}\eplus -\cM_{\ssm}\eminus\bigr) -4 \biggl[dM +\o 14\bigl(\latm\zep
- \lap\zetm +\mutp\cgm -\mutm\cgtp\bigr)\biggr]\cdot\nonumber\\
&&\mbox{}\cdot\bigl(\cM^*_{\ssp}\eplus -\cM^*_{\ssm}\eminus\bigr)-
 4\bigl(\cM^*_{\ssp}\cM_{\ssm} +\cM^*_{\ssm}\cM_{\ssp}\bigr)
\,\eplus\eminus -4\biggl[dN^*-\o 14\bigl(\mum\zep \nonumber\\
&&\mbox{}+\mutm\zetp -\latp\cgm
-\lap\cgtm\bigr)\biggr]\bigl(\cN_{\ssp}\eplus -\cN_{\ssm}\eminus\bigr)
-4\biggl[dN + \o 14\bigl(\mutp\zem +\mup\zetm -\latm\cgp \nonumber\\
&&\mbox{}-\lamm\cgtp\bigr)\biggr]\bigl(\cN^*_{\ssp}\eplus
-\cN^*_{\ssm}\eminus\bigr)-\biggl[dM\bigl(\lamm\zetp
+\latp\zem +\mutm\cgp +\mup\cgtm \bigr) +\coco\biggr]\nonumber\\
&&\mbox{}-\biggl[dN\bigl(\lamm\zetp +\latp\zem +\mutm\cgp +\mup\cgtm
\bigr) +\coco\biggr]-\o 14 \bigl(\lamm\latm\,\zep\zetp
-\lamm\latp\,\cgm\cgtp \nonumber\\
&&\mbox{}+\lamm\mum\,\zep\cgtp +\lamm\mutp\,\cgm\zetp
+\latm\mutm\,\cgp\zetp -\latm\mup\,\zep\cgtm +\mum\mutm\,\cgp\cgtp \nonumber\\
&&\mbox{}-\mutm\mutp\,\zep\zetm+\coco\bigr)+\o{\theta}{2\pi} F +
\biggl[2\cP^2+2\cQ^*\cQ -2r\cP -\bigl(s\cQ^* +s^*\cQ\bigr)\biggr]
\,\eplus\eminus \nonumber\\
&&\mbox{}-\o r2\biggl[(\latm\zetp +\mum\cgtp +\coco)\eplus +
(\lam\zep +\mutm\cgp +\coco)\eminus\biggr]\nonumber\\
&&\mbox{}+i\o s2\biggl[(\latm\cgtm -\mum\zetm)\eplus +(\lam\cgm -\mutm\zem)
\eminus\biggr]+\coco\nonumber\\
&&\mbox{}+ir\biggl[M(\zem\zetp -\cgp\cgtm)-\coco -N(\zep\cgtm +\cgm\zetp)
+\coco\biggr]\nonumber\\
&&\mbox{}+s\biggl[M\zem\cgtm +M^*\cgm\zetm -N\cgm\cgtm +N^*\zem\zetm
\biggr] +\coco
\label{nfour3}
\end{eqnarray}
By means of the usual manipulations, from eq.(\ref{nfour3}) we immediately
retrieve the
N=4 globally supersymmetric world-sheet action of the abelian vector multiplet.
It is
the following:
\begin{eqnarray}
\cL^{(ws)}_{gauge}\left ( N=4 \right ) & = &\o 12 \cF^2 -i\bigl(\lap\,\dep\lamm
+\mup\,\dep\mum
+\latp\,\dem\latm +\mutp\,\dem\mutm\bigr)\nonumber\\
&&\mbox{}+4\bigl(\dep M^*\,\dem M
+\dem M^*\,\dep M +\dep N^*\,\dem N +\dem N^*\,\dep N\bigr)\nonumber\\
&&\mbox{}+\o{\theta}{2\pi}\cF
+2\cP^2 +2\cQ^*\cQ -2r\cP-\bigl(s\cQ^* +s^*\cQ\bigr)
\label{nfour4}
\end{eqnarray}
In the above action we note the announced N=4 generalization of the Fayet
Iliopoulos term. In addition
to the $\theta$-term, proportional to the first Chern-class of the gauge field,
and to the
$r$-term linear in the real  auxiliary field $\cP$, we have term linear in the
complex auxiliary
field $\cQ$, involving a new complex parameter $s$. Differently from
the N=2 case, the only allowed self interaction of the N=4 vector multiplet is
given
by the analogue of a linear superpotential term, namely the above N=4
generalization
of the Fayet-Iliopoulos term, existing only for abelian gauge fields.
As we are going to see shortly, the
parameters $r$ and $s$ correspond, in the Lagrangian realization of the
HyperK\"ahler quotients,
to the levels of the triholomorphic momentum map. To discuss this point,
that is one of our main goals, we have to revert to the discussion of the
quaternionic
hypermultiplets. These are the N=4 analogues of the N=2 chiral multiplets.

\section{ $N=4$ Quaternionic hypermultiplets with an abelian
gauge symmetry}
\label{n4matter}
As in four-dimensions the N=2 analogue of the N=1 Wess-Zumino multiplets is
given by
the hypermultiplets that display a quaternionic structure, in the same way, in
two
dimensions, the N=4 analogues of the complex N=2 chiral multiplets are the
quaternionic
hypermultiplets that parametrize a HyperK\"ahler manifold. If this manifold is
curved we have
an N=4 $\sigma$-model, similarly to the N=2 $\sigma$-model that is constructed
on a K\"ahler
manifold. Alternatively, if the HyperK\"ahler variety is flat we are dealing
with the N=4
analogue of the N=2 Landau-Ginzburg model.
Here, however, the more stringent constraints of N=4 supersymmetry
rule out the insertion of any self-interaction driven by a holomorphic
superpotential. On the
other hand, what we can still do, just as in the N=2  case, is to couple the
flat hypermultiplets
to abelian or non-abelian gauge multiplets. In full analogy with the N=2 case,
this construction
will generate an N=4 $\sigma$-model as the  effective low-energy action
of the $gauge\,\oplus \,matter$ system.
The target manifold will be the HyperK\"ahler quotient of the flat quaternionic
manifold with
respect to the triholomorphic action of the gauge group. Hence in the present
section, we
consider quaternionic hypermultiplets minimally coupled to abelian gauge
multiplets. For
simplicity we focus on the case of one gauge-multiplet. All formulae can be
straightforwardly
generalized to the case of many abelian multiplets at the end.

Consider a set of bosonic complex fields $u^i$, $v^i$, that can be
organized in a set of quaternions
\begin{equation}
Y^{i}=\twomat{u^i}{iv^{i^*}}{iv^i}{u^{i^*}}
\label{nfour5}
\end{equation}
On these matter fields the abelian gauge group acts in a {\it triholomorphic}
fashion.
According to the discussion of section I (see eq.(\ref{triholoaction}),
the {\it triholomorphic} character of this action
corresponds to the following definition of the covariant derivatives:
\begin{eqnarray}
\dell u^i & = &d u^i + i {\cal A} \qu ij u^j\nonumber\\
\dell v^i & = &d u^i - i {\cal A} \qu ij u^j
\label{nfour6}
\end{eqnarray}
where $\qu ij$ is a hermitean matrix. Correspondingly the Bianchi identities
take the  form:
\begin{eqnarray}
\dell^2 u^i & = & + i F \qu ij u^j\nonumber\\
\dell^2 v^i & = & - i F \qu ij u^j
\label{nfour7}
\end{eqnarray}
We solve these Bianchi identities parametrizing the covariant derivatives
$\dell^2 u^i$ and
$\dell^2 v^i$ in terms of four spin 1/2 fermions, whose eight components are
given by
$\psu i  , \psut i , \psv i, \psvt i$ together with their complex conjugates
$\psus i  , \psuts i , \psvs i, \psvts i$. In the background of the abelian
gauge multiplet
(\ref{nfour2}) we obtain:
\begin{eqnarray}
\dell u^i & = &\delp u^i\,\eplus +\delm u^i\,\eminus +\psu i \zem +\psvs
i\cgm +\psut i \zetm +\psvts i\cgtm\nonumber\\
\dell v^i & = &\delp v^i\,\eplus +\delm v^i\,\eminus +\psv i \zem -\psus
i\cgm +\psvt i \zetm -\psuts i\cgtm\nonumber\\
\dell \psu i & = &\delp \psu i\,\eplus +\delm\psu i\,\eminus -\o i2 \delp
u^i\zep -\o i2 \delp v^{i^*}\cgm \nonumber\\
&&\mbox{}+i\qu ij\bigl(M u^j\,\zetp +N^* v^{j^*}\,\zetm -N^* u^j\,\cgtp
+M v^{j^*}\cgtm\bigr)\nonumber\\
\dell \psv i & = &\delp \psv i\,\eplus +\delm\psv i\,\eminus -\o i2 \delp
v^i\zep +\o i2 \delp u^{i^*}\cgm \nonumber\\
&&\mbox{}+i\qu ij\bigl(-M v^j\,\zetp +N^* u^{j^*}\,\zetm +N^* v^j\,\cgtp
+M u^{j^*}\cgtm\bigr)\nonumber\\
\dell \psut i & = &\delp \psut i\,\eplus +\delm\psut i\,\eminus -\o i2
\delm u^i\zetp -\o i2 \delm v^{i^*}\cgtm \nonumber\\
&&\mbox{}-i\qu ij\bigl(M^* u^j\,\zep +N^* v^{j^*}\,\zem -N^* u^j\,\cgp
+M^* v^{j^*}\cgm\bigr)\nonumber\\
\dell \psvt i & = &\delp \psvt i\,\eplus +\delm\psvt i\,\eminus -\o i2
\delm v^i\zetp +\o i2 \delm u^{i^*}\cgtm \nonumber\\
&&\mbox{}-i\qu ij\bigl(-M^* v^j\,\zep +N^* u^{j^*}\,\zem +N^* v^j\,\cgp
+M^* u^{j^*}\cgm\bigr)
\label{nfour8}
\end{eqnarray}
Note that the field content of the N=4 hypermultiplet is the same as the field
content
of two N=2 chiral multiplets. For each complex coordinate $u$ or $v$ we have
two complex
spin 1/2 Weyl fermions $\psu i$, $\psv i$ or $\psv i$, $\psv i$. The additional
supersymmetries
associated with the gravitinos $\chi^{\pm}$ and ${\tilde \chi}^{\pm}$ simply
mix the fields
of one N=2 chiral multiplet $u$ with the other $v$. Note also that contrarily
to the N=2
case, the rheonomic solution (\ref{nfour8}) does not involve any auxiliary
field, namely in the
N=4 case there is no room for an arbitrary interaction driven by a
Landau-Ginzburg superpotential
$U(u,v)$.
\par
{}From the Bianchi identities one gets the following fermionic equations of
motion:
\begin{eqnarray}
\o i2\delm\psu i + i\qu ij\biggl(\o 14 \lap u^j +\o 14 \mutm v^{j^*} +M\psut
j -N^*\psvts j\biggr) & = &0\nonumber\\
\o i2\delm\psv i - i\qu ij\biggl(\o 14 \lap v^j -\o 14 \mutm u^{j^*}
+M\psvt j +N^*\psuts j\biggr) & = &0\nonumber\\
\o i2\delm\psut i - i\qu ij\biggl(\o 14 \latp u^j +\o 14 \mum v^{j^*} +
M^*\psu j -N^*\psvs j\biggr) & = &0\nonumber\\
\o i2\delm\psvt i + i\qu ij\biggl(\o 14 \latp v^j -\o 14 \mum u^{j^*}
+M^*\psv j +N^*\psus j\biggr) & = &0
\label{nfour9}
\end{eqnarray}
Applying the supersymmetry transformation of parameter $\epsilon^{\ssp}$
to the first two of eq.s (\ref{nfour9}) we obtain the bosonic equations of
motion, namely:
\begin{eqnarray}
\o 18\bigl(\delp\delm +\delm\delp\bigr)\,u^i & = & \o i4\qu ij
\bigl(\lamm\psu j -\latm\psut j +\mutm\psvs j -\mum\psvts j\bigr)\nonumber\\
&&\mbox{}-\bigl(|M|^2 +|N|^2\bigr) (q^2)^i_j\,u^j+
\o 14\cP\,\qu ij\,u^j +\o i4\cQ^*\,\qu ij\, v^{j^*}\nonumber\\
\o 18\bigl(\delp\delm +\delm\delp\bigr)\,v^i & = & -\o i4\qu ij
\bigl(\lamm\psv j -\latm\psvt j -\mutm\psus j +\mum\psuts j\bigr)\nonumber\\
&&\mbox{}-\bigl(|M|^2 +|N|^2\bigr) (q^2)^i_j\,v^j
-\o 14\cP\,\qu ij\,v^j +\o i4\cQ^*\,\qu ij\, u^{j^*}
\label{nfour10}
\end{eqnarray}
The rheonomic action that yields the rheonomic parametrizations (\ref{nfour8})
and the
field equations (\ref{nfour9}) and (\ref{nfour10}) as variational equations is
given below:
\begin{eqnarray}
\lefteqn{\cL^{(rheon)}_{quatern}\,=\, \bigl(\dell u^i -\psu i\zem -\psvs i\cgm
-\psut i\zetm -\psvts
i\cgtm\bigr)\bigl(U^{i^*}_{\ssp}\,\eplus
-U^{i^*}_{\ssm}\,\eminus\bigr)}\nonumber\\
&& \mbox{}+\bigl(\dell u^{i^*} +\psus i\zep +\psv i\cgp
+\psuts i\zetp +\psvt
i\cgtp\bigr)\bigl(U^i_{\ssp}\,\eplus -U^i_{\ssm}\,\eminus\bigr)\nonumber\\
&& \mbox{}+\bigl(U^{i^*}_{\ssp}U^i_{\ssm}+U^{i^*}_{\ssm}U^i_{\ssp}\bigr)
\,\eplus\eminus\nonumber\\
&& \mbox{} +\bigl(\dell v^i -\psv i\zem +\psus i\cgm -\psvt i\zetm +\psuts
i\cgtm\bigr)\bigl(V^{i^*}_{\ssp}\,\eplus
-V^{i^*}_{\ssm}\,\eminus\bigr)\nonumber\\
&&\mbox{}+\bigl(\dell v^{i^*} +\psvs i\zep -\psu i\cgp +\psvts i\zetp -\psut
i\cgtp\bigr)\bigl(V^i_{\ssp}\,\eplus -V^i_{\ssm}\,\eminus\bigr)\nonumber\\
&&\mbox{}+\bigl(V^{i^*}_{\ssp}V^i_{\ssm}+V^{i^*}_{\ssm}V^i_{\ssp}\bigr)\,\eplus\eminus\nonumber\\
&&\mbox{}-4i\bigl(\psu i\,\dell\psus i
+\psv i\,\dell \psvs i\bigr)\,\eplus +
4i\bigl(\psut i\,\dell\psuts i +\psvt i\,\dell \psvts
i\bigr)\,\eminus\nonumber\\
&&\mbox{}+\bigl(\psu i\psus i +\psv i\psvs i\bigr)\bigl(\zep\zem
+\cgp\cgm)\nonumber\\
&&\mbox{}-
\bigl(\psut i\psuts i +\psvt i\psvts i\bigr)\bigl(\zetp\zetm
+\cgtp\cgtm)\nonumber\\
&&\mbox{}+\biggl[\bigl(\psu i\psuts i +\psv i\psvts i\bigr)\bigl(\zem\zetp
+\cgp\cgtm\bigr) +\coco\biggr]\nonumber\\
&&\mbox{} +
\biggl[\bigl(\psu i\psvt i -\psv i\psut i\bigr)\bigl(\zem\cgtp
-\cgp\zetm\bigr) +\coco\biggr]\nonumber\\
&&\mbox{} +\biggl[\dell u^{i^*}\bigl(\psu i\zem -\psut i\zetm +\psvs i\cgm
-\psvts i\cgtm\bigr)+\coco\biggr]\nonumber\\
&&\mbox{}+\biggl[\dell v^{i^*}\bigl(\psv i\zem-\psvt i\zetm -
\psus i\cgm +\psuts i\cgtm\bigr)+\coco\biggr]\nonumber\\
&&\mbox{}+
4\biggl[\psu i\qu ij\bigl(M^*\,u^{j^*}\,\zetm +N\,v^j\,\zetp -N\,u^{j^*}\,
\cgtm +M^*\, v^j\,\cgtp\bigr)\,\eplus +\coco\biggr]\nonumber\\
&&\mbox{}+4\biggl[\psv i\qu ij\bigl(-M^*\,v^{j^*}\,
\zetm +N\,u^j\,\zetp +N\,v^{j^*}\,
\cgtm +M^*\, u^j\,\cgtp\bigr)\,\eplus +\coco\biggr]\nonumber\\
&&\mbox{}+
4\biggl[\psut i\qu ij\bigl(M\,u^{j^*}\,\zem +N\,v^j\,\zep -
-N\,u^{j^*}\,
\cgm +M\, v^j\,\cgp\bigr)\,\eplus +\coco\biggr]\nonumber\\
&&\mbox{}+
4\biggl[\psvt i\qu ij\bigl(-M\,v^{j^*}\,\zem +N\,u^j\,\zep +N\,v^{j^*}\,
\cgm +M\, u^j\,\cgp\bigr)\,\eplus +\coco\biggr]\nonumber\\
&&\mbox{}+\biggl\{2i\biggl[\psu i\,\qu ij\bigl(\lamm u^{j^*} +\mup v^j\bigr)
-\coco\biggr]- 2i\biggl[\psv i\,\qu ij\bigl(\lamm v^{j^*} -\mup u^j\bigr)
-\coco\biggr]\nonumber\\
&&\mbox{}- 2i\biggl[\psut i\,\qu ij\bigl(\latm u^{j^*} +\mutp v^j\bigr)
-\coco\biggr]+ 2i\biggl[\psvt i\,\qu ij\bigl(\latm v^{j^*} -\mutp u^j\bigr)
-\coco\biggr] \nonumber\\
&&\mbox{}+8i\biggl[M^*\bigl(\psu i\,\qu ij\,\psuts j -\psv i\,\qu
ij\,\psvts j\bigr)-\coco\biggr] - 8i\biggl[N\bigl(\psu i\,\qu ij\,\psvt j
+\psv i\,\qu ij\,\psut j\bigr)-\coco\biggr] \nonumber\\
&&\mbox{}+ 8\bigl(|M|^2
+|N|^2\bigr)\biggl[u^{i^*}(q^2)^i_ju^j + v^{i^*}(q^2)^i_jv^j\biggr]
-2\cP\bigl(u^{i^*}\qu ij u^j -v^{i^*}\qu ij v^j\bigr) \nonumber\\
&&\mbox{}+2i\bigl(\cQ\,
u^i\qu ij v^j -\coco\bigr) \biggr\}\,\eplus\eminus
\label{nfour11}
\end{eqnarray}
In the above formula, the fields implementing the first order formalism for the
scalar
kinetic terms have been denoted by $U^{i}_{\pm} , V^{i}_{\pm}$. Eliminating
these fields
through their own equations and deleting the terms proportional to the
gravitinos, we obtain
the world-sheet supersymmetric Lagrangian of the N=4 quaternionic
hypermultiplets coupled
to the gauge multiplet. We write it in a basis where the $U(1)$ generator has
been diagonalised:
 $\qu ij \equiv q^i \delta ^i_{\hskip 3ptj}$:
\begin{eqnarray}
\lefteqn{\cL^{(ws)}_{quatern}\,=\,-\bigl(\delp u^{i^*}\,\delm u^i+\delm
u^{i^*}\,\delp u^i
+\delp v^{i^*}\,\delm v^i+\delp v^{i^*}\,\delm v^i\bigr)}\nonumber\\
&&\mbox{}+4i\bigl(\psu i\,\delm\psus i +\psv i\,\delm\psvs i +\psut
i \delp\psuts i +\psvt i\,\delp\psvts i\bigr)\nonumber\\
&&\mbox{}+2i\sum_i q^i\biggl\{\biggl[\psu i\,\bigl(\lamm u^{i^*} +
\mup v^i\bigr) -\coco\biggr]-\biggl[\psv i\,\bigl(\lamm v^{i^*} -
\mup u^i\bigr) -\coco\biggr]\nonumber\\
&&\mbox{}- \biggl[\psut i\,\bigl(\latm u^{i^*} +\mutp v^i\bigr)
-\coco\biggr]+ \biggl[\psvt i\,\bigl(\latm v^{i^*} -\mutp u^i\bigr)
-\coco\biggr]\biggr\} \nonumber\\
&&\mbox{}+8i\biggl[M^*\sum_i q^i\bigl(\psu i\psuts i -\psv i\,
\psvts i\bigr)-\coco\biggr] - 8i\biggl[N\sum_i q^i\bigl(\psu i\,\psvt i
+\psv i\,\psut i\bigr)-\coco\biggr] \nonumber\\
&&\mbox{}+ 8\bigl(|M|^2
+|N|^2\bigr)\sum_i (q^i)^2\bigl(|u^i|^2 + |v^i|^2\bigr)
-2\cP\sum_i q^i\bigl(|u^i|^2 -|v^i|^2\bigr) \nonumber\\
&&\mbox{}+2i\bigl(\cQ\,\sum_i q^i
u^i v^i -\coco\bigr)
\label{nfour12}
\end{eqnarray}
The most interesting feature of the action (\ref{nfour12}) is the role of the
auxilary fields.
Recalling our discussion of the HyperK\"ahler quotient in section I and
comparing with formulae (\ref{momentumcomponents}) we see that  the auxiliary
field
$\cP$ multiplies the
real component
${\cal D}^3(u^i,v^i)=\sum_{i} \, q^{i} \, \left ( |u^{i}|^2 \, - \, |v^{i}|^2
\right )$,
 while $\cQ$ multiplies the holomorphic component
${\cal D}^{\ssm}(u^i,v^i) \, = -2i\, \sum_{i} \, q^{i} \, u^{i} \,
v^{i}$ of the momentum map
 for the triholomorphic action of the gauge group. This fact is the basis for
the Lagrangian
realization of the HyperK\"ahler quotients. Indeed the vacuum of  the combined
$gauge \,\oplus \, matter$ system breaks the abelian gauge invariance giving a
mass to all
the fields in the gauge multiplet and to all the quaternionic scalars that do
not lie on the
momentum-map surface of level ${\cal D}^3 =r, \, {\cal D}^+ = s$. Integrating
on the massive modes
one obtaines an N=4 $\sigma$-model with target manifold the HyperK\"ahler
quotient. This
mechanism will be evident from the study of the scalar potential of the
combined system.

\section{ The scalar potential in the N=4 hypermultiplet-gauge system}
\label{n4pot}
As in the N=2 case the correct way of putting together the gauge and the matter
lagrangian fixed by  positivity of the energy is the following:
\begin{equation}
\cL=\cL_{gauge} -\cL_{quatern}.
\label{nfour13}
\end{equation}
As a result the bosonic scalar potential is:
\begin{eqnarray}
-U & = & 2\cP^2 +2|\cQ|^2 -2r\,\cP -\bigl(s\,\cQ^* +s^*\,\cQ\bigr)
-8\bigl(|M|^2 +|N|^2\bigr)\sum_i(q^i)^2\biggl(|u^i|^2+|v^i|^2\biggr)\nonumber\\
&&\mbox{}+2\cP\sum_i q^i\biggl(|u^i|^2-|v^i|^2\biggr) -2i\bigl(\cQ\,\sum_i
q^i u^i v^i -\coco\bigr)
\label{nfour14}
\end{eqnarray}
Varying the lagrangian in $\cP$ and $\cQ$ we obtain the algebraic equations:
\begin{eqnarray}
\cP & = &\o 12 \biggl[r-\sum_i q^i\bigl(|u^i|^2 -|v^i|^2\bigr)\biggr] \,=
\o 12 \biggl[r-\cD^3(u,v)\biggr]\nonumber\\
\cQ & = &\o 12\biggl[s -2i\sum_i q^i u^i v^i\biggr] \,=\,\o 12
 \biggl[s-{\cal D}^{\ssp}(u,v)\biggr]
\label{nfour15}
\end{eqnarray}
and substituting back eq.s (\ref{nfour15}) in eq.(\ref{nfour14}) we get the
final form
of the N=4 bosonic potential:
\begin{equation}
U\, =\,\o 12\bigl(r-{\cal D}^3\bigr)^2 +\o 12|s-{\cal D}^{\ssp}|^2
+8\bigl(|M|^2
+|N|^2\bigr)\,\sum_i (q^i)^2 \bigl(|u^i|^2 +|v^i|^2\bigr)
\label{nfour16}
\end{equation}
As we see, the parameters $r,s$ of the Fayet-Iliopoulos term are identified
with the
levels of the triholomorphic momentum-map, as we announced. In the next section
we
discuss the structure of the N=4 scalar potential extrema.

\section{Phase structure of the $N=4$ theory and reconstruction of the
associated low-energy theory}
\label{section13}
We address now the questions related with the structure of the classical vacuum
of the $N=4$ theory discussed above and with the low energy theory around this
vacuum.\par
To minimize the potential (\ref{nfour16}), which is given by a sum of squares,
we must
separately equate each addend to zero . If we compare the $N=4$ bosonic
potential
with the $N=2$ one given in eq. (\ref{ntwo19}) we note that the absence of an
$N=4$ analogue of the Landau-Ginzburg potential reduces the possibilities.
There is only an {\it $N=4$ $\sigma$-model phase}. Beside $M=0, N=0$, we must
impose
$\cD^3(u,v) = r$ and $\cD^{\ssp}=s$. Taking into account
the gauge invariance of the Lagrangian, this means that the classical vacua
are characterized by having $M=N=0$ and the matter fields $u, v$ lying on
the HyperK\"ahler quotient
\begin{equation}
\label{13.1}
\cM = \cD_3^{-1}(r)\cap\cD_{\ssp}^{-1}(s) / U(1)
\end{equation}
of the quaternionic space ${\bf H}^n$ spanned by the fields $u^i,
v^i$ with respect to the triholomorphic  the action of the $U(1)$ gauge
group (see section (\ref{intro})).
Considering the fluctuations around this vacuum, we can see that the fields
of the gauge multiplet are massive, together with the modes of the matter
fields not tangent to $\cM$. The low-energy theory will turns out to be the
$N=4$ $\sigma$-model on $\cM$.\par
Here neither we write the explicit derivation of the
general form for an $N=4$ $\sigma$-model nor we  give the $N=4$ analogue
of the recostruction of the low-energy $N=2$ $\sigma$-model discussed in
section
(\ref{cpn}). We just recall the basic fact that  a $\sigma$-model is  $N=4$
supersymmetric
only under the condition that  the target space be a hyperK\"ahler manifold.
The reason of this omission is not just to save space;  the key point  of
what happens in the $N=4$ case can be fully understood  also in an $N=2$
language.\par
Indeed  $N=4$ theories are nothing else but  particular $N=2$ theories whose
structure allows the existence of additional supersymmetries.
\par
Which kind of
$N=2$ theory is the $N=4$ gauge plus matter system described in sections
(\ref{gaugen4}, XI,\ref{n4pot})? The answer is easily given.
If we suppress the additional gravitinos $\chi^{\pm}$ and ${\tilde
\chi}^{\pm}$,
the N=4 rheonomic parametrizations (\ref{nfour2}),(\ref{nfour8}) and the N=4
action (\ref{nfour11}),(\ref{nfour3}) of $n$ quaternionic
multiplets coupled to a gauge multiplet become those of an N=2 theory
(see eq.s (\ref{ntwo6}, \ref{ntwo10}, \ref{ntwo13}, \ref{ntwo7})
containing one gauge multiplet $({\cal A}, \lap ,\lam,
\latp ,\latm , M , \cP)$ and $2n \, +\, 1$ chiral multiplets, namely
\begin{equation}
\left ( X^A, {\ps A} , {\pss A} , {\pst A} , {\psts A}\right  )~=~\cases{
\left ( u^i, {\ps i}_u , {\pss i}_u , {\pst i}_u , {\psts i}_u\right  )\cr
\left ( v^i, {\ps i}_v , {\pss i}_v , {\pst i}_v, {\psts i}_v\right  )\cr
\left ( X^0, \ps 0 ,\pss 0, \pst 0, \psts 0 \right  )\cr}
\label{n2decompo}
\end{equation}
where the index $A$ runs on $2n+1$ values, the index $i$ takes the values
$i= 1,\ldots n$ and where we have defined:
\begin{eqnarray}
X^0 &=&2 \,N\nonumber\\
\ps 0 = -\um\, \mutp \hskip 8pt & ;&\hskip 8pt\pss 0 = -\um \, \mum\nonumber\\
\pst 0 = - \o{1}{2} \, \mup \hskip 8pt & ;&\hskip 8pt
\psts 0 = - \um \, \mutm
\label{extramultiplet}
\end{eqnarray}
The match between the N=4 theory and the general form of the N=2 model is
complete if we
write the generator of the U(1) transformations on the $X^{A}$ chiral
multiplets
as the following $(2n+1) \times (2n+1)$ matrix:
\begin{equation}
q^{A}_{B} ~=~\left ( \matrix{ q^i\delta^i_j  & 0 & 0\cr 0 & - \,
q^i\delta^i_j & 0\cr 0&0&0\cr}\right )
\label{decompogenerator}
\end{equation}
and we choose as superpotential the following cubic function:
\begin{eqnarray}
W\left ( \, X^{A} \, \right ) &=&-\,\o 14 X^0 \, \left ( \, s^* \, -
 \,\cD^{\ssm} (u,v) \, \right )\nonumber\\
&=&-\,\o 14 X^0 \, \left ( \, s^* \, +2
 \, i\, \sum_i \, q^i u^{i} \, v^{i} \, \right )
\label{inducedsuperpotential}
\end{eqnarray}
where $\cD^{\ssm} (u,v) \,=-2i\,\sum_i \, q^i u^{i} \,v^{i} \,$ is the
holomorphic
part of the momentum map for the triholomorphic action of the gauge group on
${\bf H}^{n}\eq{\bf C}^{n}$. The superpotential (\ref{inducedsuperpotential})
is quasi-homogeneous of degree
\begin{equation}
d_W =1
\label{degree}
\end{equation}
if we assign the following weights to the various chiral fields:
\begin{equation}
\omega_0 ~=~1 ~~~;~~~\omega_{u^{i}}~=~\omega_{v^{i}}~=~0
\label{inducedweights}
\end{equation}
See the discussion at the end of section (\ref{n2rsym}) for the meaning of
these
assignements.\par
In particular, it is easy to check that the form (\ref{ntwo19}) of the $N=2$
bosonic potential reduces, the Landau-Ginzburg potential being given by
eq. (\ref{inducedsuperpotential}), exactly to the potential of eq.
(\ref{nfour16}):
\begin{eqnarray}
\label{section13bospot}
U &=& \o 12 \left(r-\cD^3\right)^2 +\sum_A |\partial_AW|^2 +8 |M|^2 \sum_i
(q^i)^2
\left(|u^i|^2+|v^i|^2\right)\nonumber\\
&=& \o 12 \left(r-\cD^3\right)^2 +\o 12
|s-\cD^{\ssp}|^2 +(8|M|^2 +2 |X^0|^2)\sum_i (q^i)^2
\left(|u^i|^2+|v^i|^2\right)\nonumber\\
\end{eqnarray}
{}From this $N=2$ point of view, why do we  not see two
different phases in the structure of the classical vacuum?
To answer this question let us  compare the above potential  with that of eq.
(\ref{9.2})
i.e. with the simplest of the examples considered in Witten's paper
\cite{Wittenphases}.
The crucial difference resides  in the expression of the real component
$\cD^3(u,v)$ of the momentum map, see eq. (\ref{nfour15}). It is indeed clear
that by
setting
\begin{equation}
\label{section13realeq}
r-\sum_i q^i\left(|u^i|^2-|v^i|^2\right)=0
\end{equation}
the exchange of $r>0$ with $r<0$ just corresponds to the exchange of the $u$'s
with the $v$'s. Since in all the other expressions the $u$'s and the $v$'s
play symmetric roles, the two phases $r>0$ and $r<0$ are actually the same
thing.
This is far from being accidental. The reason why the
charge of $v^i$ is opposite to the one of $u^i$ is the triholomorphicity
of the action of the gauge group, as already noted in section I.
The triholomorphicity is essential in order to have an $N=4$ theory; thus
the indistinguishability of the two phases is intrinsic to any $N=4$
theory of the type we are considering in this paper.
\vskip 0.2cm
It would be interesting to investigate in detail what happens at $r=0$,
or better in general for the values of the momentum map parameters
($r$ and $s$ here) where the hyperK\"ahler quotient degenerates
\cite{kronheimer}, \cite{newalenostro}.
This might be particularly relevant in the case of the ALE spaces
\cite{gravinstanton}, \cite{kronheimer}, four
dimensional spaces with $c_1=0$, obtained via a hyperK\"ahler construction.
Note that the supersymmetric $\sigma$-model on such spaces, because of the
vanishing of the firs Chern class, gives rise,  at the
quantum level,  to a superconformal theory.
In these cases, for certain values of the momentum map parameters the
hyperK\"ahler quotient
degenerates into an orbifold. If for these particular values of the parameters
there is no real singularity in the complete theory (the gauge plus matter
system), then we have an explicit unification of the "singular" case where
the effective theory is the superconformal theory of an orbifold space with
the case where the effective theory is a  $\sigma$-model on a ALE space.
\vskip 0.2cm
To complete the definition of the vacuum, we mut set $M=0, X^0=0$ and
require $\cD^{\ssp}(u,v)=s$.\par
We have found that considering an $N=2$ theory with a Landau-Ginzburg potential
(\ref{inducedsuperpotential}) does not introduce the possibility of a
Landau-Ginzburg phase for  the vacuum. We can understand this fact because such
a potential has a ``geometrical'' origin and at the level of the $N=4$ theory
it is related to the gauge sector; it does not come from a self-interaction
of the $N=4$ matter fields (quaternionic multiplets). This self-interaction
cannot exist, as we noted  above.
\vskip 0.2cm
To reconstruct the low-energy theory, we must follow the procedure outlined
in section \ref{cpn}. The only difference is that there we considered  the
${\bf CP}^N$ case, in which there is no Landau-Ginzburg potential. On the other
hand
here we must take into account also the constraint $\cD^{\ssp}=s$ which
comes from the potential (\ref{inducedsuperpotential}).\par
To be definite we consider,  in an extremely sketchy way,  the case that
corresponding to the obvious  $N=4$ generalization of ${\bf CP}^N$, namely  in
the above formul\ae ~  we take all the charges $q^i=1$. The spaces obtained by
means
of the hyperK\"ahler quotient procedure of ${\bf H}^n$ with respect to this
$U(1)$ action have real dimension $4(n-1)$; the K\"ahler metric metric they
inherit from the quotient construction are called Calabi metrics
\cite{calabimetrics}.\par
First of all, if we restore the gauge coupling constant (extending the
redefinitions (\ref{9.3}) to the other fields of the $N=4$ gauge multiplet)
{\it before} reducing the theory in its $N=2$ components, at the end also the
kinetic terms for $X^0$ and its fermionic partners aquire a factor $\o 1{g^2}$.
They disappear, together with the kinetic terms for the remaining $N=2$ gauge
multiplet, when we take the limit $g\rightarrow\infty$, which should correspond
to integrate over the massive fluctuations.
This matches the fact that also the fluctuations of $X^0$ and of its
partners around the vacuum are massive.\par
In analogy with section  \ref{cpn} we consider the variations of the action
with respect to the non-propagating fields. The variations in $X^0,\ps 0,
\pst 0$ are on the same footing as those in $M,\lap,\latp$. In particular
we  get fermionic constraints that, by supersymmetry, correspond
to the two momentum map equations
\begin{eqnarray}
\label{mommap}
\cD^3 =r\hskip 12pt &\Leftrightarrow &\hskip 12pt \sum_i(|u^i|^2 -|v^i|^2)
=r\\
\cD^{\ssp}=s\hskip 12pt &\Leftrightarrow &\hskip 12pt 2i\sum_i u^{i^*}
v^{i^*} =s
\end{eqnarray}
The fermionic contraints are crucial  in the technical reconstruction
of the correct form of the rheonomic lagrangian of the $N=2$ $\sigma$-model on
a space $TCP^N$ endowed with a Calabi metric,
the Calabi space. We omit all the details confining ourselves
to pointing out the essential differences with the N=2 case.
\par
Note that the holomorphic contraint $\cD^{\ssp}=s$ is not implemented in the
$N=2$ lagrangian we are starting from, eq.s (\ref{ntwo15},\ref{ntwo13},
\ref{ntwo7}), through a Lagrange
multiplier. This would be the case (by means of the auxiliary field $\cQ$) had
we chosen to utilize the $N=4$ formalism, see eq.(\ref{nfour12}), and this is
the
case for the real constraint $\cD^3=r$, through the auxiliary field $\cP$.
This fact causes no problem, as it is perfectely consistent with what happens,
from the
geometrical point of view, taking the hyperK\"ahler quotient. Indeed the
hyperK\"ahler quotient procedure is schematically represented by
\begin{equation}
\label{hikaquot}
\cS\hskip 3pt\stackrel{j^{\ssp}}{\longleftarrow}\hskip 3pt \cD_{\ssp}^{-
1}(s) \hskip 3pt\stackrel{j^3}{\longleftarrow}\hskip 3pt\cN\equiv
\cD_3^{-1}(r)\cap\cD_{\ssp}^{-1}(s)\hskip 3pt\stackrel{p}{\longrightarrow}
\hskip 3pt \cM\equiv\cN /G
\end{equation}
where we have gone back to the general case and we
have extended in an obvious way the notation of eq. (\ref{kq1}):
$j^{\ssp}$ and $j^3$ are inclusion maps and $p$ the projection on the
quotient.\par
We remarked in section (\ref{cpn}) that the surface $\cD_3^{-1}(r)$ is not
invariant under the action of the {\it complexified} gauge group $G^c$.
It is easy to verify instead that the holomorphic surface $\cD_{\ssp}^{-
1}(s)$ {\it is} invariant under the action of $G^c$. Just as in the
K\"ahler quotient procedure of section (\ref{cpn}) we can therefore replace
the restriction to $\cD_3^{-1}(r)$ and the $G$ quotient with a $G^c$
quotient, without modifying the need of taking the restriction
to $\cD_{\ssp}^{-1}(s)$. The hyperK\"ahler quotient can be realized as
follows:
\begin{equation}
\label{hikaquot2}
\cS\hskip 3pt\stackrel{j^{\ssp}}{\longleftarrow}\cD_{\ssp}^{-1}(s)\hskip 3pt
\stackrel{p^c}{\longrightarrow}\hskip 3pt\cM\equiv\cD_{\ssp}^{-1}(s) /G^c
\end{equation}
We see that,  in any case, we have to implement the constraint $\cD^{\ssp}=
s$. This does not affect  the procedure of extending the
action of the gauge group to its complexification,  which,  in our case,  is
given by:
\begin{eqnarray}
\label{section13compl}
u^i\hskip 3pt\longrightarrow\hskip 3pt e^{i\Phi}u^i \hskip 12pt &;&
\hskip 12pt v^i\hskip 3pt\longrightarrow\hskip 3pt e^{-i\Phi}v^i\nonumber\\
u^{i^*}\hskip 3pt\longrightarrow\hskip 3pt e^{-i\Phi^*}\hskip 12pt &;&
\hskip 12pt v^{i^*}\hskip 3pt\longrightarrow\hskip 3pt e^{i\Phi^*}
v^{i^*}\nonumber\\
v\hskip 3pt &\longrightarrow&\hskip 3pt v+\o i2 (\Phi -\Phi^*)
\end{eqnarray}
and of obtaining the invariance of the lagrangian under this action, by
means of the substitutions
\begin{equation}
\label{section13transf}
u^i\hskip 3pt\longrightarrow\hskip 3pt e^{-v} u^i\hskip 12pt ;\hskip 12pt
v^i\hskip 3pt\longrightarrow\hskip 3pt e^v v^i
\end{equation}
and similarly for the other fields, as it happened in eq. (\ref{9.10bis}).
\par
The variation in the auxiliary field $\cP$, that acts as a Lagrange multiplier
for the real momentum map constraints,  gives,  after the substitutions
(\ref{section13transf}), the equation $\cD^3(e^{-v}u, e^v v)=r$, that is
\begin{equation}
\label{degree2eq}
r-e^{-2v}\sum_i |u^i|^2 + e^{2v}\sum_i |v^i|^2 =0
\end{equation}
This equation is solved for $v$ as follows (we introduce the notation
$\rho^2\equiv r$):
\begin{equation}
\label{vsolut}
e^{2v}=\o{-\rho^2+\sqrt{\rho^4+4\sum_i |u^i|^2 \sum_i |v^i|^2 }}{2\sum_i
|v^i|^2}
\end{equation}
We have still to implement the holomorphic constraint $\cD_{\ssp}=s$; we
have also at our disposal the ${\bf C}^*$ gauge invariance of our lagrangian.
We can utilize this invariance choosing a gauge which can simplify the
implementation of the constraint \cite{hklr}. One can for instance, as it is
clear from
the form (\ref{section13compl}) of the ${\bf C}^*$-transformations, choose the
gauge where
$u^n = v^n$.
In this gauge the constraint
\begin{equation}
\label{holoconstraint}
\cD^{\ssm}=-2i\sum_i u^i v^i = s^*
\end{equation}
is solved by setting
\begin{eqnarray}
\label{hatteduv}
u^i &=&\sqrt{\o{is^*}{2(1+\sum_J \hat u^J\hat v^J)}} (\hat u^I,1)\nonumber\\
v^i &=&\sqrt{\o{is^*}{2(1+\sum_J \hat u^J\hat v^J)}} (\hat v^I,1)
\end{eqnarray}
where the capital indices $I,J,...$ run from $1$ to $n-1$.
The final result of the appropriate manipulations that should be made on the
lagrangian, following what was done in section  \ref{cpn}  will be the
reconstruction of the rheonomic action (\ref{twosigmod6}) for the $N=2$
$\sigma$-model
having as target space the hyperK\"ahler quotient ${\bf H}^n /U(1)$,
endowed with  that the K\"ahler metric which is naturally provided by the
hyperK\"ahler quotient construction, exactly in the same way as it happened in
the
K\"ahler quotient case of section  \ref{cpn}. The K\"ahler quotient is again
obtained through eq. (\ref{kq4}). In expressing the result, it is convenient to
assign a name to the expressions $\sum_i |u^i|^2$ and $\sum_i |v^i|^2$, that,
through eq.s (\ref{hatteduv}),
must be reexpressed in terms of the true
coordinates on the target space, the $\hat u$'s and the $\hat v$'s.
Therefore we set
\begin{eqnarray}
\label{betagamma}
\beta &\equiv& \sum_i |u^i|^2 =\o{is^*}{2|1+\sum_J \hat u^J \hat v^J|^2}
(1+\sum_I |\hat u^I|^2)\nonumber\\
\gamma &\equiv& \sum_i |v^i|^2 =\o{is^*}{2|1+\sum_J \hat u^J \hat v^J|^2}
(1+\sum_I |\hat v^I|^2)
\end{eqnarray}
We note that, differently from the ${\bf CP}^N$ case, the part of the K\"ahler
potential on the target space that comes from the restriction of the potential
for the flat metric on the  manifold ${\bf H}^n$ to the momentum-map
surface $\cD_3^{-1}(r)\cap\cD_{\ssp}^{-1}(s)$ is not an irrelevant constant.
Indeed it is given (see section  I ) by:
\begin{equation}
\label{kprestricted}
K|_{\cN}=\o 12 (e^{-2v}\sum_i |u^i|^2 + e^{2v}\sum_i |v^i|^2) =\o 12
\sqrt{\rho^4 +4\beta\gamma}
\end{equation}
The final expression of the K\"ahler potential for the Calabi metric is:
\begin{equation}
\label{kpcalabi}
\hat K = \o 12\sqrt{\rho^4 +4\beta\gamma} +\o{\rho^2}2 \log \o{-\rho^2 +
\sqrt{\rho^4+4\beta\gamma}}{2\gamma}
\end{equation}
In the case $n=2$, the target space has $4$ real dimensions and the Calabi
metric is nothing else that the Eguchi-Hanson metric, i.e. the simplest
Asymptotically Locally Euclidean gravitational instanton \cite{kronheimer}.

\section {Extension to the case where the quaternionic hypermultiplets have
several
abelian gauge symmetries }
\label{section14}
The extension of the above results to the case of several $U(1)$ multiplets is
fairly simple.
This case is relevant to implement the Kronheimer construction of the multi
Eguchi-Hanson
spaces belonging to $A_k$-series \cite{kronheimer}, \cite{newalenostro}.
Let the gauge group be $U(1)^n$ and let the corresponding
gauge fields be the 1-forms $\cA^{a}$ $(a=1,...,n)$; let the triholomorphic
action of these groups on
the hypermultiplets $Y^{i}$ be generated by the matrices $(F^{a})_{ij}$, then
the covariant derivatives of the quaternionic scalars $u^i,v^i$ will be:
\begin{eqnarray}
\label {covder}
\dell u^i &=&du^i +i\cA^a (F^a)_{ij} u^j\nonumber\\
\dell v^i &=&dv^i -i\cA^a (F^a)_{ij} v^j
\label{nfour16bis}
\end{eqnarray}
Since the group is abelian and the generators $F^a$ are commuting, the
gauge part of the action should simply be given by $n$ replicas of the $U(1)$
lagrangian; thus the world-sheet lagrangian is given by
\begin{eqnarray}
\lefteqn {\cL^{(ws)}_{gauge}\,=\, \o 12 \cF^a\cF^a -i (\lap_a\,\dep\lamm_a
+\mup_a\,\dep \mum_a +\latp\,\dem \latm_a +\mutp_a\,\dem\mutm_a) + }
\nonumber\\
&&\mbox{} +4(\dep M^*_a\dem M_a +\dem M^*_a \dep M_a +\dep N^*_a \dem N_a
+\dem N^*_a\dep N_a) +\nonumber\\
&&\mbox{}+ \o {\theta^a } {2\pi } \cF^a+2\cP^a\cP^a +2(\cQ^a)^* \cQ^a - r^a
\cP^a - \biggl( s^a (\cQ^a)^* +(s^a)^* \cQ^a \biggr)
\label{nfour17}
\end {eqnarray}
where the summation on the index $a$ enumerating the $U(1)$ generators is
understood, as usual.
Similar formulae hold for the rheonomic action.
For  the matter part of the Lagrangian, note that the covariant derivatives
(\ref{nfour1})
are nearly identical to the ones  utilized in the case of one multiplet
(\ref{nfour6})
 We just have to take into account the
substitution $\qu ij \longrightarrow (F^a)_{ij}$ and the summation over the
index $a$.
The modification of the rheonomic parametrizations and of the action are
almost trivial, substantially because of the abelian nature of the group
that we consider. Let us therefore quote here only the spacetime lagrangian:
%
\begin{eqnarray}
\lefteqn{\cL^{(ws)}_{quatern}\,=\,-\bigl(\delp u^{i^*}\,\delm u^i+\delm
u^{i^*}\,\delp u^i
+\delp v^{i^*}\,\delm v^i+\delp v^{i^*}\,\delm v^i\bigr)}\nonumber\\
&&\mbox{}+4i\bigl(\psu i\,\delm\psus i +\psv i\,\delm\psvs i +\psut
i \delp\psuts i +\psvt i\,\delp\psvts i\bigr)\nonumber\\
&&\mbox{}+2i\sum_i\sum_a (f^a)^i_j\biggl\{\biggl[\psu i\,\bigl(\lamm_a u^{j^*}
+
\mup_a v^j\bigr) -\coco\biggr]-\biggl[\psv i\,\bigl(\lamm_a v^{j^*} -
\mup_a u^j\bigr) -\coco\biggr]\nonumber\\
&&\mbox{}- \biggl[\psut i\,\bigl(\latm_a u^{j^*} +\mutp_a v^j\bigr)
-\coco\biggr]+ \biggl[\psvt i\,\bigl(\latm_a v^{j^*} -\mutp_a u^j\bigr)
-\coco\biggr]\biggr\} \nonumber\\
&&\mbox{}+8i\sum_a\biggl[M_a^*\sum_i (F^a)^i_j
\bigl(\psu i\psuts j -\psv i\,
\psvts j\bigr)-\coco\biggr]\nonumber\\
&&\mbox{} - 8i\biggl[\sum_a N\sum_i (F^a)^i_j
\bigl(\psu i\,\psvt j
+\psv i\,\psut j\bigr)-\coco\biggr] +\nonumber\\
&&\mbox{}+ 8\sum_a\bigl(|M_a|^2
+|N_a|^2\bigr)\sum_i (f^a\, f^a)^i_j\bigl(u^{i^*} u^j
 + v^{i^*} v^j\bigr)\nonumber\\
&&\mbox{}-2\sum_a\cP^a\sum_i (F^a)^i_j\bigl(u^{i^*}u^j -v^{i^*}v^j\bigr)
\nonumber\\
&&\mbox{}+2i\sum_a\bigl(\cQ^a\,\sum_i (F^a)^i_j
u^i v^j -\coco\bigr)
\label{nfour19}
\end{eqnarray}
As expected, the auxiliary fields $\cP^a, \cQ^a$ multiply
the $a^{th}$ component of the momentum-map, respectively the real part ${\cal
D}^3$ and
 the anti-holomorphic part  ${\cal D}^{\ssm}$.\par
The complete bosonic potential takes therefore the following direct sum form:
\begin{equation}
U=\sum_a \,\left [ \,\o 12\bigl(r_{a}-{\cal D}^3_{a}\bigr)^2 +
\o 12|s_a-{\cal D}^{\ssp}_a|^2 +8\bigl(|M_a|^2
+|N_a|^2\bigr)\,\sum_{i,j} \, (F^{a}F^{a})_{ij}\,  \bigl( u^{i^*} u^j\, +
\, v^{i^*} v^{j}\bigr) \, \right ]
\label{nfour20}
\end{equation}

\section{Quaternionic notation for the $N=4$ gauge-matter system  and
identification of
the  N=4 $R$-symmetries}
\label{quatn4}
The above construction of the N=4 $gauge\, \oplus \, matter$ system can be
recast in a more
compact quaternionic notation that allows a simple identification of a
$U(2)_{L}$ and a
$U(2)_R$ global R-symmetry group, respectively acting on the left-moving and
right-moving
degrees of freedom. As it is well known \cite{ellipgenera}, in the N=2 case the
R-symmetries
play an important role in the identification of the superconformal theories
emerging at the
critical point and in the calculation of the so called {\it elliptic genus}, a
very interesting
type of genus one path integral with  twisted boundary conditions that has a
topological
meaning as the index of one of the supersymmetry charges. Therefore the
identification of the
N=4 R-symmetries is particularly valuable, their $SU(2)$ subgroups will turn
into
the $SU(2)_L$ and $SU(2)_R$ currents of the N=4 superalgebras for the
left-moving
and right-moving sectors, respectively. Let us then introduce the quaternionic
formalism.
Setting $\omega = 0$, we can write the super-world-sheet structure equations as
follows:
\begin{eqnarray}
d\eplus & = & \o i4 \tr (\dag Z\, Z)\nonumber\\
d\eminus & = & \o i4 \tr (\dag {\Zt}\, \Zt)
\label{nfour21}
\end{eqnarray}
where
\begin{equation}
Z=\twomat{\zem}{i\cgp}{i\cgm}{\zep} \hskip 0.5cm ;\hskip 0.5cm
\Zt=\twomat{\zetm}{-i\cgtp}{-i\cgtm}{\zetp}
\end{equation}
To describe the abelian gauge multiplet we
group the gauginos into quaternions, according to :
\begin{equation}
\La=\twomat{\latm}{-i\mutp}{-i\mum}{\latp} \hskip 0.5cm ;\hskip 0.5cm
\Lat=\twomat{\lam}{i\mup}{i\mutm}{\lap}
\label{nfour22}
\end{equation}
and the gauge scalars, according to:
\begin{equation}
\Sigma=\twomat {M}{iN}{iN^*}{M^*}
\label{nfour23}
\end{equation}
It is also useful, although we do not use such a notation in the Lagrangian,
 to group the field strength $\cF $and the auxiliary fields $\cP,\cQ $ into
another quaternion:
\begin{equation}
{ f}=\twomat{\o {\cF}2+i\cP}{-i\cQ}{-i\cQ^*}{\o {\cF}2-i\cP}
\hskip 0.3cm ;\hskip 0.3cm
\tilde f=\twomat{\o {\cF}2-i\cP}{-i\cQ}{-i\cQ^*}{\o {\cF}2+i\cP}
\label{nfour24}
\end{equation}
Then the rheonomic parametrizations (\ref{nfour2}) can be written as follows:
\begin{eqnarray}
\label{paraquat}
F&=&\tr { f} \eplus\eminus -\o i2 \tr (\dag\Lat Z)\,\eminus + \o i2 \tr
(\dag{\La}\Zt)\,\eplus +\tr(Z\sigma_3\dag{\Zt}\Sigma)\nonumber\\
d\Sigma&=&\dep \Sigma\,\eplus +\dem \Sigma\,\eminus -\o 14
\La\sigma_3\dag Z +\o 14 \Zt \sigma_3 \Lat\nonumber\\
d\La&=&\dep\La\,\eplus +\dem\La\,\eminus + \Zt f
+ 2i \dep\Sigma Z\sigma_3\nonumber\\
d\Lat&=&\dep\Lat\,\eplus +\dem\Lat\,\eminus +Z\tilde f
+ 2i\dem \dag \Sigma \Zt\sigma_3 \nonumber\\
d{f}&=&\dep f\,\eplus+\dem f\,\eminus +\o i2 \dep\dag{\Lat}Z -\o i2
\dag\Zt\dem \La
\label{nfour24bis}
\end{eqnarray}

These parametrizations (\ref{nfour24bis}) are invariant under the following
{\it left-moving} and {\it right-moving} $R$-symmetries, where
 $U_L,\, U_R\in U(2)$ are arbitrary unitary $2 \, \times \, 2$ matrices:
\begin{equation}
\label{nfour25}
\begin{array}{l}
Z\,\longrightarrow U_L\, Z\\
\Zt\,\longrightarrow U_R\,\Zt
\end{array}
\hskip 0.3cm ;\hskip 0.3cm
\begin{array}{l}
\Lat\,\longrightarrow U_L\,\Lat\\
\La\,\longrightarrow U_R\,\La
\end{array}
\hskip 0.3cm ;\hskip 0.3cm
\Sigma\,\longrightarrow U_R\, \Sigma\, U_L^{-1}
\end{equation}
The rheonomic action (\ref{nfour3}) can also be rewritten in this notation as
it follows:
\begin{eqnarray}
\lefteqn{\cL^{(ws)}_{gauge}\,=\, -\o 12 \cF^2 + \cF\biggl[F
+\o i2 \tr (\dag\Lat Z)\,\eminus- \o i2 \tr
(\dag{\La}\Zt)\,\eplus -\tr(\Sigma\dag\Zt\sigma_3 Z)\biggr] }\nonumber\\
&&\mbox{}- \o i4 \tr(\dag{\Lat}\, d\Lat )\eminus +\o i4 \tr(\dag {\La} \,d\La
)\eplus \nonumber\\
&&\mbox{}- 4\tr\biggl\{ \biggl[d\dag{\Sigma} +\o 14 Z\sigma_3\dag\La
-\o 14 \Lat\sigma_3\dag \Zt\biggr] (S_{\ssp}\eplus-S_{\ssm}\eminus)
+(\dag S_{\ssp}\,S_{\ssm} +\dag S_{\ssm}\,S_{\ssp})
\eplus\eminus\biggr\}\nonumber\\
&&\mbox{} +\tr (d\Sigma\,\Lat\sigma_3 \dag{\Zt}
+d\dag{\Sigma}\,\La\sigma_3\dag Z) -\o 14 \tr (\dag\La\Zt\sigma_3
\dag{\Lat})\sigma_3\nonumber\\
&&\mbox{}+\o 12\tr\biggl\{\twomat {-r}{s^*}{s}{r}[\dag\La\Zt\,\eplus
+\dag{\Lat} Z\,\eminus]\biggr\}
+ 2i\tr\biggl\{\twomat r{s^*}{-s}{r} \dag\Zt\Sigma Z\biggr\}\nonumber\\
&&\mbox{}
\biggl\{
+\o{\theta}{2\pi} F +\biggl[2\cP^2 +2\cQ^*\cQ -2r\cP -(s\cQ^*
+s^*\cQ)\biggr]\biggr\}\eplus\eminus
\label{nfour26}
\end{eqnarray}
Written in this form, the superspace Lagrangian is invariant {\it by
inspection}
against the $R$-symmetries (\ref{nfour25}).
\par
The hypermultiplets are rewritten in quaternionic notation as follows:
\def\Ps#1{\Psi^{#1}}
\def\Pst#1{\tilde\Psi^{#1}}
\begin{equation}
Y^i=\twomat{u^i}{iv^{i^*}}{iv^i}{u^{i^*}}\hskip 0.2cm ;\hskip 0.2cm
\Ps i =\twomat{\psu i}{-i \psvs i}{-i\psv i}{\psus i}\hskip 0.2cm
;\hskip 0.2cm
\Pst i =\twomat{\psut i}{i \psvts i}{i\psvt i}{\psuts i}
\label{nfour27}
\end{equation}
The Bianchi Identities take the form:
\begin{equation}
\nabla^2 Y^i =i F\qu ij \sigma_3\,Y^j
\label{nfour28}
\end{equation}
and the rheonomic parametrizations (\ref{nfour8}) become:
\begin{eqnarray}
\dell Y^i &=&\delp Y^i\eplus +\delm Y^i\eminus +\sigma_3,\Ps i\,Z\, +
\Pst i\,\Zt \,\sigma_3\nonumber\\
\dell\Ps i &=&\delp\Ps i\eplus +\delm\Ps i\eminus -\o i2\sigma_3\,\delp Y^i
\dag Z
+ iY^j\,\qu ij\sigma_3\,\dag{\Zt}\,\Sigma\nonumber\\
\dell\Pst i &=&\delp\Pst i\eplus +\delm\Pst i\eminus -\o i2
\,\delm Y^i\sigma_3\,\dag\Zt
+ iY^j\,\qu ij\sigma_3\,\dag Z\,\dag\Sigma
\label{nfour29}
\end{eqnarray}
These parametrizations are invariant under the left- and right-moving
R-symmetries provided the transformations (\ref{nfour25})
are adjoined to the following ones:
\begin{equation}
\label{nfour30}
\Ps i\longrightarrow \Ps i U^{-1}_L \hskip 0.3cm ;\hskip 0.3cm
\Pst i\longrightarrow \Pst i U^{-1}_R
\end{equation}
The rheonomic action (\ref{nfour11}) is rewritten as follow in quaternionic
notation:
\begin{eqnarray}
\label{nfour31}
\cL^{(rheon)}_{quatern}\, &=& \,\tr\,\biggl\{\,(\dell Y^i +\sigma_3\Ps i Z
+\sigma_3\Pst i\Zt)({\dags Yi}_{\ssp}\eplus -{\dags Yi}_{\ssm}\eminus)
+{\dags Yi}_{\ssp}Y^i_{\ssm}\eplus\eminus \nonumber \\
&&\mbox{} -4i(\dags {\Psi}{i})\,\dell\Ps i\,\eplus
-\dags{\tilde \Psi}{i}\,\dell\Pst i\,\eminus) +\dags{\tilde\Psi}{i}
\sigma_3\Ps i\,Z
\sigma_3 \dag\Zt \nonumber\\
&&\mbox{}+ \o 12 (\dags{\Psi}{i}\sigma_3\Ps i\,\tr\dag Z Z -\o 12
\dags{\tilde\Psi}{i}
\Pst i\,\dag{\Zt}\Zt) -\dell \dags Yi(\sigma_3\Ps i Z - \Pst i \Zt\sigma_3)
\nonumber\\
&&\mbox{}-4\Ps i\dag\Sigma\Zt\qu ij\sigma_3 \dags Yj\,\, \eplus -
4\Pst i\Sigma Z\qu ij \dags Yj\sigma_3\,\,\eminus \nonumber\\
&&\mbox{}-\o 12 \twomat{\cD^3}{-i\cD^{\ssm}}{-i\cD^{\ssp}}{-\cD^3}
[\dag\La\Zt\eplus +\dag\Lat Z\eminus] + 2i \twomat{\cD^3}{-i\cD^{\ssm}}
{i\cD^{\ssm}}{\cD^3}\dag\Zt\Sigma Z \biggl\}\nonumber\\
&&\mbox{}+\biggl\{2i\tr\,[\qu ij\dags Yj \Ps i\Lat +\qu ij\sigma_3 Y^j
\sigma_3\La\dags{\tilde \Psi}{i}] -8i\tr\,\Pst i\Sigma\qu ij
\dags{\Psi}{j})\nonumber \\
&&\mbox{}+8\tr\,(\dag{\Sigma}\Sigma \dags Yi)(q^2)^i_{\hskip 3pt
j}Y^j)- 2\cP\cD^3 +i[\cQ\cD^{\ssp} -\cQ^*\cD^{\ssm}]
\biggr\}\eplus\eminus
\end{eqnarray}
Written in this form, also the hypermultiplet action  is invariant by
inspection
with respect to the $R$-symmetries (\ref{nfour25},\ref{nfour30}).

\section{ The A and B Topological Twists of the N=2 and N=4 Theories}

We now discuss the two possible topolological twists (A and B models, in
Witten's nomenclature)
of the above reviewed N=2 Landau-Ginzburg theories with local gauge symmetries.
The topological twists of
the corresponding N=4 theories are covered by this discussion, since, as we
have shown,
they can
be regarded as special instances of N=2 theories with a specific field-content
and a special
choice of the superpotential.
\par
We focus on the formal aspects of the topological twist procedure, relying on
the clarification
of the  involved steps, recently obtained in the D=4 case \cite{FreAnselmi},
and making
preparatory remarks for our planned extension of the whole procedure to the
case of locally N=2
supersymmetric theories ({\it matter-coupled 2D topological gravity}).
\par
As discussed at length in \cite{FreAnselmi}, the topological twist extracts,
from any N=2
supersymmetric theory, a topological field-theory that is already {\it
gauge-fixed}, namely where
the BRST-algebra already contains the antighosts, whose Slavnov variation is
proportional to
the gauge-fixings. The appropriate {\it instanton conditions} that play the
role of gauge-fixings
for the topological symmetry are thus automatically selected when the
topological field-theory
is obtained via the topological twist. This latter consists of the following
steps:
{\it
\par
i) First one BRST quantizes the ordinary N=2 theory. ({\rm This step is
relevant when
the ordinary N=2 theory is locally supersymmetric (supergravity) and/or it
contains
gauge-fields as in our specific case. For rigid N=2 theories containing only
matter
multiplets as the N=2 $\sigma$-model or the rigid N=2 Landau-Ginzburg models,
this step is
empty})
\par
ii) Then one redefines the spins of all the fields taking as new Lorentz group
the diagonal
of the old Lorentz group with the internal automorphism group of N=2
supersymmetry.
\par
iii) After this redefinition, one recognizes that at least one component of the
N=2 multiplet of supercharges, say $Q_0$ has spin zero, is nilpotent and
anticommutes
with the old BRST-charge  $Q^{(old)}_{BRST} \, Q_0 \, + \, Q_0
\,Q^{(old)}_{BRST} \, =\, 0 $.
Then one defines the new BRST-charge as $Q^{(new)}\, = \,Q_0 \, + \,
Q^{(old)}_{BRST}$
\par
iv) Next one redefines the ghost-number $gh^{(new)} \, =Ê\, gh^{(old)} \, + \,
F $ where $F$ is
some appropriate fermion number in such a way that the operator
$(-1)^{gh^{(new)}}$ anticommutes
with the new BRST-charge, in the same way as the operator $(-1)^{gh^{(old)}}$
did anticommute
with the old BRST-charge. In this way all the fields of the BRST-quantized N=2
theory acquire
a new well-defined ghost-number.
\par
v) Reading the ghost-numbers one separates the physical fields from the ghosts
and the
antighosts, the BRST-variation of these latter yielding the gauge fixing
instanton equations.
The gauge-free BRST algebra (that involving no-antighosts) \cite{brsformalism}
should, at this point, be
recognizable as that associated with a well defined topological symmetry: for
instance
the continuous
deformations of the vielbein (topological gravity), the continuous deformations
of the gauge-connection (topological Yang-Mills theory), the continuous
deformations
of the embedding functions (topological $\sigma$-model) and so on}
\vskip 0.2cm
{\sl STEP 1}
\par
The first step is straightforward. The case of interest to us just involves an
ordinary
gauge symmetry.
Hence we just make the shift:
\begin{equation}
{\cal A} \, \longrightarrow \, {\hat {\cal A}} \,= \, {\cal A} \, + \,
c^{(gauge)}
\label{toptwist1}
\end{equation}
where $c^{(gauge)}$ are ordinary Yang-Mills ghosts. Imposing the BRST-rheonomic
conditions:
\begin{eqnarray}
{\hat F} \, & \eqdef & \,{\hat d} \, {\hat {\cal A}} \, + \, {\hat {\cal A}} \,
\wedge \,
{\hat {\cal A}} \nonumber\\
&=&(d+s) \,\left ( {\cal A} \, +\,c^{(gauge)} \right ) \, + \,
\left ( {\cal A} \, +\,c^{(gauge)} \right ) \, \wedge \,
\left ( {\cal A} \, +\,c^{(gauge)} \right )\nonumber\\
&=&\cF\,\eplus\eminus -\o i2\bigl(\lap\zem +\lamm\zep\bigr)\,\eminus
+\o i2\bigl(\latp\zetp +\latm\zetp \bigr)\,\eplus + M\,\zem\zetp
- M^*\,\zep\zetm\nonumber\\
\label{toptwist2}
\end{eqnarray}
we obtain the ordinary BRST algebra of an N=2 supersymmetric gauge theory.  We
do not dwell
on this trivial point.
\par
{\sl STEP 2}
\par
The second step is the delicate one. In two dimensions the Lorentz group is
$O(1,1)$ which
becomes $O(2)$ after Wick-rotation. Let us name $J_S$ the Lorentz generator:
the eigenvalues
$s^{i}$ of this operator are the spins of the various fields $\varphi^{i}$.
The number $s^{i}$ appears in the Lorentz covariant derivative of the field
$\varphi^{i}$:
\begin{equation}
\nabla \, \varphi^{i} \, = \, d\varphi^{i} \, - \,s^{i}\, \omega \, \varphi^{i}
\label{toptwist3}
\end{equation}
The automorphism group of the supersymmetry algebra that can be used to
redefine the Lorentz
group is the R-symmetry group $U(1)_L \, \otimes \, U(1)_R$.
Denoting by $J_L$ , $J_R$ the two R-symmetry generators,
we redefine the Lorentz generator according to the formula:
\begin{equation}
J^{'}_{S} ~= ~J_S ~+~\o{1}{2} \,\left [ \, J_R ~\pm~J_L \, \right ]
\label{toptwist4}
\end{equation}
Correspondingly the new spin quantum number is given by:
\begin{equation}
s^{'} ~= ~s ~+~\o{1}{2} \,\left [ \, q_R ~\pm~q_L \, \right ]
\label{toptwist5}
\end{equation}
The choice of sign in eq.s (\ref{toptwist4}), (\ref{toptwist5}) corresponds to
the existence of
two different topological twists for the same N=2 theory. Following Witten
\cite{Wittenmirror}
they will be named
the A-twist, leading to the A-model (upper choice of the sign) and the B-twist,
leading
to the B-model (lower choice of the sign). It might seem arbitrary to restrict
the possible
linear combinations of the operators $J_S$, $J_L$ and $J_R$ to those in eq.s
(\ref{toptwist4}), (\ref{toptwist5}), but, actually, these are the only
possible ones if we take
into account the following requirements. In the gravitational sector the spin
redefinition must
transform N=2 supergravity into topological gravity, hence the spins of the
vielbein $e^{\pm}$
must remain the same before and after the twists: this fixes the coefficient of
$J_S$ to be equal
to one as in eq. (\ref{toptwist4}). Furthermore, of the four gravitino 1-forms
$\zep , \, \zem , \,
\zetp , \, \zetm$, two must acquire spin $s=1$ and $s=-1$, respectively, and
the other two must
have spin zero. This is so because two of the gravitinos have to become the
topological
ghosts corresponding to continuous deformations of the vielbein (so they must
have the
same spins as the vielbein) while the other two must be the gauge fields of
those supersymmetry
charges that, acquiring spin zero, can be used to redefine the BRST-charge.
These
constraints have two solutions: indeed they fix the coefficients of $J_L$ and
$J_R$ to the values
displayed in eq.s (\ref{toptwist4}), (\ref{toptwist5}), the choice of sign
distinguishing the
two solutions.
\par
{\sl STEP 3}
\par
 Naming $Q_{BRS}$ the BRST-charge of the original gauge theory and $Q^{\pm}$,
${\tilde Q}^{\pm}$
the supersymmetry charges generating the transformations of parameters
$\varepsilon^{\pm}$,
${\tilde \varepsilon}^{\pm}$, whose corresponding gauge fields are the
gravitinos $\zeta^{\pm}$, ${\tilde \zeta}^{\pm}$, we realize that in the
A-twist the spinless
supercharges are $Q^{-}$ and ${\tilde Q}^{+}$ while, in the B-twist they are
$Q^{+}$ and ${\tilde Q}^{+}$. In both cases the two spinless supercharges
anticommute among
themselves and with the BRST-charge so that we can define the new BRST-charge
of the topological
theory according to the formula:
\begin{equation}
Q^{'}_{BRS} ~=~Q_{BRS} ~\mp~Q^{\mp} ~+~{\tilde Q}^{+}
\label{toptwist6}
\end{equation}
The choice of signs in eq. (\ref{toptwist6}) is just a matter of convention:
once more the upper choice of sign corresponds to the A-twist while the lower
corresponds to the
B-twist. The physical states of the topological theory are the cohomology
classes of the operators
(\ref{toptwist6}).
\par
{\sl STEP 4}
\par
What matters in the definition of the ghost number are the differences of ghost
numbers for the
fields related by a BRST-transformation. Indeed ghost number is one of the two
gradings in
a double elliptic complex. Hence to all the fields we must assign an integer
grading which has to
be increased of one unit by the application of the BRST-charge (or Slavnov
operator). In other
words $Q^{'}_{BRS}$ must have ghost-number $gh=1$.  These requirements are
satisfied if,
for the redefinition of the ghost number $gh^{'} = gh + F$, we
use the generator $F$ of some $U(1)$ symmetry of the original N=2 theory
with respect to which all the fields have integer charges and in particular
the new BRST-generator (\ref{BRS1}) has charge $q_{BRST}=1$: furthermore the
two gravitinos
that acquire the same spin as the vielbein and become the ghosts of topological
gravity
must have $gh^{'} =1$. In this case, the action, being invariant under
the chosen symmetry, has ghost number $gh^{'} =0$.
This is the situation that can be realized in the A-twist. In the B-twist the
situation
is more complicated since there is no symmetry of the original theory that
satisfies
all the requirements: yet ghost-numbers can be consistently assigned to all the
fields
in such a way that $(-1)^{gh^{'}}$ does anticommute with the new BRST-charge.
The action
however has no fixed ghost number, rather it is the sum of terms having
 different values of $gh^{'}$. However
modulo BRST-exact terms, ghost-number is conserved since, modulo these
terms, the action has a fixed ghost number $gh_{action}=-2$.
\par
We examine first the situation for the A-twist. In this case,
naming ${\#} gh$ the ghost number of the original gauge theory, we fulfill all
the
desired properties if we define the ghost number of the topological theory
according to the formula:
\begin{equation}
{\#} gh^{'} ~=~{\#}  gh ~+~ q_L ~-~q_R
\label{pippus}
\end{equation}
In this case the $U(1)$ symmetry utilized to redefine the ghost-number is
generated by
$F=J_L \, - \,J_R$ and it is a subgroup of the R-symmetry group $U(1)_L \otimes
U(1)_R$.
\par
In the B-twist case, the new ghost numbers are defined as follows. For the
fields
belonging to the N=2 gauge multiplet we set:
\begin{equation}
{\#} gh^{'} ~=~{\#} gh ~-~ q_L ~-~q_R
\label{pippusbis}
\end{equation}
while for the fields in the chiral matter multiplet, we put:
\begin{eqnarray}
\# gh \left [ X^{i} \right ] &=&0\nonumber\\
\# gh \left [ X^{i^*} \right ] &=&0\nonumber\\
\# gh \left [ \pss i \, + \, \psts i \right ] &=&1\nonumber\\
\# gh \left [ \pss i \, - \, \psts i \right ] &=&-1\nonumber\\
\# gh \left [ \ps i  \right ] &=&1\nonumber\\
\# gh \left [ \pst i  \right ] &=&1
\label{pippustris}
\end{eqnarray}
The spin and charge assignments, before and after the twist,
in the N=2 abelian gauge theory are summarized in table I.  In this table
the fermions $\pss i$ and $\psts i$ appear with {\it non diag.} on the
ghost-number column
of the B-twist because their $gh$-number is undefined:
indeed as it appears from eq. (\ref{pippustris}) only their sum and difference
have a well-defined
ghost-number. The same will happen for the corresponding fields of the N=4
theory.
\par
As a preparation to {\sl STEP 5}, namely the identification of the topological
BRST-algebras
and theories generated by the twists, we consider the explicit form of the
BRST-transformations
of all the fields. In view of a very simple and powerful fixed point theorem
due to Witten
\cite{Wittenmirror} we also recall that the topological theory, besides being
BRST-invariant
with respect to the supercharge (\ref{toptwist6}) has also a supergroup $(0|2)$
of fermionic
symmetries commuting with the BRST-transformations and generated by the two
spinless supercharges
utilized to redefine the BRST-charge. Hence while writing the topological
BRST-transformations we write  also the $(0|2)$-transformations. As Witten
pointed out the
topological functional integral is concentrated on those configurations that
are a fixed point
of the $(0|2)$-transformations: these are the true {\it instantons} of our
theory and
can be read from the formulae we are going to list.
\par
In the A-twisted case the BRST-charge is given by:
\begin{equation}
Q^{(A)}_{BRS} ~=~Q^{(gauge)}_{BRS} ~-~Q^{-}~+~{\tilde Q}^{+}
\label{BRS1}
\end{equation}
Correspondingly we rename the supersymmetry parameters as follows:
\begin{eqnarray}
-\varepsilon^{-} ~&=&~\al\nonumber\\
{\tilde \varepsilon}^{+} ~&=&~\alp\nonumber\\
\alA~&=&~{\tilde \varepsilon}^{+}~=~\varepsilon^{-}~=~\alg
\label{BRS2}
\end{eqnarray}
where $\alg$ is the nilpotent BRST-parameter associated with the original gauge
symmetry and
$\alA$ is the BRST-parameter of the A-twisted model. The parameters $\al$ and
$\alp$ correspond
to the two fermionic nilpotent transformations, commuting with the BRST
transformations and
generating the $(0|2)$ supergroup of exact symmetries of the topological
action.
Using the above conventions the form of the BRST-transformations and of the
$(0|2)$ symmetries in
the A-twisted version of the
N=2 gauge coupled Landau-Ginzburg model is given by the following formulae:
\begin{eqnarray}
\delta \cAp &=&\alA \, \left ( \,-\o{i}{2}  \, \latm \, + \,
\partial_{+} c^{gauge} \, \right ) ~=~-\o{i}{2}\, \al \, \latm \,  + \,\alg \,
\partial_{+} c^{gauge} \nonumber\\
\delta \cAm &=&\alA \, \left ( \, -\o{i}{2}\,  \lap \, + \,
\partial_{-}c^{gauge} \, \right )~=~-\o{i}{2}\, \alp \, \lap \, + \, \alg \,
\partial_{-}c^{gauge} \,\nonumber\\
\delta M &=&0\nonumber\\
\delta M^{*} &=& \o{1}{4} \, \alA \, \left ( \, \latp \, + \, {\tilde
\lambda}^{-}\, \right )~=~
\o{1}{4} \, \left ( \,  \al \,\latp \, + \, \alp \, {\tilde  \lambda}^{-}\,
\right )
\nonumber\\
\delta \latp &=& \alA \, \left ( \, \o {\cF}{2} \, - \, i \, \cP \, \right )~=~
\alp \, \left ( \,\o{\cF}{2} \, - \, i \, \cP \, \right )\nonumber\\
\delta \latm &=& - 2i \, \alA \, \partial_{+} \, M~=~- 2i \, \al \,
\partial_{+} \, M \nonumber\\
\delta \lap &=& - 2i \, \alA \, \partial_{-} \, M~=~- 2i \, \alp \,
\partial_{-} \, M \nonumber\\
\delta \lam &=&  \, \alA \,\left ( \,\o {\cF}{2} \, - \, i \, \cP \, \right )~=
{}~\, \al \,\left ( \, \o {\cF}{2} \, - \, i \, \cP \, \right )\nonumber\\
\delta \cP &=&\alA \, \o{1}{4} \Biggl [ \, - \partial_{+} \, \lap \, + \,
\partial_{-} \latm
\, \Biggr ]~=~\o{1}{4} \Biggl [ \, -\, \al  \, \partial_{+} \, \lap \, +
\, \alp \, \partial_{-} \latm\, \Biggr ]\nonumber\\
\delta X^{i} &=& \alA \, \left ( \ps i \, +
\, i \,  c^{gauge} \, q^{i}_{j} X^{j} \, \right ) ~=~
 \left ( \, \al \ps i \, +
\, i \, \alg c^{gauge} \, q^{i}_{j} X^{j} \, \right )\nonumber\\
\delta X^{i^*} &=&\alA \, \left ( \psts i \, -
\, i \,  c^{gauge} \, q^{i}_{j} X^{j^*} \, \right ) ~=~
 \left ( \, \alp \psts i \, -
\, i \, \alg c^{gauge} \, q^{i}_{j} X^{j^*} \, \right )\nonumber\\
\delta \ps i &=& \alA \, \left ( \, i \, M \, q^{i}_{j} X^{j} \,
+ \, i \, c^{gauge} \, q^{i}_{j} \ps j \, \right )\nonumber\\
&=&\left ( \, i \, \alp  M \, q^{i}_{j} X^{j} \,
+ \, i \, \alg \,  c^{gauge} \, q^{i}_{j} \ps j \, \right )\nonumber\\
\delta \pst i &=& \alA \, \left ( \,\o{i}{2} \nabla_{-} X^{i} \, +\,
\eta^{ij^*} \, \partial_{j^*} {W^*} \,
+ \, i \, c^{gauge} \, q^{i}_{j} \pst j \, \right )\nonumber\\
&=&\left ( \,\alp \, \o{i}{2} \nabla_{-} X^{i} \, +\,
\al \, \eta^{ij^*} \, \partial_{j^*} {W^*} \,
+ \, i \,\alg \, c^{gauge} \, q^{i}_{j} \pst j \, \right )\nonumber\\
\delta \pss i &=& \alA \, \left ( -\,\o{i}{2} \nabla_{+} X^{i} \, +\,
\eta^{i^* j} \, \partial_{j} {W} \,
- \, i \, c^{gauge} \, q^{i^*}_{j^*} \pss j \, \right )\nonumber\\
&=&\left ( -\,\al \,\o{i}{2} \nabla_{+} X^{i} \, +\,
\alp \, \eta^{i^* j} \, \partial_{j} {W} \,
- \, i \, \alg \, c^{gauge} \, q^{i^*}_{j^*} \pss j \, \right )\nonumber\\
\delta \psts i &=& \alA \, \left ( \, i \, M \, q^{i^*}_{j^*} X^{j^*} \,
+ \, i \, c^{gauge} \, q^{i^*}_{j^*} \psts j \, \right )\nonumber\\
&=&\left ( \,- i\, \al  \, M \, q^{i^*}_{j^*} X^{j^*} \,
+ \, i \,\alg \, c^{gauge} \, q^{i^*}_{j^*} \psts j \, \right )\nonumber\\
\label{BRS3}
\end{eqnarray}

On the other hand in the B-twisted version of the same N=2 theory, the
BRST-charge is given by:
\begin{equation}
Q^{(A)}_{BRS} ~=~Q^{(gauge)}_{BRS} ~+~Q^{+}~+~{\tilde Q}^{+}
\label{BRS4}
\end{equation}
In view of eq. (\ref{BRS4}) and of our previous discussion of the ghost number,
in the B-twist case, we rename the supersymmetry parameters as follows:
\begin{eqnarray}
\o{1}{2} \left (\varepsilon^{+}\,+ \,{\tilde \varepsilon}^{+}\right )
{}~&=&~\al\nonumber\\
\o{1}{2} \left (\varepsilon^{+}\,-\, {\tilde \varepsilon}^{+}\right )
{}~&=&~\alp\nonumber\\
\alB~&=&~{\tilde \varepsilon}^{+}~=~\varepsilon^{+}~=~\alg
\label{BRS5}
\end{eqnarray}
$\alB$ being the new BRST-parameter and $\al$, $\alp$ the parameters of the
$(0|2)$ fermionic
supergroup relevant to this case. With these notations the BRST-transformations
and $(0|2)$
symmetries of the B-model are the following:
\begin{eqnarray}
\delta \cAp &=&\alB \, \left ( \,-\o{i}{2}  \, \latm \, + \,
\partial_{+} c^{gauge} \, \right ) ~=~-\o{i}{2}\, (\al -\alp) \, \latm \,  +
\,\alg \,
\partial_{+} c^{gauge} \nonumber\\
\delta \cAm &=&\alB \, \left ( \, \o{i}{2}\,  {\tilde  \lambda}^{-}\, + \,
\partial_{-}c^{gauge} \, \right )~=~-\o{i}{2}\, (\al -\alp) \, {\tilde
\lambda}^{-}\, + \, \alg \,
\partial_{-}c^{gauge} \,\nonumber\\
\delta M &=&\alB \o{1}{4} \latm ~=~(\al +\alp) \o{1}{4} \latm \nonumber\\
\delta M^{*} &=&\alB \, \o{1}{4} \,{\tilde  \lambda}^{-} ~=~
(\al -\alp) \o{1}{4} \, {\tilde  \lambda}^{-}\nonumber\\
\delta \latp &=& \alB \,
\Biggl [ \,
\left ( \, \cF \, - \, i \, \cP \, \right )
\, - \, 2i \partial_{+} M^* \,
\Biggr ]
\nonumber\\
&=&
\Biggl [
\,(\al -\alp) \, \left ( \, \cF \, - \, i \, \cP \, \right )
\, - \, 2i \, (\al +\alp) \, \partial_{+} M^* \,
\Biggr ]
\nonumber\\
\delta \latm &=&0\nonumber\\
\delta \lap &=& \alB \,
\Biggl [
\left ( \, \cF \, + \, i \, \cP \, \right )
\, - \, 2i \partial_{-} M \,
\Biggr ]
{}~=~
\Biggl [
\,(\al +\alp) \, \left ( \, \cF \, - \, i \, \cP \, \right )
\, - \, 2i \, (\al -\alp) \, \partial_{-} M \,
\Biggr ]
\nonumber\\
\delta {\tilde  \lambda}^{-}&=& 0\nonumber\\
\delta \cP &=& \alB \, \o{1}{4} \Biggl [ \, - \partial_{+} \, {\tilde
\lambda}^{-}\, +
\, \partial_{-} \latm
\, \Biggr ]~=~\o{1}{4} \Biggl [ \, -\, (\al +\alp) \, \partial_{+} \, {\tilde
\lambda}^{-}\, +
\, (\al -\alp) \, \partial_{-} \latm\, \Biggr ]\nonumber\\
\delta X^{i} &=& \alB \, \Biggl [
\, i \,  c^{gauge} \, q^{i}_{j} X^{j} \, \Biggr ] ~=~
 \Biggl [
\, i \, \alg c^{gauge} \, q^{i}_{j} X^{j} \, \Biggr ]\nonumber\\
\delta X^{i^*} &=&\alB \, \Biggl [ \pss i \, + \, \psts i\, +
\, i \,  c^{gauge} \, q^{i}_{j} X^{j^*} \, \Biggr ] \nonumber\\
&=&\Biggl [ \, (\al + \alp) \pss i \, + \, (\al - \alp) \psts i \, +
\, i \, \alg c^{gauge} \, q^{i}_{j} X^{j^*} \, \Biggr ]\nonumber\\
\delta \ps i &=& \alB \, \Biggl [\, - i\o{1}{2}\nabla_{+}X^{i} \,+\, i \, M \,
q^{i}_{j} X^{j} \,
+ \, i \, c^{gauge} \, q^{i}_{j} \ps j \, \Biggr ]\nonumber\\
&=&\Biggl [ \,- i\,(\al +\alp)\o{1}{2}\nabla_{+}X^{i} \,+\,
 i \,(\al - \alp ) M \, q^{i}_{j} X^{j} \,
+ \, i \, \alg \,  c^{gauge} \, q^{i}_{j} \ps j \, \Biggr ]\nonumber\\
\delta \pst i &=& \alB \, \Biggl [\, - i\o{1}{2}\nabla_{-}X^{i} \,
-\, i \, M^* \, q^{i}_{j} X^{j} \,
+ \, i \, c^{gauge} \, q^{i}_{j} \pst j \, \Biggr ]\nonumber\\
&=&\Biggl [ \,- i\,(\al -\alp)\o{1}{2}\nabla_{-}X^{i} \,+\,
 i \,(\al + \alp ) M \, q^{i}_{j} X^{j} \,
+ \, i \, \alg \,  c^{gauge} \, q^{i}_{j} \ps j \, \Biggr ]\nonumber\\
\delta \pss i &=& \alB \, \Biggl [
\eta^{i^* j} \, \partial_{j} {W} \,
- \, i \, c^{gauge} \, q^{i^*}_{j^*} \pss j \, \Biggr ]\nonumber\\
&=&\Biggl [
(\al -\alp )\, \eta^{i^* j} \, \partial_{j} {W} \,
- \, i \, \alg \, c^{gauge} \, q^{i^*}_{j^*} \psts j \, \Biggr ]\nonumber\\
\delta \psts i &=& \alB \, \Biggl [ \, - \,\eta^{i^* j} \, \partial_{j} {W} \,
+ \, i \, c^{gauge} \, q^{i^*}_{j^*} \psts j \, \Biggr ]\nonumber\\
&=&\Biggl [ \, (\al + \alp) \, \eta^{i^* j} \, \partial_{j} {W} \,
+ \, i \,\alg \, c^{gauge} \, q^{i^*}_{j^*} \psts j \, \Biggr ]\nonumber\\
\label{BRS7}
\end{eqnarray}
To discuss the topological twists of the N=4 matter coupled gauge theory it
might seem necessary
to write down the analogues of eq.s (\ref{BRS5} ) and (\ref{BRS7}) as they
follow from the
rheonomic parametrizations of the N=4 theory (see eq.s (\ref{nfour2}) and
(\ref{nfour8})). Actually
this is not necessary since the N=4 model is just a particular kind of N=2
theory so that
the BRST-transformations relevant to the N=4 case can be obtained with a
suitable specialization
of eq.s (\ref{BRS5} ) and (\ref{BRS7}), according to what shown
in section XIII.
\par
Using eq.s (\ref{degree}) and (\ref{inducedweights}) in the general formulae
(\ref{n2gaugeRsym}) for
the R-symmetries of the N=2 theory we obtain a result that coincides with
the assignments of R-charges given in table II. In this table the charge
assignements were deduced by restricting the non-abelian $U(2)_L \otimes
U(2)_R$
R-symmetry group of the N=4 theory to its abelian subgroup $U(1)_L \otimes
U(1)_R$, generated
by the two third components of the $SU(2)_L \otimes SU(2)_R$ isospin
generators. As
a consequence the twisted spins and twisted ghost numbers displayed in table II
can be alternatively deduced from the N=4 formulae (\ref{nfour25}) or from the
N=2 formulae
(\ref{n2gaugeRsym}), upon use of the special values of $d$ and $\omega_A$ given
in eq. (\ref{degree})
and (\ref{inducedweights}).
\par
In this way we have completly reduced the twisted N=4 models to special
instances of the twisted
N=2 models, the crucial point being the identification of the superpotential
 (\ref{inducedsuperpotential}). From now on we discuss the structure of the
twisted models
in the N=2 language. The next point is the analysis of {\sl STEP 5}: we devote
the next section
to this.
\section{Identification of the topological systems described by the  A and B
models}
In this section we consider the interpretation of the topological
field-theories described by
the A and B models.
\par
{\sl THE A-MODEL}
\par
We begin with the A model. To this effect we start by recalling the structure
of a pure topological Yang-Mills theory \cite{topolgauge}.
In any space-time dimensions the field-content
of this theory is given by table III,
where $ A= A_\mu \, dx^\mu  $ is the gauge-field, $\psi = \psi_\mu \, dx^\mu$
the ghost
of the topological symmetry, $c=c^{gauge}$ the ghost of the ordinary gauge
symmetry and
$\phi$ the ghost for the ghosts (indeed the ghost 1-form $\psi$ is, by itself a
gauge field).
These fields enter the {\it gauge-free} topological BRST-algebra that has the
following form:
\begin{eqnarray}
s \, A &=& - \left ( \, Dc \, + \, \psi \, \right )\nonumber\\
s \, F &=& D\psi \, - \, \left [ \, c \, , \, F \, \right ]\nonumber\\
s \, c &=& \phi \, - \, \um \, \left [ \, c \, , \, c \, \right ]\nonumber\\
s \, \psi &=& D\phi \, - \, \left [ \, c \, , \, \phi\, \right ] \nonumber\\
s \, \phi &=& - \left [ \, c \, , \, \phi \, \right]
\label{topolYM1}
\end{eqnarray}
The above algebra follows from the ghost-form Bianchi identities:
\begin{equation}
{\hat d} \, {\hat F} \, + \, \left [ \, {\hat A} \, , \, {\hat F} \, \right ]
\, = \, 0
\label{topolYM3}
\end{equation}
where
\begin{eqnarray}
{\hat A} &=&A \, + \, c\nonumber\\
{\hat F}&=&{\hat d} {\hat A} \, + \, \um Ê\, \left[ \, {\hat A}  \, ,
\, {\hat A} \, \right ]\nonumber\\
{\hat d}& = & d \, + \, s
\label{topolYM2}
\end{eqnarray}
 by removing the BRST-rheonomic conditions
\begin{equation}
{\hat F}\, = \, F_{ab} \, V^{a} \, \wedge \, V^{b}
\label{BRSTrheonomic}
\end{equation}
that characterize the BRST quantization of the ordinary gauge-theory.
 Indeed if we write the decomposition:
\begin{equation}
{\hat F} ~=~F_{(2,0)} \, + \, F_{(1,1)} \, + \, F_{(0,2)}
\label{topolYM4}
\end{equation}
and we remove the BRST-rheonomic conditions (\ref{BRSTrheonomic}) that imply
$F_{(1,1)} \, =\, F_{(0,2)}\, = \, 0$ we see that  eq.s (\ref{topolYM1}) follow
from eq.
(\ref{topolYM3}) upon use of the identifications:
\begin{eqnarray}
\psi &=& - F_{(1,1)}\nonumber\\
\phi &=& - F_{(0,2)}\nonumber\\
\label{topolYM5}
\end{eqnarray}
The other fields appearing in table III  are either antighosts or auxiliary
fields. Indeed the complete, {\it gauge-fixed} topological BRST-algebra is
obtained by
adjoining to eq.s (\ref{topolYM1}) the following ones:
\begin{eqnarray}
s \, {\bar c} &=& b\nonumber\\
s \, {\bar \psi} &=& T \nonumber\\
s \, {\bar \phi} &=& {\bar \eta} \nonumber\\
\label{topolYM6}
\end{eqnarray}
where ${\bar c}$, ${\bar \psi}$, ${\bar \phi}$  are the antighosts and $T$ and
${\bar \eta}$ are
the auxiliary fields, as displayed in table III. Actually rather than ${\bar
\psi}$
and $T$  it
is more convenient to use, as  antighost and auxiliary field the functionals
${\bar \chi}$ and
$B$ defined by
the following equations
\begin{eqnarray}
{\bar \chi}\, &=& dx^\mu \, \wedge \, dx^{\nu} \, {\bar \chi}_{\mu\nu} \, = \,
D \,
 {\bar \psi}\nonumber\\
B&=&dx^\mu \, \wedge \, dx^{\nu} \, B_{\mu\nu} \, = \, - D \, T \, - \, \left [
\, Dc \, ,
 \, {\bar \psi} \, \right ] \, - \, \left [ \, \psi \, , \, {\bar \psi} \,
\right ]
\label{topolYM7}
\end{eqnarray}
in such a way that:
\begin{equation}
s \, {\bar \chi} ~=~B
\label{topolYM8}
\end{equation}
is an identity.
\par
In the quantum action, $b$ is  the Lagrangian multiplier for the
gauge-fixing of  the  ordinary gauge transformations, while $T_\mu$ (or rather
its
functional $B_{\mu\nu}$) is the Lagrangian multiplier associated with
the gauge-fixing of the topological symmetry. Finally ${\bar \eta}$ is
utilized to gauge fix the gauge invariance of the topological ghost $\psi_\mu$.
Indeed the quantum action has the form:
\begin{equation}
S_{quantum} \, = \, S_{class} \, + \, \int_{M_4} \, s \, \left (  \,
\Psi_{topol} \,
+ \, \Psi_{gauge} \, +\, \Psi_{ghost} \right )
\label{topolYM9}
\end{equation}
where the gauge fermion is the sum of a gauge fermion fixing the topological
symmetry
$(\Psi_{topol}) $, plus one fixing the ordinary gauge symmetry
$(\Psi_{gauge} $), plus a last one fixing the gauge of the ghosts.
In D=4 the classical action is the integral of the first Chern-class
$S_{class} \, = \, \int \, tr \, \left ( \, F \, \wedge \, F \, \right )$,
while in D=2
as classical action one takes the integral of the field strength in the
direction of
the center of the gauge Lie-algebra $S_{class} \, = \, \o {\theta}{2\Pi} \,
\int \, F_{center}$.
 The topological gauge-fixing must break the invariance under
continuous deformations of the connection still preserving ordinary
gauge-invariance.
In four-dimensions a convenient gauge condition that satisfies this requirement
is
provided by enforcing self-duality of the field strength (the instanton
condition).
Setting:
\begin{equation}
{\cal G}_{\mu\nu}^{\pm} ~=~F_{\mu\nu} \,  \pm \,  \um \, \varepsilon_{\mu \nu
\rho\sigma}
 F^{\rho\sigma}
\label{topolYM10}
\end{equation}
the four-dimensional topological gauge-fixing can be chosen to be:
\begin{equation}
{\cal G}_{\mu\nu}^{+} \, = \, 0
\label{topolYM11}
\end{equation}
In two-dimensions, eq.(\ref{topolYM10}) has no meaning and the topological
gauge-fixing
(\ref{topolYM11}) can just be replaced by the condition of constant curvature
(${\cal F}_{center}=const$) where, as already stated, ${\cal F}_{center}$
denotes the field
strenght of
the gauge group restricted to the center of the Lie-algebra, namely the only
components of the
field-strength that, being gauge invariant, can be given a constant value.
This makes sense under the assumption that, within the set of all
gauge-connections characterized
 by the same first Chern-class $\int \, F_{center}$ (value of the classical
action),
there is always at least one that has a constant field strength in the
center-direction and a
vanishing field-strength in the other directions (almost flat connections).
 Hence we set:
 \begin{eqnarray}
 \Psi_{topol} &=& tr \, \left \{  \,{\bar \chi}_{\rho\sigma} \, \left ( \,
{\cal F}_{\mu \nu} \,
 + \, const \,\varepsilon_{\mu\nu} \, \right ) \, g^{\rho\mu} \, g^{\sigma\nu}
\, \right \}
\nonumber\\
 \Psi_{gauge} &=& tr \, \left \{  \, {\bar c} \, \left ( \, {\partial }_{\rho }
\, A_\mu \,
 g^{\rho\mu}
 \,  + \, b \, \right ) \, \right \} \nonumber\\
 \Psi_{ghost} &=& tr \, \left \{  \, {\bar \phi} \, {\partial }_{\rho } \,
\psi_\mu \,
 g^{\rho\mu} \,  \right \}
 \label{topolYM13}
 \end{eqnarray}
fixing the ordinary gauge symmetry of the physical gauge-boson $A_\mu$
and of the topological ghost $\psi_\mu$ by means of the Lorentz gauge.
As we see ${\bar c}$ is the antighost of ordinary gauge-symmetry, while ${\bar
\phi}$ is the
antighost for the gauge symmetry of the topological ghost $\psi$.
\par
In the case the topological Yang-Mills theory is coupled to some topological
matter
system, the gauge-fixing of the topological gauge-symmetry can be achieved by
imposing that
the field-strength ${\cal F} \, = \, F_{+-}^{(center)}$
be equal to some appropriate function of the matter fields:
\begin{equation}
{\cal F} ~=~2i\,\cP (X)
\label{topolYM14}
\end{equation}
In this case we can also suppress the auxiliary field $B$ and replace the
antighost part
of the BRST-algebra with the equations:
\begin{eqnarray}
s \, {\bar c} &=& b\nonumber\\
s \, {\bar \chi}_{+-} &=& \left ( \, \o{1}{2} \,{\cal F}\, - \, i\, \cP  \right
)\nonumber\\
s \, {\bar \phi} &=& {\bar \eta}
\label{topolYM15}
\end{eqnarray}
that substitute eq.s (\ref{topolYM6}).
Correspondingly, the gauge-fermion $\Psi_{topol}$ of eq. (\ref{topolYM13}) can
be replaced with:
\begin{equation}
\Psi_{topol}~=~2 {\bar \chi}_{+-} \, \left ( \, \o{1}{2} \, {\cal F} \, + \, i
\, \cP \, \right )
\label{topolYM16}
\end{equation}
 \par
It is now worth noting that, for consistency with the BRST algebra
(\ref{topolYM1}), if we define
the 2-form $\Theta^{(2)} \, = \,2i \cP (X,X^*) \, e^+ \, \wedge \, e^-$, we
must have
$s \, \Theta^{(2)} \, = \, d \psi_{(center)}$. Indeed, by restriction to the
center of
the Lie-algebra we obtain an abelian topological gauge theory, for which $s \,
F \, = \, d\psi$.
Reconsidering the supersymmetry transformation rules of the gauge multiplet
(\ref{ntwo6}) and the rules of A-twisting, we realize that the property
required for the function
$\cP (X,X^*)$ is satisfied by the auxiliary field $\cP$ of the gauge multiplet,
provided
we identify $\psi = \o{i}{2} \, \left ( {\tilde  \lambda}^{-}\, e^{-} \, + \,
\latp \, e^{+} \,
\right )$. This is correct since, by:
looking at table I  and at eq.s (\ref{BRS3}) we recognize that a subset of the
fields does indeed describe a topological Yang-Mills theory upon the
identifications:
\begin{eqnarray}
\psi &=& \o{i}{2} \, \left ( {\tilde  \lambda}^{-}\, e^{-} \, + \, \latp \,
e^{+} \,
\right )\nonumber\\
\phi &=& M \nonumber\\
{\bar \phi}&=& M^{*}\nonumber\\
{\bar \eta} &=&\o 12 \, \left ( \lap \, + \, \latp \, \right )\nonumber\\
{\bar \chi}_{+-} &=& \o 12 \, \left ( \lap \, - \, \latm\right )
\label{topolYM15bis}
\end{eqnarray}
We also see that the descent equations:
\begin{eqnarray}
s \Theta^{(2)} &=& d \Theta^{(1)}\nonumber\\
s \Theta^{(1)} &=& d \Theta^{(0)}\nonumber\\
s \Theta^{(0)} &=&0
\label{descent}
\end{eqnarray}
are solved by the position:
\begin{eqnarray}
\Theta^{(2)} & = & 2i \cP \, e^+ \, \wedge \, e^- \,
= \, 2i \cP (X,X^*) \, e^+ \, \wedge \, e^-\nonumber\\
\Theta^{(1)} & = &\psi \, = \, \o{i}{2} \, \left ( {\tilde  \lambda}^{-}\,
e^{-} \, +
 \, \latp \, e^{+} \,
\right )\nonumber\\
\Theta^{(0)} &=& \phi \, = \, M
\label{Aobservable}
\end{eqnarray}
so that the quantum action of the topological gauge-theory (\ref{topolYM9}) can
be topologically
deformed by:
\begin{equation}
S_{quantum} \, \longrightarrow \, S_{quantum} \, - i \, r \, \int \,
\Theta^{(2)}
\label{Adeformation}
\end{equation}
Altogether we see that the classical action
 $S_{class} \, = \, \o {\theta}{2\Pi} \, \int \, F_{center}$, plus the
topological deformation
$- i \, r \, \int \, \Theta^{(2)}$ constitute the Fayet-Iliopoulos term, while
the
remaining terms in the action (\ref{ntwo9}) are BRST-exact and come from the
gauge-fixings:
\begin{equation}
s \, \int \, \left [ \, {\bar \chi}_{+-} \, \left ( \o {\cal F}{2} \, - \, i\cP
\, \right ) \, + \,
{\bar \phi} \, \left ( \partial_{+} \psi_{-} \, + \, \partial_{-}
\psi_{+}\right ) \, \right ]
\label{exactpartaction}
\end{equation}
\par
On the other hand the matter multiplets with their fermions span a topological
$\sigma$-model
\cite{topolsigma}
coupled to the topological gauge-system. The topological symmetry, in this
case, is the
possibility of deforming the embedding functions $X^{i}(z,\bz)$ in an arbitrary
way. Correspondingly,
in the absence of gauge couplings, the {\it gauge-free} topological
BRST-algebra is very simple:
\begin{eqnarray}
s \, X^{i} &=& c^{i}\nonumber\\
s \, X^{i^*} &=& c^{i^*}\nonumber\\
s \, c^{i} &=& 0\nonumber\\
s \, c^{i^*} &=& 0
\label{topolYM16bis}
\end{eqnarray}
$c^{i}$ and $c^{i^*}$ being the ghost of the deformation-symmetry.
In the presence of a coupling to a topological gauge-theory, defined by the
covariant derivative:
\begin{equation}
\nabla \, X^{i} \, = \, d \, X^{i} \, - i \, {\cal A} \, q^{i}_{j} \, X^{j}
\label{topolYM17}
\end{equation}
the {\it gauge-free} BRST-algebra of the matter system becomes:
\begin{eqnarray}
s \, X^{i} &=& c^{i} \, - i \, c^{g}\, q^{i}_{j} \, X^{j}\nonumber\\
s \, X^{i^*} &=& c^{i^*} \, + i \, c^{g}\,  q^{i}_{j} \, X^{j^*}\nonumber\\
s \, c^{i} &=& i\, q^{i}_{j} \, \left ( \, c^j \, c^g \, + \, X^{j} \, \phi \,
\right ) \nonumber\\
s \, c^{i^*} &=&i\, q^{i}_{j} \, \left ( \, c^{j^*} \, c^g \, +
 \, X^{j^*} \, \phi \, \right )
\label{topolYM16tris}
\end{eqnarray}
the last two of eq.s (\ref{topolYM16tris}) being uniquely fixed by the
nilpotency $s^2=0$ of the
Slavnov operator.
\par
Comparing with eq.s (\ref{BRS3}) we see that, indeed, eq.s (\ref{topolYM16})
are reproduced if
we make the following idenfications:
\begin{eqnarray}
c^{i}&=&\psi^{i}\nonumber\\
c^{i^*}&=&{\tilde \psi}^{i^*}
\label{topolYM17tris}
\end{eqnarray}
The remaining two fermions are to be identified with the antighosts:
\begin{eqnarray}
{\bar c}^{i}&=&{\tilde \psi}^{i}\nonumber\\
{\bar c}^{i^*}&=&\psi^{i^*}
\label{topolYM18}
\end{eqnarray}
and their BRST-variation, following from eq.s (\ref{BRS3}) yields the
topological gauge-fixing
of the matter sector:
\begin{eqnarray}
s \,{\bar c}^{i}&=&i \, q^{i}_j \, {\bar c}^{j} \, c^g \, +Ê\,
 \eta^{ij^*} \, \partial_{j^*} \, W^* \, + \, \o{i}{2} \, \nabla_{-}
X^{i}\nonumber\\
s \,{\bar c}^{i^*}&=&-i \, q^{i}_j \, {\bar c}^{j^*} \, c^g \, +Ê\,
 \eta^{i^* j} \, \partial_{j^*} \, W^* \, - \, \o{i}{2} \, \nabla_{+} X^{i}
\label{topolYM18bis}
\end{eqnarray}
Following Witten \cite{Wittenmirror} and \cite{Wittenphases} we easily recover
the interpretation
of the "{\it instantons}" encoded in the topological gauge-fixings dictated by
 eq.s (\ref{topolYM18bis}) and (\ref{topolYM15}). Indeed we just recall that
the functional
integral is concentrated on those configurations that are a fixed point of the
$(0|2)$
supergroup transformations. Looking at eq.s (\ref{BRS3}) we see that such
configurations
have all the ghosts and antighosts equal to zero while the bosonic fields
satisfy the following
conditions:
\begin{eqnarray}
\eta^{ij^*}\, \partial_{j^*} \, W^{*}(X^*) &=&0\nonumber\\
\eta^{i^* j}\, \partial_{j} \, W(X) &=&0\nonumber\\
\nabla_{-} \, X^{i} &=&0\nonumber\\
\nabla_{+} \, X^{i^*} &=&0\nonumber\\
{\cal F}&=& 2i\cP \, = \, -i \biggl[{\cal D}^{\bf X}(X,X^*) \, - \, r
\biggr]
\label{topolYM19}
\end{eqnarray}
where ${\cal D}^{\bf X}\left ( X,X^*\right ) \,= \,
\sum_i q^i |X^i|^2$ is the momentum map function defined in section V.
Hence the instantons are holomorphic maps from the world-sheet to a locus in
${\bf C}^n$
characterized by the equations $\eta^{i^* j}\, \partial_{j} \, W(X) =0$. In the
case
chosen by Witten and reviewed in section IX where $W(X) \, =\, P \, {\cal
W}(S^{i})$, this locus
is the hypersurface ${\cal W}(S^{i})=0$, ($P=0$) in a weighted projective space
$WCP_{n-2}$,
the weights of the homogeneous coordinates $S^{i}$ being their charges. In
other words the
instantons are holomorphic solutions of the corresponding $N=2$ $\sigma$-model.
The value
of the action on these {\it instantons} has been calculated by Witten and the
results is
easily retrieved in our notations. Indeed the Lagrangian (\ref{ntwo15})
restricted to the bosonic fields of zero ghost-number is given by:
\begin{eqnarray}
\cL_{noghost}&=& \o 12 \cF^2 + 2\cP^2 +\o{\theta}{2\pi}\cF\nonumber\\
& -&\bigl(\delp X^{i^*}\delm X^i + \delm X^{i^*}\delp X^i
\bigr) \, + \, 2 \, \cP \, {\cal D}(X,X^*) \, - \, 2 \, r \, \cP
\label{topolYM20}
\end{eqnarray}
Using eq.s (\ref{topolYM19}) and $\left [ \, \nabla_{-} \, , \, \nabla_{+} \,
\right ] \, X^{i} \,=
\, i \,{\cal F} \, q^{i}_{j} \, X^{j}$, we obtain:
\begin{equation}
\int \, \cL_{noghost} \, =\,\left ( \o {\theta}{2\pi} \, + \, i\,r \, \right )
\,
\int \,{\cal F} \,=\, 2\pi i \, t \, N
\label{topolYM21}
\end{equation}
where $N$ is the winding number and the parameter $t$ was defined in eq.
(\ref{ntwo8tris}).
\par
\par
{\sl THE B-MODEL}
\par
In order to identify the system described by the B-model we discuss the
structure of
a topological Landau-Ginzburg theory \cite{topolLGliterature}
 coupled to an ordinary abelian gauge theory. To this effect we
begin with the structure of a {\it topological rigid Landau-Ginzburg theory}.
The rigid
Landau-Ginzburg model was defined in section VII and it is described by the
action
(\ref{rigidLGaction}). It has the R-symmetries (\ref{rigidLGRsym}) and it is
N=2 supersymmetric
under the transformations following from the rheonomic parametrizations
(\ref{rigidLGparam}).
The rigid topological Landau-Ginzburg model has the same action
(\ref{rigidLGaction}),
but the spin of the fields is changed, namely it is that obtained by
B-twisting:
the scalar fields $X^{i}$
and $X^{i^*}$ mantain spin-zero as in the ordinary model, while the spin 1/2
fermions acquire
either spin zero or spin $\pm 1$. Specifically $\pss i$, $\psts i$ have both
spin zero,
while $\ps i$ and $\pst i$ have spin $s=1$ and $s=-1$, respectively. In view of
this fact it
is convenient to introduce the new variables:
\begin{eqnarray}
C^{i^*} &=& \pss i \, + \psts i\nonumber\\
{\bar C}^{i} &=& \left ( \,{\bar C}^{i}_{+}  \, e^+ \, + \,{\bar C}^{i}_{-} \,
e^- \, \right )
\nonumber\\
             &=& \left ( \, \ps i \, e^+ \, + \, \pst i \, e^- \, \right
)\nonumber\\
\theta^{i^*}&=&\pss i \, - \psts i
\label{topolLG1}
\end{eqnarray}
and rewrite the action (\ref{rigidLGaction}) in the form:
\begin{eqnarray}
\cL_{(topolLG)}&=& - \, \left ( \, \partial_{+} X^{i^*}\, \partial_{-} X^{i} \,
+
\, \partial_{-} X^{i^*}\, \partial_{+} X^{i} \, \right )\nonumber\\
&+&2i \, \left ( \, {\bar C}^{i}_{+} \, \partial_{-}C^{i^*} \, + \,
{\bar C}^{i}_{-} \, \partial_{+}C^{i^*} \, \right )\nonumber\\
&+&2i \, \left ( \, {\bar C}^{i}_{+} \, \partial_{-}\theta^{i^*} \, - \,
{\bar C}^{i}_{-} \, \partial_{+}\theta^{i^*} \, \right )\nonumber\\
&+&8 \, {\bar C}^{i}_{+} \,  {\bar C}^{i}_{-} \, \partial_{i}\partial_{j} {\cal
W}\,
+ \, 4 \, C^{i^*} \, \theta^{j^*} \, \partial_{i^*} \partial_{j^*}
 {\bar {\cal W}}\nonumber\\
&+& 8 \,\partial_{i}{\cal W} \, \partial_{i^*} {\bar {\cal W}}
\label{topolLG2}
\end{eqnarray}
If we denote by $\left [ \Omega \right ]_{*}=\Omega_{+} e^+ -\Omega_{-} e^-$
the Hodge-dual of the 1-form $\Omega=\Omega_{+} e^+ +\Omega_{-} e^-$, then the
action
(\ref{topolLG2}) can be rewritten in the following more condensed form:
\begin{eqnarray}
S_{topolLG}&=&\int \,  \cL_{(topolLG)} \, e^+ \wedge e^- \nonumber\\
&=&\int \, \Bigg  \{ dX^{i} \, \wedge \, \left [dX^{i^*}\right ]_{*} \, + \,
2i \, {\bar C}^{i} \, \wedge \, \left [dC^{i^*}\right ]_{*} \nonumber\\
&+&4 \,{\bar C}^{i} \, \wedge \, {\bar C}^{j} \, \partial_{i}\partial_{j} {\cal
W}\,
+ \, 2i \, d{\bar C}^{i}  \, \theta^{i^*}\nonumber\\
&+&4 \left \{ \, C^{i^*} \, \theta^{j^*} \, \partial_{i^*}\partial_{j^*}
{\bar {\cal W}}\, +\, 2 \,
  \partial_{i}{\cal W} \, \partial_{i^*} {\bar {\cal W}} \right \} \, e^+
\wedge e^-
\Bigg \}
\label{topolLG3}
\end{eqnarray}
and it is closed under the following BRST-transformations:
\begin{eqnarray}
s \, X^{i} &=& 0\nonumber\\
s \, X^{i^*}&=& C^{i^*}\nonumber\\
s \, C^{i^*}&=&0\nonumber\\
s \, \theta^{i^*}&=& 2 \eta^{i^* j} \, \partial_{j} {\cal W}\nonumber\\
s \, {\bar C}^{i}&=& - \o i2 \, dX^{i}
\label{topolLG4}
\end{eqnarray}
where we have assigned ghost number $\# gh=0$ to
{\it the physical fields} $X^{i}$ and $X^{i^*}$, $\# gh=1$ to {\it the ghost}
$C^{i^*}$ and $\# gh=-1$ to  {\it  the antighost}, namely the 0-form
$\theta^{i^*}$ and the 1-form  ${\bar C}^{i}$.
In this way the {\it gauge-free BRST-algebra} is given by the first three of
eq.s (\ref{topolLG4}):
it quantizes a symmetry which corresponds to a deformation of the complex
structure of the target
coordinates $X^{i} \, , X^{i^*}$. The variation of the antighosts  defines the
gauge-fixings:
\begin{eqnarray}
\partial_{i} {\cal W} \left ( X\right ) &=&0\nonumber\\
dX^{i}&=&0
\label{topolLG5}
\end{eqnarray}
that select, as ``{\it instantons}", the constant maps ($dX^{i}=0$) from the
world-sheet to
the critical points ($\partial_{i} {\cal W} \left ( X_0\right )=0$) of the
superpotential
${\cal W}$. The action (\ref{topolLG3}) is the sum of a BRST non-trivial part:
\begin{equation}
\Omega_{(-2)}\left [ W \right ]~=~\int \,  \left [ \,4 \,{\bar C}^{i} \, \wedge
\, {\bar C}^{j} \,
\partial_{i}\partial_{j} {\cal W}\,+ \,    2i \,{\bar C}^{i} \, \wedge \,
d\theta^{i^*}\,
 \right ]
\label{topolLG6}
\end{equation}
that is closed ($s\Omega_{(-2)} \, =\,0$), but not exact ($\Omega_{(-2)} \ne s
({something})$)
and has ghost-number $\# gh =-2$, plus two BRST exact terms:
\begin{eqnarray}
K^{(Kin)}_{(0)}&=&\int \, \left \{ dX^{i} \, \wedge \, \left [dX^{i^*}\right
]_{*} \, + \,
2i \,{\bar C}^{i} \, \wedge \, \left [dC^{i^*}\right ]_{*} \, \right
\}\nonumber\\
 &=&s \, \int \, \Psi^{(Kin)}~=~
 s \,\int  \, 2i\,  {\bar C}^{i} \,\wedge \,  \left [dX^{i^*}\right ]_{*} \,
 \nonumber\\
K^{(W)}_{(0)}&=&\int \, 4 \left \{ \, C^{i^*} \, \theta^{j^*} \, \partial_{i^*}
\partial_{j^*}
{\bar {\cal W}}\, +\, 2 \,
  \partial_{i}{\cal W} \, \partial_{i^*} {\bar {\cal W}} \right \} \, e^+
\wedge e^- \nonumber\\
&=&s \, \int \, \Psi^{(W)}~=~s \, \int \,4 \, \partial_{j^*} {\bar {\cal W}} \,
\theta^{j^*}
 \,  e^+ \wedge e^-
\label{topolLG7}
\end{eqnarray}
that have ghost-number $\# gh=0$ and
correspond to the BRST-variation of the gauge-fermions associated with the two
gauge-fixings
(\ref{topolLG5}). As already pointed out the {\it rigid topological
Landau-Ginzburg model} has
been extensively studied in the literature. Here we are interested
in the case where the {\it topological Landau-Ginzburg model} is
{\it coupled to an  ordinary abelian gauge
theory}. Under this circonstance the BRST-algebra (\ref{topolLG4}) is replaced
by:
\begin{eqnarray}
s \,{\cal A}^{'} &=& dc^{g}\nonumber\\
s\, F^{'}&=&0\nonumber\\
s \, c^{g}&=&0\nonumber\\
s \, X^{i}&=& i \, c^{g} \, q^{i}_{j} \, X^{j}\nonumber\\
s \, X^{i^* }&=&C^{i^*} \, - \,  i \, c^{g} \, q^{i}_{j} \, X^{j^*}\nonumber\\
s \, C^{i^*}&=&i \, c^{g} \, q^{i}_{j} \, C^{j^*}\nonumber\\
s \, \theta^{i^*}&=& 2 \eta^{i^* j} \, \partial_{j} W \, - \,
 i \, c^{g} \, q^{i}_{j} \,\theta^{j^*}\nonumber\\
s \, {\bar C}^{i}&=& - \o i2 \, \nabla X^{i} \, +
\, i \, c^{g} \, q^{i}_{j} \,{\bar C}^{j}
\label{topolLG8}
\end{eqnarray}
where $\nabla (...)^{i}\, = \, d(...)^{i}\, + \, i \, A^{'} \, q^{i}_{j} \,
(...)^{j}$
denotes the gauge
covariant derivative and the superpotential ${\cal W}(X)$
of the rigid theory has been replaced by $W(X)$, namely the superpotential of
the
gauged-coupled model. The action (\ref{topolLG3}) is also replaced by a similar
expression where
the ordinary derivatives are converted into covariant derivatives.
\par
The topological system emerging from the B-twist of the N=2 model discussed in
the present article
is precisely a Landau-Ginzburg model of this type: in particular, differently
from the case of
the A-twist, there is no {\it topological gauge theory}, rather an ordinary
gauge theory plus a
topological massive vector. The identification is better discussed at the level
of the BRST-algebra
comparing eq.s (\ref{topolLG8}) with eq.s (\ref{BRS7}) after setting:
\begin{eqnarray}
{\cal A}^{'} &=&\left [ \, {\cal A}_{+} \, + \, 2i \, M\, \right ] \, e^+ \,
+Ê\,
 \left [ {\cal A}_{-} \, - \, 2i \, M^{*} \right ] \, e^- \nonumber\\
{\cal B}&=&M \, e^+ \, + \, M^* \, e^-\nonumber\\
\psi^{(mass)}&=& \o 14 \left ( \lambda^{-} \, e^{+} \,
+ \, {\tilde \lambda}^{-} \, e^{-} \right )\nonumber\\
{\bar \chi}&=& \lambda^{+} \, + \, {\tilde \lambda}^{+}\nonumber\\
{\bar \chi}^{(mass)}&=&\o i2 \, \left [ \,  \lambda^{+} \, -
\, {\tilde \lambda}^{+} \, \right ]\nonumber\\
C^{i^*}&=&\pss i \, + \,  \psts i\nonumber\\
\theta^{i}&=&\pss i \, - \,  \psts i\nonumber\\
{\bar C}^{i}&=&\ps i \, e^+ \, + \, \pst i \, e^-
\label{topolLG9}
\end{eqnarray}
With these definitions the BRST-transformations of eq.s (\ref{BRS7}) become
indeed identical with
those of eq.s (\ref{topolLG8}) plus the following ones:
\begin{eqnarray}
s \, {\cal B} &=& \psi^{(mass)}\nonumber\\
s \,  \psi^{(mass)}&=&0\nonumber\\
s \, {\bar \chi}^{(mass)}&=& \cP (X,X^*) \,  +  \,
\left ( \partial_{+}{\cal B}_{-} \, - \, \partial_{-}{\cal B}_{+} \right
)\nonumber\\
s \, {\bar \chi}&=&{\cal F}^{\, '}
\label{topolLG10}
\end{eqnarray}
The first two  of eq.s (\ref{topolLG10}) correspond to the {\it gauge-free}
BRST-algebra of the topological
massive vector, $\psi^{(mass)}$ being the 1-form ghost associated with the
continuous deformation
symmetry of the vector ${\cal B}$. The second two of eq.s (\ref{topolLG10}) are
BRST-transformations
of antighosts and the left hand side defines the gauge-fixings of the massive
vector and gauge
vctor, respectively, namely:
\begin{eqnarray}
\cP (X,X^*) \,  +  \,
\left ( \partial_{+}{\cal B}_{-} \, - \, \partial_{-}{\cal B}_{+} \right
)&=&0\nonumber\\
{\cal F}^{\, '}\, = \, \partial_{+} {\cal A}_{-} \, - \, \partial_{+}{\cal
A}_{+}&=&0
\label{topolLG11}
\end{eqnarray}
 Actually, looking at eq.s (\ref{BRS7}) we realize that the configurations
corresponding to
a {\it fixed point} of the $(0|2)$ supergroup are characterized by all the
fermions ( = ghosts
+ antighosts) equal to zero and by:
\begin{eqnarray}
M=M^{*}&=&0 ~~\Longrightarrow~~{\cal B}~= ~0\nonumber\\
{\cal F}&=&0~~\Longrightarrow~~{\cal F}^{\, '}~=~0\nonumber\\
\cP(X,X^*)&=&0\nonumber\\
 \eta^{i^* j} \, \partial_{j} W(X)&=&0\nonumber\\
dX^{i}&=&0
\label{topolLG12}
\end{eqnarray}
Hence in the B-twist the functional integral is concentrated on the constant
maps from the
world-sheet to the extrema of the classical scalar potential (\ref{ntwo18}). As
we have seen, in the
A-twist the functional integral was concentrated on the holomorphic maps to
such extrema:
furthermore, in the A-twist the classical extrema were somewhat modified by the
winding number effect
since the equation $\cP \, = \, 0$ was replaced by $\cP \, = \, - \o i2 \,
{\cal F}$. In the
B-twist no {\it instantonic} effects modifies the definition of classical
extremum.
The extrema of the scalar potential can be a point (Landau-Ginzburg phase) or a
manifold
($\sigma$-model phase). The B-twist selects the constant maps in either case,
and the
A-twist selects the holomorphic maps in either case.  However, in the
Landau-Ginzburg
phase the holomorphic maps  to a point are the same thing as the constant maps,
so that,
in this phase the instantons of the A-model coincide with those of the B-model.
\par
 In the case of those N=2 theories that are actually N=4 theories, there is
only the
$\sigma$-model phase, as we have already pointed out, and the above coincidence
does
not occur.
\par

\section{Topological Observables of the A and B models and HyperK\"ahler
quotients}
Having identified the topological theories produced by the A and B twists, let
us
consider their meaning in relation with the HyperK\"ahler quotient
construction.
\par
We recall that in any 2D topological field-theory the key objects are
 the solutions of the descent equations
(\ref{descent}). Indeed they provide the means to deform the topological action
according
to the generalization of eq. (\ref{Adeformation}) :
\begin{equation}
S_{quantum} \, \longrightarrow \, S_{quantum} \, + \, \sum_A \, t_A \, \int \,
\Theta^{(2)}_A
\label{topoldeformation}
\end{equation}
$ \Theta^{(2)}_A$ being a complete base of solutions to eq. (\ref{descent}),
and to study
the deformed correlation functions:
\begin{equation}
c_{A_1,A_2,...A_N}(t) \, = \, < \, \Theta^{(0)}_{A_1},....,\,
\Theta^{(0)}_{A_N} \,
\exp\left [\, \sum_A \, t_A \, \int \, \Theta^{(2)}_A \, \right ] \, >
\label{correlators}
\end{equation}
In the case of the A-twist we have seen that a solution of the descent eq.
(\ref{descent}) is
associated with each abelian factor of the gauge group and it is given by
eq.(\ref{Aobservable}).
A set
of topological deformations of the action are therefore proportional, in the
A-model, to the
r-parameters of the N=2 Fayet-Iliopoulos terms. In the N=4 case where the
effective
$\sigma$-model target space ${\cal M}_{target}$, namely
the locus of the scalar potential extrema
\begin{equation}
M=0 ~; ~{\cal D}^{3}\, \left ( X \, X^* \right ) ~=~r~;~\partial_{i} W (X)~=~0
\, \longrightarrow \, \cases {N~=~0 \cr {\cal D}^{+}(u,v)~= \, s\cr}
\label{extremum}
\end{equation}
is equal to the HyperK\"ahler quotient ${\cal D}^{-1}(\zeta)/ G$ of flat space
with respect
the triholomorphic action of
the gauge group $G$, the topological observables of the
A-model associated with the $r$-parameters correspond to the K\"ahler structure
deformations of
of ${\cal M}_{target}$. To see this it suffices to recall the way the K\"ahler
potential
of the quotient manifold ${\cal D}^{-1}(\zeta)/G$ is determined (see eq. s
(\ref{kq4} ),  (\ref{kq2bis}).
Let $K \,=\,\o 12\sum_i \,( {u}^{i^*} \, u^{i} \, + \, {v}^{i^*}
\, v^{i})$ be the
K\"ahler potential of flat space and ${\cal D}^{3}(u,{\bar u},v,{\bar v})$ be
the real,
non-holomorphic
part of the momentum map. By definition both $K$ and ${\cal D}^{3}$ are
invariant under the
action of the isometry group $G$ but not under the action of its
complexification
$G^{c}$. On the other hand the holomorphic part ${\cal D}^{+}(u,v)$
of the momentum-map
is invariant not only under $G$, but also under $G^{c}$: furthermore one shows
that the quotient
manifold ${\cal D}^{-1}(\zeta)/G$ is the same thing as the quotient
${\cal H}\, = \, \o {D^{+}(u,v)=s}{G}$ of the holomorphic
hypersurface $D^{+}(u,v)=s$ ($s=\zeta^1 \, + \, i \zeta^2$, being the complex
level parameters)
modded by the action
of $G^{c}$. Naming $e^V \, \in \, G^{c}$ an element of
this complexified group
eq. (\ref{kq2bis}) specializes in our case to the following equation:
\begin{equation}
{\cal D}^3\, \left ( \,  e^{- V} \{ u,v \}  \, \right ) ~=
{}~r
\label{complessificatio}
\end{equation}
and it is true equation on the hypersurface ${\cal H}\, = \,
\o {D^{+}(u,v)=s}{G^c}$.
Then  in agreement with eq. (\ref{kq4}) the K\"ahler potential
of the HyperK\"ahler quotient manifold ${\cal D}^{-1}(\zeta)/G$ is :
\begin{equation}
{\hat K}~= ~K |_{\cal H} \, \left ( \,  e^{- V} \{ u,v \}  \, \right ) ~
+~ r \, V
\label{Kpotential}
\end{equation}
Consequently a variation of the $r$ parameters uniquely affects the K\"ahler
potential,
the quotient ${\cal H} \, = \, \o {D^{+}(u,v)=s}{G^c}$, as an analytic
manifold,
being insensitive
to such a variation. Summarizing the A-model is a cohomological theory in the
moduli space
of K\"ahler structure deformations.\par
 Before addressing the structure of observables in the
B-model it is also worth discussing the general form of observables in any
topological
theory described by the BRST-algebra (\ref{topolYM16tris}) and coupled to a
topological
gauge-theory (\ref{topolYM1}). In a topological $\sigma$-model the observables,
namely
the solutions of the descent equations (\ref{descent}) are in correspondence
with the
cohomology classes of the target-manifold. If
$\omega^{(n)}\, =\, \omega_{i_1 , ..., i_n}(X) \, dX^{i_1} \, \wedge \,
.....\, dX^{i_n}$ is a closed n-form $d \omega^{(n)}=0$, we promote it to a
ghost-form
${\hat \omega}^{(n)}$ by substituting $d\,  \longrightarrow \, d \, + \, s$,
$dX^{i}\,  \longrightarrow \, dX^{i} \, + \, c^{i}$ then by expanding this
ghost-form into addends of
definite ghost number ${\hat \omega}^{(n)}\, = \, \sum_{g=0}^{n} \, {\hat
\omega}^{(n-g)}_{(g)}$
we solve the descent equations by setting $\Theta^{(2)}\, = \,{\hat
\omega}^{(2)}_{(n-2)}$,
$\Theta^{(1)}\, = \,{\hat \omega}^{(1)}_{(n-1)}$, $\Theta^{(0)}\, = \,{\hat
\omega}^{(0)}_{(n)}$.
(Note that in this discussion, for simplicity, we do not distinguish
holomorphic and
antiholomorphic indices). In a similar way in the topological model described
by eq.s
 (\ref{topolYM16tris}), (\ref{topolYM1}), the solutions of the descent
equations are in
correspondence with the antisymmetric constant tensors $a_{i_1,....,i_n}$ that
are
invariant under the action of the gauge-group, namely that satisfy the
condition:
\begin{equation}
a_{p,[i_2,....,i_n} \, q^{p}_{i_1]} ~=~0
\label{Ainvarianttensor}
\end{equation}
Indeed, setting
\begin{eqnarray}
{\hat \nabla}\,&=&{\hat d} \,- i \, q \,  {\hat {\cal A}}\nonumber\\
&=&(d \, +  \, s ) \, - i \, q\, (A \, + \, c^g)\nonumber\\
&=&\nabla_{(1,0)} \, + \, \nabla_{(0,1)}\nonumber\\
&=&(d \, - \,i \, {\cal A} \, q ) \, +\,( s \, - i \, c^g \, q)
\label{hatcovariant}
\end{eqnarray}
we obtain ${\hat \nabla}^{2} ~= ~-i\,  {\hat F}\, q \,$ and to every invariant
antisymmetric
tensor we can associate the ${\hat d}$-closed ghost-form
${\hat \omega}~=~a_{i_1,....,i_n} {\hat \nabla}X^{i_1}\,......\,{\hat
\nabla}X^{i_n}$. Expanding
it in definite ghost-number parts, the solution of the descent equations is
obtained in the
same way as in the $\sigma$-model case.
\par
Let us now turn our attention to the B-model, which describes a topological
gauge-coupled
Landau-Ginzburg theory. Here the topological observables are in correspondence
with the
symmetric invariant tensors, rather than with the antisymmetric ones. To see it
we
recall the solutions of the descent equations in the case of the topological
rigid
Landau-Ginzburg model: in this case the topological observables are in
correspondence
with the elements of the local polynomial ring of the superpotential
${\cal R}_{\cal W}~=~\o{{\bf C}(X)}{\partial {\cal W}}$. Indeed, let
$P(X)\, \in  \,{\cal R}_{\cal W}$
be some non trivial polynomial of the local ring, a solution of the descent
equations
(\ref{descent}) is obtained by setting:
\begin{eqnarray}
\Theta_{P}^{ } &=& P(X) \nonumber\\
\Theta_{P}^{(1)} &=&  -2 \,{\rm i} \pa_i \,P
\, {\bar C}^{i}\nonumber\\
\Theta_{P}^{(2)} &=&  - 2 \,\pa_i \pa_j P\,
{\bar C}^{i} \wedge {\bar C}^{j}\,- \,4 \, \left [  \pa_k P \, \pa_{l^{*}}
{\bar W} \,
\eta^{kl^{*}}
\right ] e^{+}
\wedge  e^{-}
\label{Bobservables}
\end{eqnarray}
The reason why $P(X)$ has to be a non trivial element of the local ring is
simple.
If $P(X)$ were proportional to the vanishing
relations ({\it i.e.} if $P(X)=\sum_ip^{i} (X) \o {\pa W}{\pa X_i}$), then
using the BRST transformations (\ref{topolLG4}),
 one could  see that $P(X) = s \, K $ and
so $\Theta_{P}^{ }$ would be exact. (For the proof
it suffices to set $K = p^{i}(X)
\o{1}{2} \theta^{j^*} \eta_{ij^{*}}\ .$)
In our case where the Landau-Ginzburg theory is gauged-coupled and the
BRST-transformations
are given by eq.s (\ref{topolLG8}), the solution of the descent equations has
the same form
as in eq.(\ref{Bobservables}), provided the polynomial $P(X)$ has the form
\begin{equation}
P(X) ~=~s_{i_1,...,i_n}\, X^{i_1}...... X^{i_n}
\label{Stensor}
\end{equation}
the symmetric tensor $s_{i_1,...,i_n}$ being gauge invariant:
\begin{equation}
s_{p,\{i_2,....,i_n} \, q^{p}_{i_1\}} ~=~0
\label{Binvarianttensor}
\end{equation}
and such that $P(X)$ is a non-trivial element of the the ring ${\cal R}_W$.
Consider now the case of N=4 theories, where the
superpotential is given by eq. (\ref{inducedsuperpotential}), and consider the
polynomial:
\begin{equation}
P_s(X^{A})~=~{\rm cost} ~ n
\label{npolynomial}
\end{equation}
which is gauge-invariant ($n$ is neutral under the gauge-group) and non-trivial
with respect
to the vanishing relations $\o {\partial}{\partial X^{A}} \, W(X^{A})\approx
0$. The corresponding
two-form is easily calculated from eq.s (\ref{Bobservables}). We obtain:
\begin{equation}
\Theta^{(2)}_{P_s}~=~2\, cost \, \left ( \, s^* \, - \, i \, {\cal D}^{-}\left
( {\bar u}\, , \,
{\bar v} \, \right ) \, \right ) \, e^+ \, \wedge \, e^-
\label{Qtwoform}
\end{equation}
Hence a topological deformation of the action is given by :
\begin{equation}
S_{quantum} \, \longrightarrow \, \delta s \,  \int \, \Theta^{(2)}_{P_s}
\label{sdeformation}
\end{equation}
For a convenient choice of the constant ${\rm const}$ this deformation is
precisely the
variation of the action (\ref{nfour13}),(\ref{nfour14}),(\ref{nfour15}) under a
shift
$s \, \longrightarrow \, s \, + \, \delta s $ of the $s$ parameters of the
triholomorphic
momentum-map, namely of the N=4 Fayet-Iliopoulos term. These parameters define
the
complex structure of the HyperK\"ahler quotient manifold.
\par
Summarizing, we have seen that the three parameters $r=\zeta^3, s=\zeta^1 \,
+\, i\, \zeta^2$
of the N=4 Fayet-Iliopoulos term are on one hand identified with the
momentum-map levels in the
geometrical HyperK\"ahler quotient construction and, on the other hand, are the
{\it coupling constants} of two topological field-theories: the A-twist selects
the parameters
$r$ that play the role of moduli of the K\"ahler structure, while the B-twist
selects
the $s$ parmaeters that play the role of moduli of the complex structure. It is
an obvious
programme to apply now the topological field-theory framework to the
investigation of the
moduli space of interesting HyperK\"ahler quotient manifolds like the ALE
spaces \cite{kronheimer},
\cite{newalenostro}. This is left to future publications.

\eject
\section{Tables}

{\sl TABLE I }
\par
{\sl N=2 THEORY: SPIN and CHARGES
 before and after the TWISTS}
\vskip 0.2cm
\begin{center}
\begin{tabular}
{||c||c|c|c|c||c|c||c|c||}\hline
\hline
 $ ~$   &   $~$   &  $~$&$~$& $ ~$ &  $ ~$  & $~$  & $~$     & $~$  \\
$ ~$    & $Un-$   &   $~$   &  $~$& $Un-$ &  $ A$  & $A$  & $B$     & $B$  \\
$Field $&$twisted$&$q_{L}$&$q_{R}$ &$twisted$&$ Twist $&$Twist $&$Twist
$&$Twist $  \\
$~$ & $Spin$ & $~$ & $~$& $gh$ & $Spin$ & $gh$ &  $Spin$ & $gh$  \\
$ ~$    & $ ~$   &   $~$   &  $~$& $ ~$ &  $ ~$  & $~$  & $~$     & $~$  \\
\hline
\hline
$c^{gauge}$   &    $0  $ & $ 0 $ & $ 0 $   &$1$&  $0 $ & $ 1 $ &
$ 0 $ & $ 1 $  \\
\hline
${\cal A}_{+}$   & $ -1$ & $ 0 $ & $ 0 $   &$0$& $-1$ & $ 0 $ &
 $ 1 $ & $ 0 $  \\
\hline
${\cal A}_{-}$   & $ +1$ & $ 0 $ & $ 0 $    &$0$& $+1$ & $ 0 $ &
 $-1 $ & $ 0 $  \\
\hline
$\lambda^{+}$   & $-1/2$ & $ 0 $ & $ 1 $   &$0$& $0 $ & $-1 $ &
 $0  $ & $ -1$  \\
\hline
$\lambda^{-}$   & $-1/2$ & $ 0 $ & $ -1$   &$0$& $-1$ & $1  $ &
 $-1 $ & $ 1 $  \\
\hline
$\latp$         & $ 1/2$ & $ 1 $ & $ 0 $   &$0$& $1 $ & $1  $ &
 $0  $ & $-1 $  \\
\hline
$\latm$         & $ 1/2$ & $-1 $ & $ 0 $   &$0$& $0 $ & $-1 $ &
 $1  $ & $ 1 $  \\
\hline
$ M    $        & $ 0  $ & $ 1 $ & $-1 $   &$0$& $0 $ & $ 2 $ &
 $-1 $ & $ 0 $  \\
\hline
$M^* $      & $ 0  $ & $ -1$ & $ 1 $   &$0$& $0 $ & $-2 $ &
 $ 1 $ & $ 0 $  \\
\hline
$\cP     $      & $ 0  $ & $0  $ & $ 0 $   &$0$& $0 $ & $0  $ &
 $ 0 $ & $ 0 $  \\
\hline
$X^{i}$         & $ 0 $  &$-\omega^{i}/d$ &$-\omega^{i}/d$
&$0$& $-\omega^{i}/d$  & $ 0 $ & $ 0 $ & $0  $  \\
\hline
$X^{i*}$    & $ 0  $ & $\omega^{i}/d$ &$\omega^{i}/d$
&$0$&$\omega^{i}/d$ & $ 0 $ & $ 0 $ & $0  $  \\
\hline
$ ~$    & $ ~$   &   $~$   &  $~$&$~$& $ ~$ &  $ ~$  & $~$  & $~$\\
${\ps i}$       &$-1/2$&$(d-\omega^{i})/d$& $-\omega^{i}/d$&
$ 0 $    & $-\omega^{i}/d$& $ 1 $ & $-1 $ & $-1 $  \\
$ ~$    & $ ~$   &   $~$   &  $~$&$~$& $ ~$ &  $ ~$  & $~$  & $~$     \\
\hline
$ ~$    & $ ~$   &   $~$   &  $~$&$~$& $ ~$ &
  $ ~$  & $~$  & $~$\\
${\pst i}$      &$ 1/2$&$-\omega^{i}/d$& $(d-\omega^{i})/d$
& $ 0 $ & $1-\omega^{i}/d  $& $ -1 $& $1 $ & $-1 $  \\
 $ ~$   &   $~$   &  $~$&$~$& $ ~$ &  $ ~$  & $~$  & $~$     & $~$  \\
\hline
$ ~$   &   $~$   &  $~$&$~$& $ ~$ &  $ ~$  & $~$  & $~$     & $non$  \\
${\pss i}$      &$1/2$&$(\omega^{i}-d)/d$& $ \omega^{i}/d$&
$ 0 $ & $-1+\omega^{i}/d$ & $-1 $ & $0$   & $diag.$  \\
$ ~$   &   $~$   &  $~$&$~$& $ ~$ &  $ ~$  & $~$  & $~$     & $~$  \\
\hline
$ ~$   &   $~$   &  $~$&$~$& $ ~$ &  $ ~$  & $~$  & $~$     & $non$  \\
${\psts i}$     &$-1/2$&$\omega^{i}/d$& $(\omega^{i}-d)/d$&
 $ 0 $   & $\omega^{i}/d$& $ 1 $   & $0$ & $diag.$  \\
$ ~$   &   $~$   &  $~$&$~$& $ ~$ &  $ ~$  & $~$  & $~$     & $~$  \\
\hline
\hline
\end{tabular}
\end{center}

\eject

{\sl TABLE II}
\par
{\sl N=4 THEORY: SPIN and CHARGES before and after the TWISTS}
\vskip 0.2cm
\begin{center}
\begin{tabular}
{||c||c|c|c|c||c|c||c|c||}\hline
\hline
$ ~$    & $ ~$   &   $~$   &  $~$    & $ ~$ &
  $ ~$  & $~$  & $~$     & $~$  \\
$ ~$    & $ ~$   &   $~$   &  $~$    & $ ~$ &
  $ A$  & $A$  & $B$     & $B$  \\
$Field $&$Untwisted$&$q_{L}$&$q_{R}$&$Untwisted$&$ Twist $&$Twist
$&$Twist $&$Twist $  \\
$~$ & $Spin$ & $~$ & $~$ & $gh$ & $Spin$ & $gh$ &  $Spin$ & $gh$  \\
$ ~$    & $ ~$   &   $~$   &  $~$    & $ ~$ &  $ ~$  & $~$  & $~$
& $~$  \\
\hline
\hline
$c^{gauge}$   &    $0  $ & $ 0 $ & $ 0 $ & $1 $ &
  $0 $ & $ 1 $ & $ 0 $ & $ 1 $  \\
\hline
${\cal A}_{+}$   & $ -1$ & $ 0 $ & $ 0 $ & $ 0 $ &
 $-1$ & $ 0 $ & $ 1 $ & $ 0 $  \\
\hline
${\cal A}_{-}$   & $ +1$ & $ 0 $ & $ 0 $ & $ 0 $ &
 $+1$ & $ 0 $ & $-1 $ & $ 0 $  \\
\hline
$\lambda^{+}$   & $-1/2$ & $ 0 $ & $ 1 $ & $ 0 $ &
 $0 $ & $-1 $ & $0  $ & $ -1$  \\
\hline
$\lambda^{-}$   & $-1/2$ & $ 0 $ & $ -1$ & $ 0 $ &
 $-1$ & $1  $ & $-1 $ & $ 1 $  \\
\hline
$\latp$         & $ 1/2$ & $ 1 $ & $ 0 $ & $ 0 $ &
 $1 $ & $1  $ & $0  $ & $-1 $  \\
\hline
$\latm$         & $ 1/2$ & $-1 $ & $ 0 $ & $ 0 $ &
 $0 $ & $-1 $ & $1  $ & $ 1 $  \\
\hline
$\mup $         & $-1/2$ & $0  $ & $-1 $ & $ 0 $ &
 $-1$ & $ 1 $ & $-1 $ & $ 1 $  \\
\hline
$\mum $         & $-1/2$ & $0  $ & $ 1 $ & $ 0 $ &
 $0 $ & $-1 $ & $0  $ & $-1 $  \\
\hline
$\mutp $        & $ 1/2$ & $-1 $ & $ 0 $ & $ 0 $ &
 $0 $ & $-1 $ & $1  $ & $ 1 $  \\
\hline
$\mutm $        & $ 1/2$ & $ 1 $ & $ 0 $ & $ 0 $ &
 $1 $ & $ 1 $ & $0  $ & $-1 $  \\
\hline
$ M    $        & $ 0  $ & $ 1 $ & $-1 $ & $ 0 $ &
 $0 $ & $ 2 $ & $-1 $ & $ 0 $  \\
\hline
$M^* $      & $ 0  $ & $ -1$ & $ 1 $ & $ 0 $ &
 $0 $ & $-2 $ & $ 1 $ & $ 0 $  \\
\hline
$ N    $        & $ 0  $ & $ -1$ & $-1 $ & $ 0 $ &
 $-1 $& $ 0 $ & $ 0 $ & $0  $  \\
\hline
$N^* $      & $ 0  $ & $ 1$ &  $ 1 $ & $ 0 $ &
 $ 1 $& $ 0 $ & $ 0 $ & $0 $  \\
\hline
$\cP$           & $ 0  $ & $ 0$ &  $ 0$  & $ 0 $ &
 $ 0 $& $ 0 $ & $ 0 $ & $0  $  \\
\hline
$\cQ$           & $ 0  $ & $ 0$ &  $ 0$  & $ 0 $ &
 $ 0 $& $ 0 $ & $ 0 $ & $0  $  \\
\hline
$u^{i}$         & $ 0  $ & $ 0$ &  $ 0$  & $ 0 $ &
 $ 0 $& $ 0 $ & $ 0 $ & $0  $  \\
\hline
$v^{i}$         & $ 0  $ & $ 0$ &  $ 0$  & $ 0 $ &
 $ 0 $& $ 0 $ & $ 0 $ & $0  $  \\
\hline
$u^{i*}$    & $ 0  $ & $ 0$ &  $ 0$  & $ 0 $ &
 $ 0 $& $ 0 $ & $ 0 $ & $0  $  \\
\hline
$v^{i*}$    & $ 0  $ & $ 0$ &  $ 0$  & $ 0 $ &
 $ 0 $& $ 0 $ & $ 0 $ & $0  $  \\
\hline
$ ~$    & $ ~$   &   $~$   &  $~$    & $ ~$ &
  $ ~$  & $~$  & $~$     & $~$  \\
${\ps i}_{(u,v)}$&$-1/2$ & $1 $ &  $ 0$  &
 $ 0 $ & $ 0 $& $ 1 $ & $-1
$ & $-1 $  \\
$ ~$    & $ ~$   &   $~$   &  $~$    & $ ~$ &
  $ ~$  & $~$  & $~$     & $~$  \\
\hline
$ ~$    & $ ~$   &   $~$   &  $~$    & $ ~$ &
  $ ~$  & $~$  & $~$     & $~$  \\
${\pst i}_{(u,v)}$&$1/2$ & $0 $ &  $1 $  &
 $ 0 $ & $1  $& $-1 $ & $ 1 $
& $-1 $  \\
$ ~$    & $ ~$   &   $~$   &  $~$    & $ ~$ &
  $ ~$  & $~$  & $~$     & $~$  \\
\hline
$ ~$    & $ ~$   &   $~$   &  $~$    & $ ~$ &
  $ ~$  & $~$  & $~$     & $non$  \\
${\pss i}_{(u,v)}$&$-1/2$& $-1$ &  $0 $  &
 $ 0 $ & $-1 $& $-1 $ & $ 0 $ & $diag.$  \\
$ ~$    & $ ~$   &   $~$   &  $~$    & $ ~$ &
  $ ~$  & $~$  & $~$     & $~$  \\
\hline
$ ~$    & $ ~$   &   $~$   &  $~$    & $ ~$ &
  $ ~$  & $~$  & $~$     & $non$  \\
${\psts i}_{(u,v)}$&$1/2$& $0 $ &  $-1$  &
 $ 0 $ & $ 0 $& $ 1 $ & $ 0 $ & $diag.  $  \\
$ ~$    & $ ~$   &   $~$   &  $~$    &
 $ ~$ &  $ ~$  & $~$  & $~$     & $~$  \\
\hline
\hline
\end{tabular}
\end{center}
\eject
{\sl TABLE III}
\par
{\sl Field Content of Pure Topological Y.M. theory}
\begin{center}
\begin{tabular}{||c||c|c|c||}\hline
$ ~$   & $ ~$  &$~$&$~$    \\
$~~~~~~~~~~Form-degree$ & $0$ & $1$ & $2$ \\
$ ~$   & $ ~$  &$~$&$~$    \\~
$Ghost-number~~~~~~~~~~$   & $ ~$  &$~$&$~$    \\~
$ ~$   & $ ~$  &$~$&$~$    \\
\hline
\hline
$ ~$   & $ ~$  &$~$&$~$    \\
$-2$ & ${\bar \phi}$ & $~$ & $~$ \\
$ ~$   & $ ~$  &$~$&$~$    \\
\hline
$ ~$   & $ ~$  &$~$&$~$    \\
$-1$ & ${\bar  c} ~, ~ {\bar \eta} $ & ${\bar \psi}_\mu $ & ${\bar
\chi}_{\mu\nu}$ \\
$ ~$   & $ ~$  &$~$&$~$    \\
\hline
$ ~$   & $ ~$  &$~$&$~$    \\
$0$ & $ b \, , \, L $ & $A_\mu ~, ~ T_\mu $ & $F_{\mu\nu} ~ , ~ B_{\mu\nu}$ \\
$ ~$   & $ ~$  &$~$&$~$    \\
\hline
$ ~$   & $ ~$  &$~$&$~$    \\
$1$ & $c~,~\eta$ & $\psi_\mu $ & $ \chi_{\mu\nu} $ \\
$ ~$   & $ ~$  &$~$&$~$    \\
\hline
$ ~$   & $ ~$  &$~$&$~$    \\
$2$ & $\phi$ & $~$ & $~$ \\
$ ~$   & $ ~$  &$~$&$~$    \\
\hline
\hline
\end{tabular}
\end{center}
\eject


\begin{thebibliography}{}

\bibitem{topolfield} For a review see D. \ Birmingham, M.\ Blau and M.\
Rakowski,
Phys.\ Rep. \ 209 (1991) 129.
\bibitem{Wittenphases} E. \ Witten, Preprint IASSNS-HEP-93/3
\bibitem{topolLGliterature}C. \ Vafa, Mod. Phys. Lett.{\bf A6} (1991) 337;
R.\ Dijkgraaf, E. \ Verlinde and H. \ Verlinde, Nucl.\ Phys.\ B352
(1991) 59,
B.\ Block and A. \ Varchenko, Prepr. IASSNS-HEP-91/5;
E. \ Verlinde and N.\ P.\ Warner, Phys. \ Lett.\ 269B (1991) 96;
A.\ Klemm, S.\ Theisen and M.\ Schmidt, Prepr. TUM-TP-129/91,
KA-THEP-91-00, HD-THEP-91-32;
Z.\ Maassarani, Prepr. USC-91/023;
P.\ Fre' , L.\ Girardello, A.\ Lerda, P.\ Soriani, Preprint SISSA/92/EP.
\bibitem{newalenostro} D. \ Anselmi, M. \ Billo', P. \ Fre', L. \ Girardello,
A. \ Zaffaroni, preprint SISSA/44/93/EP, to appear on IJMP
\bibitem{gravinstanton} T.\ Eguchi and P.G.O. \ Freund, Phys. Rev. Lett. {\bf
37}
(1976) 1251; S.W. \ Hawking, Phys. Lett. {\bf 60A} (1977) 81;
 G.W.Gibbons, S.\ Hawking, Commun. Math. Phys. {\bf
66} (1979) 291;
T.Eguchi, A.J. \ Hanson, Phys. Lett. {\bf 74B} (1978)
249, Ann. Phys. {\bf 120} (1979) 82;
 S.W. \ Hawking, C.N. \ Pope, Nucl. Phys. {\bf B146}
(1978) 381;
 N.J. \ Hitchin, Math. Proc. Camb. Phyl. Soc. {\bf 85}
(1979) 465;
T. \ Eguchi, P.B. \ Gilkey, A.J. \ Hanson, Phys. Rep. {\bf
66} (1980) 213.
\bibitem{CastellaniDauriaFre} L.\ Castellani, R.\ D'Auria, P.\ Fr\`e,
``Supergravity and
Superstrings'', World Scientific, 1991.
\bibitem{FreAnselmi}
D. \ Anselmi, P. \ Fr\`e, Nucl.Phys. {\bf B329} (1993) 401;
D. \ Anselmi, P. \ Fr\`e, Nucl.Phys. {\bf B404} (1993) 288;
D.\ Anselmi, P.\ Fr\`e, preprint, SISSA 73/93/EP,
July, 1993; submitted to Nucl.\ Phys.\ B.
\bibitem{Wittenmirror} E. \ Witten, "Mirror Manifolds and Toplogical Field
Theory" in
S.- T. Yau, ed. ,{\it Essays On Mirror Manifolds} (International Press, 1992).
\bibitem{hklr} N.J.Hitchin, A.Karlhede, U.Lindstr\"om, M.Ro\u{c}ek, Commun.
Math. Phys. {\bf 108} (1987) 535.
\bibitem{lr} U.Lindstr\"om, M.Ro\u{c}ek, Nucl. Phys. {\bf B222} (1983) 285
\bibitem{kronheimer} P.B.Kronheimer, J. Differ. Geometry {\bf 29} (1989) 665;
 P.B.Kronheimer, J. Differ. Geometry {\bf 29} (1989) 685.
\bibitem{Galicki} K.\ Galicki, Comm. Math. Phys. {\bf 108} (1987) 117;
\bibitem{FerPorKounGirar} S. \ Ferrara, C. \ Kounnas, L. \ Girardello, M. \
Porrati,
Phys. Lett.{\bf B194} (1987) 368
\bibitem{calabimetrics} E. \ Calabi, Metriques K\"ahleriennes et fibres
holomorphes,
Ann. Scie. Ec.  Norm. Sup. 12,  269 (1979).
\bibitem{LGliterature} S.\ Cecotti, L.\ Girardello and A. \ Pasquinucci, Nucl.\
Phys.\
B328 (1989) 701 and IJMP A6 (1991) 2427;
N.\ P.\ Warner, Lectures at Trieste Spring school 1988, World
Scientific, Singapore;
E.\ Martinec, Phys.\ Lett. 217B (1989) 431.
For a review see also "Criticality, Catastrophe and
Compactification", V.\ G.\ Knizhnik memorial volume, (1989 );
W.\ Lerche, C.\ Vafa and N.\ P. \ Warner, Nucl.\ Phys.\ B324 (1989) 427;
B.\ Greene, C. \ Vafa and N.\ P.\ Warner, Nucl.\ Phys.\ B324 (1989) 371.
\bibitem{brsformalism} L.\ Beaulieu, M.\ Bellon, Nucl.\ Phys.\ B294 (1987) 279.
\bibitem{ellipgenera} E. \ Witten, Comm. Math. Phys. {\bf 109} (1987) 525;
E. \ Witten, preprint IASSNS-HEP-93/10
P. \ Di Francesco, S. \ Yankielowicz,  preprint SPhT 93/049-TAUP 2047-93
hep-th/9305037;
T. \ Kawai, Y. \ Yamada, S.K. \ Yang, preprint KEK-TH-362, KEK 93-51;
P. \ Di Francesco, O. \ Aharony,  S. \ Yankielowicz,
preprint SPhT 93/068-TAUP 2069-93 hep-th/9306157
\bibitem{topolgauge} E.\ Witten, Comm.\ Math.\ Phys.\ 117 (1988) 353;
 S.\ K.\ Donaldson, J.\ Diff.\ Geom.\ 18 (1983) 279;
 L.\ Beaulieu, I.\ M.\ Singer, Nucl.\ Phys.\ B
(Proc.\ Suppl.) 5B (1988) 12;
\bibitem{topolsigma} E. \ Witten, Comm. Math. Phys. {\bf 118} (1988) 411;
 L. \ Baulieu and E. M. \ Singer, Comm. Math. Phys.{\bf 125} (1989) 125.
\end{thebibliography}
\end{document}